\documentclass[a4paper,11pt]{article}
\usepackage{geometry}
\usepackage{amsmath}
\usepackage{soul}

\usepackage{amssymb}
\usepackage{graphicx}
\graphicspath{{./Figures/}}
\usepackage{color}
\usepackage[colorlinks=true,citecolor=blue,linkcolor=blue,urlcolor=blue]{hyperref}

\usepackage[square,numbers,comma,sort&compress]{natbib}
\usepackage{relsize}
\usepackage{caption}
\usepackage{subcaption}
\usepackage{tikz-feynman}
\tikzfeynmanset{compat=1.1.0}

\def\varabstract{ }
\def\varkeywords{ }
\def\vararxivnumber{ }
\def\vartitle{ }
\def\varpreprint{ }
\renewcommand{\title}[1]{\gdef\vartitle{#1}}
\renewcommand{\abstract}[1]{\gdef\varabstract{#1}}
\newcommand{\keywords}[1]{\gdef\varkeywords{#1}}
\newcommand{\arxivnumber}[1]{\gdef\vararxivnumber{#1}}

\newtoks\authtoks
\renewcommand{\author}[2][]{%
	\authtoks=\expandafter{\the\authtoks#2$^{#1}$\ }%
}
\newtoks\affiltoks
\newcommand{\affiliation}[2][]{%
    \affiltoks=\expandafter{\the\affiltoks{\item[$^{#1}$]#2}}%
}
\newtoks\emailtoks\newcounter{emailcounter}%
\newcommand{\emailAdd}[1]{%
\ifnum\theemailcounter>0\emailtoks=\expandafter{\the\emailtoks, \typeemail{#1}}%
\else\emailtoks=\expandafter{\typeemail{#1}}%
\fi
\stepcounter{emailcounter}}
\newcommand{\typeemail}[1]{\href{mailto:#1}{\tt #1}}

\renewcommand\maketitle{
	\newgeometry{margin=2cm}
	\pagestyle{empty}\setcounter{page}{0}
	\if!\varpreprint!\else\begin{flushright}\varpreprint\end{flushright}\fi
	{\LARGE\flushleft\sffamily\bfseries\vartitle\par}
\vskip6ex
{\large\bfseries\raggedright\sffamily\the\authtoks\par}
\vskip2ex
\begin{list}{}{%
\setlength{\leftmargin}{0.28cm}%
\setlength{\labelsep}{0pt}%
\setlength{\itemsep}{-3pt}%
\setlength{\topsep}{-\parskip}}
\itshape\small%
\the\affiltoks
\end{list}
\vskip2ex
\noindent\hspace{0.28cm}\begin{minipage}[l]{\textwidth}
\begin{flushleft}
\textit{E-mail:} \the\emailtoks
\end{flushleft}
\end{minipage}
\vskip5ex
\noindent{\renewcommand\baselinestretch{.9}\textsc{Abstract:}}\ \varabstract
\vskip5ex 
\if!\varkeywords!\else\noindent{\textsc{Keywords:}}\ \varkeywords \vskip2ex\fi
\newpage
	\newgeometry{margin=2.3cm,top=2.1cm,bottom=2.1cm}
\pagestyle{plain}

\setcounter{footnote}{0}
\restoregeometry
} 

\begin{document}
\title{Cosmic phase transitions: their applications and experimental signatures}
\author[a]{Anupam Mazumdar}
\author[b]{Graham White}
\affiliation[a]{Van Swinderen Institute, University of Groningen, 9747 AG, Groningen, The Netherlands}
\affiliation[b]{TRIUMF, 4004 Wesbrook Mall, Vancouver, British Columbia V6T 2A3, Canada}
\emailAdd{anupam.mazumdar@rug.nl}
\emailAdd{gwhite@triumf.ca}
\date{\today}
\abstract{The study of cosmic phase transitions are of central interest in modern cosmology. In the standard model of cosmology the Universe begins in a very hot state, 
right after the end of inflation via the process of reheating/preheating, and cools to its present temperature as the Universe expands. Both new and existing physics at any scale can be responsible for catalyzing either first, second or cross over phase transition, which could be either thermal or non-thermal with a potential observable imprint. Thus this field prompts a rich dialogue between gravity, particle physics and cosmology. It is all but certain that at least two cosmic phase transitions have occurred - the electroweak and the QCD phase transitions. The focus of this review will be primarily on phase transitions above such scales. We review different types of phase transitions that can appear in our cosmic history,  and their applications and experimental signatures in particular in the context of 
exciting gravitational waves, which could potentially be constrained by LIGO/VIRGO, Kagra, and eLISA.
}

\arxivnumber{1809.xxxxx}

\maketitle

 \section{Introduction}
 
Cosmic phase transitions \cite{Linde:1978px,Kibble:1980mv,Vachaspati:1993pj,Vilenkin:2000jqa} are macroscopic cosmic events so dramatic that they are capable of leaving imprint via  non-adiabatic vacuum fluctuations and creation of particles~\cite{Kofman:1995fi, Tkachev:1998dc, Khlebnikov:1998sz}, formation of defects~\cite{Rajantie:2003xh,Kawasaki:2014sqa,Vachaspati:2006zz,Vilenkin:2000jqa}, generation of magnetic field~\cite{Vachaspati:1991nm,Durrer:2013pga}, generation of baryonic asymmetry~\cite{Kuzmin:1985mm,Shaposhnikov:1987tw,McLerran:1990zh,Farrar:1993sp,Rubakov:1996vz}, and the gravitational wave background \cite{PhysRevLett.69.2026}, and yet their properties are determined by particle physics. Indeed all the macroscopic properties such as the order of the phase transition \cite{Linde:1978px,Kibble:1980mv}, the length of the phase transition \cite{Turner:1992tz}, the latent heat \cite{Enqvist:1991xw}, and the amount of supercooling that occurs \cite{Enqvist:1991xw,Turner:1992tz,Profumo:2007wc} are all controlled by the quantum properties such as effective mass, effective coupling, finite temperature effects  whose masses are near the temperature scale at which the phase transition occurs, for extensive reviews on these topics, see~\cite{Rubakov:1996vz}.

Furthermore, the particle physics responsible for a cosmic phase transition can potentially occur at any scale: from the QCD scale $\sim 10^{-1}$ GeV \cite{RAJAGOPAL1993395} right up to the GUT scale $\sim 10^{15}$ GeV \cite{Dimopoulos:1981zb}. Thus, any cosmic phase transition might shed light on particle physics occurring at scales that are potentially out of reach to both present and future particle colliders. Besides, thermal phase transitions, non-thermal phase transitions can also occur. Usually, they occur after the end of inflation, prompted by the inflationary sector or some hidden sector~\cite{Kofman:1995fi, Tkachev:1998dc, Khlebnikov:1998sz}. Such phase transitions do not require initial conditions to be set by thermal initial conditions.

It is likely that at least two transitions have occurred: the QCD transition between a quark gluon plasma and a hadron gas \cite{RAJAGOPAL1993395}, and the electroweak transition \cite{PhysRevD.9.3357} in which electroweak symmetry was spontaneously broken allowing for the Standard Model (SM) particles to acquire gauge invariant masses \cite{PhysRevLett.13.508,PhysRev.130.439}. The latter phase transition may be involved in generating baryon asymmetry of the Universe via the electroweak (EW) baryogenesis mechanism \cite{Farrar:1993sp,Farrar:1993hn,PhysRevD.36.581,1402-4896-1991-T36-019,KUZMIN198536}. Both of these phase transitions can leave an observable signature in the gravitational wave background \cite{Kamionkowski:1993fg,Witten:1984rs,Caprini:2015zlo} as well as affecting both the cold and hot dark matter backgrounds \cite{Boyle:2005se}. Other phase transitions are also possible: examples of which include the spontaneous breaking of the gauge symmetries of Grand Unified Theories (GUT) giving rise to inflation \cite{SATO198166,PhysRevLett.48.1220,Guth:1980zm}, and the production of dark matter through a strongly first order phase transition \cite{Seto:2003kc}.

The nature of these phase transitions not only affect any observable relic footprints they might leave, but also their utility. For example, if a relic gravitational background suggests that the electroweak phase transition occurred through bubble nucleation then that could be a sign that the baryon asymmetry of the Universe was produced during this phase transition. First order phase transitions also generate primordial magnetic field during the turbulence phase of the plasma and bubble collision~\cite{Caprini:2006jb}, see a very nice review~\cite{Durrer:2013pga}, and in some cases they may generate domain walls and strings~\cite{Vilenkin:2000jqa,Kofman:1995fi,Kasuya:1998td,Hindmarsh:2016lhy}, as it happens in the case of Next to Minimal Supersymmetric Standard Model (NMSSM)~\cite{Ellwanger:2009dp,Kawasaki:2014sqa}.

Furthermore the ingredients behind cosmic phase transitions have many discovery avenues. Observation of proton decay would give strong evidence of grand unification \cite{Lazarides:1980nt,LANGACKER1981185} which is an ingredient in a GUT phase transition, discovery of new scalar particles would give more information about the electroweak phase transition or new massive gauge bosons could indicate a dark phase transition.  Finally the principles behind cosmic phase transitions can in principle be tested in condensed matter systems which can imitate cosmological situations for a given Lagrangian \cite{0953-8984-25-40-400301}. 

 \section{Basic cosmological overview}

Let us briefly summarize the early Universe cosmology in chronological order. How the Universe began remains a profound question, for which we do not have direct experimental evidence yet. Nonetheless, we can speculate based on sound physical arguments and the observations confirmed by the detection of cosmic microwave background (CMB) radiation~\cite{Hinshaw:2012aka,Aghanim:2018eyx}.


\subsection{ How did the Universe begin?}

Einstein's theory of gravity (GR) is extremely successful in the infrared (IR) matching of all possible observables~\cite{Will:2014kxa}, including the recent discovery of gravitational waves from mergers of two blackholes~\cite{Abbott:2016blz}, and binary neutron star mergers~\cite{TheLIGOScientific:2017qsa}. However at short distances and small time scales, i.e. in the ultraviolet (UV), GR has pathologies, besides being a non-renormalizable theory, GR introduces cosmological and blackhole singularities, see~\cite{Hawking:1973uf}, and in some cases naked singularities, see~\cite{Joshi:2008zz}. In GR, our Universe has a distinct starting point, a singular spacetime - as long as all the standard energy conditions are always satisfied, i.e. strong, weak, and null energy conditions, see ~\cite{Hawking:1973uf}. It is possible to address the cosmological singularity problem without violating the matter energy conditions by weakening the gravitational interaction in the 
UV. This can happen in ghost free infinite derivative gravity inspired from string field theory~\cite{Biswas:2005qr, Biswas:2011ar}. There could be two consequences for such study; one could be a realization of a  non-singular bounce~\cite{Biswas:2010zk,Biswas:2012bp}, and the other scenario would be that Universe could be frozen in time in the UV, such that the Universe becomes conformal as  $t\rightarrow 0$~\cite{Biswas:2006bs}. Bouncing cosmologies and cosmological density perturbations have been reviewed in this nice review~\cite{Brandenberger:2016vhg,Battefeld:2014uga}. There is a strong indication that this non-singular initial phase of the Universe has a  key role to play towards understanding the subsequent phases of the Universe such as cosmic inflation, horizon, homogeneity and isotropy of the Universe , to create appropriate initial conditions for the Universe . In this review we will not discuss any further the very genesis of the Universe , we will merely assume that Universe is homogeneous and isotropic from the very beginning. We will  discuss inflationary cosmology very briefly, but inflationary cosmology has its own limitations, when it comes to explaining the initial condition problem - it cannot solve or address the initial singularity problem~\cite{Borde:2001nh}. 


\subsection{Cosmic inflation to reheating}

\subsubsection{\bf  Primordial inflation:}
A phase of primordial inflation addresses some of the key challenges of  the hot big bang cosmology, such as the flatness and the horizon problems, i.e. generating the large scale structures on roughly $4000$~Mpc scale, and  the age of the Universe $\sim 13.8$~Gyrs~\cite{Guth:1980zm,Linde:1981mu,Albrecht:1982wi,Sato:1981ds,Starobinsky:1982ee,Starobinsky:1980te}. Cosmic inflation is a very successful paradigm, and we know a lot more about inflationary predictions than the pre-inflationary phase of the Universe , such as big bounce or freeze-in phase of the Universe . Within the inflationary paradigm, with the help of Planck satellite we could see only the 8 e-folding of inflation around the pivot scale $0.05{\rm Mpc}^{-1}$~\cite{Aghanim:2018eyx}.
 
 One of the key predictions of inflation remains that of stretching the long wavelength quantum fluctuations on dark matter on scales larger than the size of the Hubble patch during inflation~\cite{Mukhanov:1990me,Mukhanov:1981xt,Bardeen:1983qw,Kodama:1985bj}, in order to match the current observations, in the temperature anisotropy in CMB, which has been observed quite  precisely by a number of space based missions, starting from  COBE~\cite{Smoot:1992td}, WMAP~\cite{Hinshaw:2012aka}, and now by Planck~\cite{Aghanim:2018eyx}. The other key predictions for inflation is that by inflating the scale factor of the Universe , it makes the spatial curvature of the Universe flatter and flatter. As a consequence the quantum fluctuations in the scalar field becomes very close to the Gaussian random fluctuations, with almost scale invariant perturbations. Indeed, the epoch of inflation has to come to an end, and this yields a slight departure from the scale invariant perturbations, which we we have inferred from flat-$\Lambda CDM$ paradigm, where flat signifies the spatial flatness of the Universe , $\Omega_k$, see \cite{Guth:1980zm}.  Note that $\Lambda $ stands for late cosmological constant - the fact that the current Universe appears to be accelerating, as it is evident from dark energy surveys~\cite{Abbott:2017wau}, and CDM stands for the cold dark matter paradigm, which is a non-relativistic, potentially non-baryonic form of matter. The CDM is required to form the very first stars and structures in the Universe , effectively with zero equation of state parameter, i.e., pressure-less fluid. There are compelling paradigms other than $\Lambda CDM$, such as modified Newtonian gravity~\cite{Famaey:2011kh}, which might as well lead to structure formation~\cite{Dayal:2018hft} and explain some of the outstanding observations of the baryonic physics in the neighborhood of our galaxy, but here in this review we will limit ourselves to $CDM$ paradigm, other variants of CDM are  warm dark matter (WDM)~\cite{Dayal:2014nva}, or fuzzy dark matter scenarios~\cite{Hui:2016ltb}, which can ameliorate some of the problems related to CDM in the structure formation simulations versus observations, see~\cite{Weinberg:2013aya}.

Cosmic inflation within GR can be explained in a very simple manner. Inflation happens along a flat direction with a non-negligible slope of the potential, given by the order parameter, $\phi$, as an inflaton, and its potential, $V(\phi)$. Inflation could last eternally in future due to stochastic fluctuations of the scalar field in almost deSitter background~\cite{Starobinsky:1986fx,Starobinsky:1994bd}, nevertheless, in a given observable Hubble patch, it has to come to an end. Note that inflation cannot be past eternal unless the singularity problem can be resolved~\cite{Biswas:2009fv}. The current data, from Planck~\cite{Akrami:2018odb}, at best can probe the second derivative of the potential and not beyond that. The inflationary predictions~\cite{Akrami:2018odb} are compatible with slow-roll inflation, which assumes that the potential dominates over the kinetic energy  $\dot\phi^2 \ll V(\phi)$, and $\ddot \phi \ll V^{\prime}(\phi)$,  where dots are w.r.t physical time, $t$, therefore the Friedmann and the Klein-Gordon equations are approximated as:
\begin{eqnarray}
\label{slowr1}
H^2 \approx  \frac{V(\phi)}{3M_{\rm P}^2}\,,~~~~~~~~~~~~~~~~
3H\dot\phi \approx -V^{\prime}(\phi)\,,
\end{eqnarray}
where prime denotes derivative with respect to $\phi$. 
The slow-roll conditions, which parametrize the shape of the potential,
are given by:
\begin{eqnarray}
\label{ep1}
\epsilon(\phi)\equiv \frac{M_{\rm P}^2}{2}\left(\frac{V^{\prime}}{V}\right)^2
\ll 1\,,~~~
|\eta(\phi)| \equiv {M_{\rm P}^2}\left |\frac{V^{\prime\prime}}{V}\right| \ll 1\,.
\end{eqnarray}
Note that the  slow-roll conditions are violated when
$\epsilon \sim 1$, and $\eta \sim 1$, which marks the end of inflation. The number of e-foldings can be defined 
between, $t$, and the end of inflation, $t_{end}$:
\begin{equation}
N\equiv \ln\frac{a(t_{end})}{a(t)}=\int_{t}^{t_{end}}Hdt\approx
\frac{1}{M_{\rm P}^2}\int_{\phi_{end}}^{\phi}\frac{V}{V^{\prime}}d\phi\,,
\end{equation}
where $\phi_{end}$ is defined by $\epsilon(\phi_{end})\sim 1$, provided
inflation comes to an end via a violation of the slow-roll conditions.
The number of e-foldings can be related to the Hubble crossing
mode $k=a_{k}H_{k}$ by comparing with the present Hubble length $a_{0}H_{0}$.
The final result is \cite{Liddle:2003as,Burgess:2005sb}
\begin{eqnarray}
\label{efoldsk}
N(k)= 62-\ln\frac{k}{a_0H_0}-\ln\frac{10^{16}{\rm GeV}}{V_{k}^{1/4}}+
\ln\frac{V_{k}^{1/4}}{V_{end}^{1/4}}
-\frac{1}{3}\ln\frac{V_{end}^{1/4}}
{\rho_{R}^{1/4}}\,,
\end{eqnarray}
where the subscripts $end$ ($\rm R$) refer to the end of inflation (end of
reheating). Today's Hubble length would correspond to $N_Q\equiv N(k=a_0H_0)$ number of 
e-foldings, whose actual value would depend on the equation of state, i.e. $\omega=p/\rho$ ($p$ denotes the pressure, 
$\rho$ denotes the energy density), from the end of 
inflation to radiation and matter dominated epochs.  A high scale inflation with a prompt reheating with relativistic species would yield
approximately, $N_Q\approx 50-60$. A significant modification in the number of e-foldings can take place.  If the scale of inflation is low, and if 
the reheat temperature is as low as that of $1$~MeV, roughly the temperature before the Big Bang Nucleosynthesis (BBN), for a review~\cite{Cyburt:2015mya}, the number of e-foldings 
to explain the horizon and the flatness problem could be as low as $\sim 25$, see~\cite{Lyth:1999ty,Mazumdar:1999tk,Mazumdar:2001ya,ArkaniHamed:1999gq,Green:2002wk}.

For a single field slow-roll inflation there exists a
late time attractor behaviour, such that  the evolution of a scalar field after sufficient e-foldings become
independent of the initial conditions \cite{Salopek:1990jq,Kofman:2002cj}. This particular initial condition is solely related to the homogeneous
inflaton and its initial velocity, and has nothing to do with the initial homogeneity and isotropy of the Universe . Inflation as such does not
solve these problems, there are obstructions to that within GR due to focusing theorems due to Raychaudhury~\cite{Raychaudhuri:1953yv}, and Hawking-Penrose
singularity theorems~\cite{Penrose:1964wq,Hawking:1973uf}. The initial patch of the Universe should be homogeneous on scales larger than the inflating 
Hubble patch~\cite{Vachaspati:1991nm}, similarly, in order to inflate such a patch, the patch should be already isotropic. In this regard inflation assumes homogeneity and isotropy of spacetime within GR~\cite{Farhi:1986ty,Albrecht:1985yf}.


\subsection{Primordial perturbations}\label{SP}

\subsubsection{\bf Scalar perturbations:}

The small inhomogeneities in the inflaton field can be recast as, $\phi(\vec x,t) =\phi(t)+\delta\phi(\vec x,t)$, where $\delta\phi\ll\phi$,
is the inflaton perturbations in the background metric. During inflation $\delta \phi$ are stretched outside the Hubble patch, because $m^{2}\sim V^{''}\ll H^{2}$.
These fluctuations can then be tracked from a sub-Hubble to that of a super-Hubble length scales right when the wave numbers 
have crossed the Hubble patch, these fluctuations are random Gaussian, and can be given by:
\begin{equation}
\label{pertphi}
\langle |\delta\phi_{k}|^2\rangle=({H(t_{\ast})^2}/{2k^3})\,,
\end{equation}
where $t_\ast$ denotes the instance of Hubble crossing. One can define a power spectrum for
the perturbed scalar field
\begin{equation}
\label{spect}
{\cal P}_{\phi}(k)=\frac{k^3}{2\pi^2}\langle |\delta\phi_{k}|^2\rangle=
\left[\frac{H(t_{\ast})}{2\pi}\right]^2 \equiv
\left. \left[\frac{H}{2\pi}\right]^2\right|_{k=aH}\,.
\end{equation}
Note that the phase of $\delta\phi_{k}$ can be arbitrary, and therefore,
inflation has generated a Gaussian perturbation. We  can calculate the power spectrum for the metric perturbations, this is what we observe in the CMB, translated 
into temperature anisotropy. Since, the separation between the background metric
and a perturbed metric is not unique, a choice of gauge, or a choice of a particular coordinate system becomes necessary to simplify the 
metric perturbations. One particular choice would be to fix the gauge where the non-relativistic limit of the full perturbed Einstein equation can be recast as
a Poisson equation with a Newtonian gravitational potential, $\Phi$. The induced metric perturbations 
can be written  in GR as, e.g.~\cite{Mukhanov:1990me}:
\begin{equation}
\label{gauge}
ds^2=a^2(\tau)\left[(1+2\Phi)d\tau^2-(1-2\Psi)\delta_{ik}dx^{i}dx^{k}\right]\,,
\end{equation}
and when the spatial part of the energy momentum tensor is diagonal,
i.e. $\delta T^{i}_{j}=\delta^{i}_{j}$, it follows that $\Phi=\Psi$. For a critical density Universe , i.e. for a flat Universe , $\delta_{k}\equiv \left.\frac{\delta\rho}{\rho}\right|_{k}$, the power spectrum is given by;
\begin{equation}
\label{pspect}
\delta_{k}^2\equiv \frac{4}{9}{\cal P}_{\Phi}(k)\, =\frac{4}{9}\frac{9}{25}
\left(\frac{H}{\dot\phi}\right)^2\left(\frac{H}{2\pi}\right)^2\,,
\end{equation}
where the right hand side can be evaluated at the time of horizon exit
$k=aH$. It is convenient for the observers to express the perturbations  in a different gauge, known as the 
comoving gauge, where on the comoving hypersurface the energy flux vanishes, and the amplitude is denoted by 
$\zeta_{k}$~\cite{Lukash:1980iv,Mukhanov:1990me}. The comoving curvature perturbation, 
$\zeta_{k}$ is a conserved quantity for the super-Hubble modes, $k\rightarrow 0$, and $\zeta_{k}=-(5/3)\Phi_{k}$. Therefore, 
$\delta_{k}$ can also be expressed in terms of the curvature perturbations \cite{Liddle:2000cg}
$\delta_{k}=\frac{2}{5}\left(\frac{k}{aH}\right)^2{\zeta}_{k}$.
The corresponding power spectrum
$\delta_{k}^2=({4}/{25}){\cal P}_{\zeta}(k)=({4}/{25})(H/\dot\phi)^2(H/2\pi)^2$.
With the help of the slow-roll equation $3H\dot\phi=-V^{\prime}$,
and the critical density formula $3H^2M^2_{\rm P}=V$, one obtains
\begin{eqnarray}\label{eq:powerspectrum1}
\delta_{k}^2 \approx \frac{1}{75\pi^2 M_{\rm P}^6}\frac{V^3}{V^{\prime 2}}\,
=\frac{1}{150\pi^2 M_{\rm P}^4}\frac{V}{\epsilon}\,,
{\cal P}_{\zeta}(k) = \frac{1}{24\pi^2M_{\rm P}^4}\frac{V}{\epsilon}\,,
\end{eqnarray}
where we have used the slow-roll parameter
$\epsilon\equiv (M_{\rm P}^2/2)(V^{\prime}/V)^2$.
If we assume that the primordial 
spectrum can be approximated by a power law, see~\cite{Akrami:2018odb}
\begin{equation}\label{eq:primordialpowerspectrum}
{\cal P}_{\zeta}(k)\simeq (3.044\pm 0.014)\times 10^{-10}\left(\frac{k}{k_0}\right)^{n_s-1}~,
\end{equation}
where $n_s$ is called the spectral index (or spectral tilt), the reference 
scale is: $k_0=7.5a_{0}H_{0}\sim 0.002~{\rm Mpc}^{-1}$, and the error bar on the 
normalization is given by the characterization of polarization at low and high multipoles, Planck temperature, polarization, 
and lensing data yields at $68\%$ CL~\cite{Akrami:2018odb}
\begin{equation}
 n_s(k_0)=0.9649\pm 0.0042
\end{equation}
In the slow-roll approximation, this tilt can be expressed in terms of the 
slow-roll parameters and at first order:
\begin{equation}
\label{spectind4}
n_s-1=-6\epsilon +2\eta +\mathcal{O}(\epsilon^2,\eta^2,\epsilon\eta,\xi^2)~,
\end{equation}
where
\begin{equation}
\xi^2\equiv M_{\rm P}^4\frac{V^{\prime}(\mathrm{d}^3V/\mathrm{d}\phi^3)}{V^2}\,,
\quad\sigma^3\equiv M_{\rm P}^6\frac{V^{\prime 2}(\mathrm{d}^4V/\mathrm{d}\phi^4)}{V^3}\,.
\end{equation}
The running of these parameters are given by~\cite{Salopek:1990jq}. Since the
slow-roll inflation requires that $\epsilon \ll 1, |\eta|\ll 1$, therefore naturally predicts small
variation in the spectral index within $\Delta \ln k\approx 1$~\cite{Kosowsky:1995aa}
\begin{equation}
\frac{\mathrm{d} n(k)}{\mathrm{d}\ln k}=-16\epsilon\eta+24\epsilon^2+2\xi^2\,.
\end{equation}
There is no evidence of scale dependence of $n_s$ has been found by the latest Planck data~\cite{Akrami:2018odb}.

\subsubsection{\bf Tensor perturbations:}

Like scalar field induced metric perturbations during inflation, we would also expect pure stochastic gravitational 
waves \cite{Grishchuk:1974ny,Grishchuk:1989ss,Allen:1987bk,Sahni:1990tx,Starobinsky:1979ty}. 
For reviews on gravitational waves, see
\cite{Mukhanov:1990me,Maggiore:1999vm}.
The gravitational wave perturbations are described by a line element
$ds^2+\delta ds^2$, where
\begin{equation}
{d}s^2=a^2(\tau)(\mathrm{d}\tau^2-\mathrm{d}x^{i}\mathrm{d}x_{i})\,,\quad \delta 
{d}s^2=-a^2(\tau)h_{ij} \mathrm{d}x^{i} \mathrm{d}x^{j}\,.
\end{equation}
The  $3$-tensor $h_{ij}$ is symmetric,
traceless $\delta^{ij}h_{ij}=0$, and divergenceless $\nabla_{i}h_{ij}=0$   
($\nabla_{i}$ is a covariant derivative), and also gauge and conformally invariant. Massless spin
$2$ gravitons have two transverse  degrees of freedom ({\it d.o.f})
For the Einstein's GR, the gravitational wave equation of motion follows
that of a massless Klein Gordon equation \cite{Mukhanov:1990me}. Especially,
for a flat Universe 
\begin{equation}
\ddot h^{i}_{j}+3H\dot h^{i}_{j}+\left({k^2}/{a^2}\right)h^{i}_{j}=0\,,
\end{equation}
As  any massless field, the gravitational waves also feel the quantum
fluctuations in an expanding background. The spectrum mimics that of
Eq.~(\ref{spect})
\begin{equation}
{\cal P}_{\rm grav}(k)=\left.\frac{2}{M_{\rm P}^2}\left(\frac{H}{2\pi}\right)^2
\right|_{k=aH}\,.
\end{equation}
The corresponding
spectral index can be expanded in terms of the slow-roll parameters at first 
order as
\begin{equation}
r\equiv \frac{{\cal P}_{\rm grav}}
{ {\cal P}_{\zeta}}=16\epsilon~,\quad 
n_t=\frac{\mathrm{d}\ln{\cal P}_{\rm grav}(k)}{\mathrm{d}\ln k}\simeq -2\epsilon,\,.
\end{equation}
Note that the tensor spectral index is negative, in some sense gravitational waves spectrum is solely determined by the the Hubble expansion rate during 
inflation and the initial vacuum condition. Relaxing the initial vacuum condition may lead to different predictions in the value of tensor-to-scalar ratio, $r$, see~\cite{Ashoorioon:2012kh,Ashoorioon:2014nta}. A classical initial condition
can also produce $r$, albeit the magnitude will be very tiny~\cite{Ashoorioon:2012kh}. So, non detection of primordial gravitational waves does not confirm the quantum nature of gravitons in CMB based experiments. The latest constraint on the tensor to scalar ratio is given by the Planck upper limit $95\%$CL is $r < 0.1$, which is further tightened by BICEP2/Keck Arracy BK14 data $r <0.064$~\cite{Akrami:2018odb}.


\subsection{Reheating phase}

There is no dearth of models of inflation which can potentially match the current set of observations in CMB, for a review see~\cite{Mazumdar:2010sa}. Within particle physics, typically inflation is assumed to be driven by SM gauge singlets, either driven by a single or multiple fields, such as hybrid~\cite{Linde:1993cn}, or infinitely many, assisted inflation~\cite{Liddle:1998jc,Copeland:1999cs}. However, well motivated particle physics models are SM driven Higgs inflation~\cite{Bezrukov:2007ep}, which requires unnatural coupling between the SM Higgs and the Ricci scalar, which leaves the model very similar to the Starobisnky's model of inflation of $R+\alpha R^2$~\cite{Starobinsky:1980te}, after one-loop computation~\cite{Salvio:2015kka}. Note that in the original paper the relative sign difference was negative, and the motif was to obtain a bouncing Universe , but with a ghost in the spin-2 sector. Amongst the well-motivated models of inflation, Starobinsky's model remains very minimal in content and driven purely within the gravitational sector. Its UV completion has been given in~\cite{Biswas:2005qr,Biswas:2013dry,Koshelev:2016xqb}. There are also other well-motivated models of inflation within particle physics, where the inflaton can be recognized by the supersymmetric partners of quarks and leptons, namely the gauge invariant combination of squarks and sleptons carrying the SM gauge charges within minimal supersymmetric standard model (MSSM)~\cite{Allahverdi:2006iq,Allahverdi:2006we}, for a review of MSSM see~\cite{Haber:1984rc,Chung:2003fi}. However, there has been no evidence of 
supersymmetric partners at the LHC, which has constrained the scale of inflation above the 3-4 TeV scale~\cite{Wang:2013qti}. These models, i.e. MSSM and Higgs inflation, are also known as visible sector models of inflation, because the inflaton directly decays into visible sector d.o.f. All the Yukawas and gauge couplings are well known in these models.
In this review we will not delve into model building of inflation any further, as this is not so relevant for this discussions below, but we will now divert to how the inflaton decays and thermalizes the Universe .

Typically, inflation ends via smooth phase transition as we had discussed, i.e., by violation of slow roll conditions.  Inflation could also end with tachyonic instability~\cite{Linde:1993cn}, in some cases inflation can be driven by tachyons as well~\cite{Mazumdar:2001mm}, or via tunneling from the inflationary vacuum to the SM type vacuum~\cite{Guth:1980zm}.  Unfortunately, in the Guth's model the bubble never thermalizes, 
the bubble wall is still expanding in the deSitter background, and inside the bubble the SM Universe is super cooled. The SM vacuum needs to be thermalized by the collision of $2$ or more bubbles, which never takes place if the bubble nucleation rate is smaller than 
the Hubble expansion rate of the deSitter. Depending on the gauge group, the phase where inflation comes to an end can create topological defects, see~\cite{Vilenkin:1994,Vachaspati:1993pj,Vachaspati:2006zz}, such as cosmic strings, or domain walls, etc., however, these are very much model dependent. The phase after inflation leads to reheating and preheating. Non-peturbative preheating can give rise to $1^{st}$ order phase transitions~\cite{Khlebnikov:1998sz,Kofman:1995fi}, gravitational waves~\cite{Khlebnikov:1997di,GarciaBellido:1998gm,GarciaBellido:2007af,GarciaBellido:2007dg,Mazumdar:2008up,Dufaux:2010cf,Fenu:2009qf,Kusenko:2008zm,Kusenko:2009cv}, magnetic fields~\cite{DiazGil:2008tf}, topological defects~\cite{Tkachev:1998dc,Rajantie:2003xh,Kasuya:1998td,Kawasaki:2014sqa}, and non-topological solitons~\cite{Enqvist:2002si,Enqvist:2002rj}.

Topological defects are another consequences of phase transitions, it was  Kibble~\cite{Kibble:1976sj,Kibble:1980mv} and Zurek~\cite{Zurek:1985qw}  who independently postulated the formation of topological defects during cosmic phase transition. The topological defect is also known as a solitonic solution in quantum field theory which are homotopically distinct from the vacuum solution. In topology, two continuous functions if they can be continuously deformed into each other, then such deformations are known as homotopy between the two functions. If not, then they are homotopically distinct functions. The latter produces defects, kinks, cosmic strings, cosmic textures,  and also Dirac monopoles, for a review see~\cite{Vilenkin:2000jqa,Vachaspati:2006zz}.   Besides, topological solitons, there are also non-topological solutions in any interacting field theory where the boundary conditions at infinity are the same as that of the vacuum state~\cite{Lee:1991ax}, for example Q-ball, a detailed review of Q-balls, see~\cite{Enqvist:2003gh,Dine:2003ax}.


\subsubsection{\bf Perturbative decay and thermalization phase:}
\label{PDT}

During inflation, the Universe is cold and devoid of any thermal entropy. It is thus paramount to create a thermal bath, which can at least achieve 
local thermodynamical equilibrium (LTE), means that the species can be in thermal equilibrium as long as $\Gamma \geq H(t)$, where $\Gamma$ denotes 
the interaction rate and $H(t)$ is the Hubble expansion rate. Note that $\Gamma$ is solely determined by the particle physics interaction rate at a given energy, temperature,
while $H(t)$ is the Hubble expansion rate of the Universe . For the species in LTE, the energy density,
$\rho$, and the number density, $n$, for relativistic particles are
given by~\cite{Kolb:1988aj}
\begin{eqnarray} \label{full}
\rho &=& \left({\pi^2/30}\right) T^4\,, ~~~~~~~~~~n = 
\left({\zeta(3)/\pi^2}\right)T^3\,, ~~~~~~~~~~~~ ({\rm Boson}) \, , 
\nonumber \\ 
\rho &=& \left({7/8}\right) \left({\pi^2/30}\right) T^4\,, ~~~ 
n = \left({3/4}\right) \left({\zeta(3)/\pi^2}\right) T^3\,, ~~~~ 
({\rm Fermion}) \, ,
\end{eqnarray}
where $T$ is the temperature of an ambient bath, shared by all the species present in the bath. Typically, the average energy of every species 
will be shared $\langle E \rangle \sim \rho^{1/4}$, and $n \sim \rho^{3/4}$ hold, with $\langle E \rangle = \left(\rho/n
\right) \simeq 3 T$ being the average particle energy.

On the other hand, right after the inflaton has decayed, the energy density
of the Universe is determined by the total decay width, $\Gamma_{\rm d}$, of the inflaton to the relativistic species, 
$\rho \approx 3 \left(\Gamma_{\rm d} M_{\rm P}\right)^2$. The ambient plasma has a thermal entropy, 
given by: $\langle E \rangle \approx m_{\phi} \gg \rho^{1/4}$. Then, 
 the total number density is roughly given by $n \approx \left(\rho/m_{\phi}\right) \ll \rho^{3/4}$.
Note that the initial energy density $\rho$ is always bounded below the energy density of the inflaton energy, i.e. $\rho \leq 3H^2M_{\rm p}^2$.
Therefore, the decay products which creates the ambient plasma results in a very dilute plasma, the number density of the decayed products is very tiny,
though the energy of the decayed particles can be as large as that of the inflaton mass, i.e. $m_\phi$. This suggests that the initial plasma is 
far from full thermal equilibrium initially~\cite{Turner:1983he,Davidson:2000er,Allahverdi:2005mz,Kofman:1997yn,Kofman:1994rk,Shtanov:1994ce,Traschen:1990sw,Abbott:1982hn}. 

Reaching full equilibrium requires re-distribution of the energy among
different particles, {\it kinetic equilibrium}, as well as increasing
the total number of particles, {\it chemical equilibrium}. Therefore,
both the number-conserving and the number-violating reactions must be taken into account.
Kinetic equilibrium can be achieved  by $2 \rightarrow 2$ scatterings with gauge boson exchange
in the $t$-channel~\cite{Davidson:2000er,Allahverdi:2005mz}. While the chemical equilibrium is achieved by changing
the number of particles in the reheat plasma. It was recognized in~\cite{Davidson:2000er}, see 
also~\cite{Allahverdi:2005mz}, that the most
relevant processes are $2 \rightarrow 3$ scatterings with gauge-boson
exchange in the $t-$channel. The latter is the inelastic scattering, when this become efficient, the scattering rate exceeds that of 
the Hubble expansion rate, and the number of particles also
increases very rapidly~\cite{Enqvist:1993fm}, due to the fact that the produced gauge bosons
subsequently participate in similar $2 \rightarrow 3$ scatterings.
During this phase, decays of particles can also be considered, but they do not play an important role, they cannot
 increase the number of particles to the required level. 
The full thermal LTE is established
shortly after $2 \rightarrow 3$ scatterings become efficient. For
this reason, to a very good approximation, one can use the rate for
inelastic scatterings as a thermalization rate of the Universe 
$\Gamma_{\rm thr}$.  If the inflaton decay products have SM like
gauge interactions, i.e. relatively large gauge interactions, then the Universe reaches full thermal 
equilibrium quite quickly, the main reason is that the $2
\rightarrow 3$ scatterings with gauge boson exchange in the
$t-$channel are indeed very efficient, see~\cite{Davidson:2000er,Allahverdi:2005mz,Croon:2015naa}.
During this phase of thermalization one can produce massive
long--lived or stable weakly interacting massive particles (WIMPS), or long lived feebly interacting massive particles (FIMPS)
\cite{Chung:1998zb,Chung:1998ua,Chung:1998rq,Chung:2001cb,Giudice:2000dp,Giudice:2000ex,Dev:2013yza,Mazumdar:2013gya}.

A rough estimate of the reheat temperature can be made. The release of the inflaton energy density
into the thermal bath of relativistic particles take place
when $H(a)=\sqrt{(1/3M_{\rm P}^2)\rho_{i}}(a_{i}/a)^{3/2}\approx \Gamma_{\phi}$.
The energy density of the thermal bath is determined by the reheat
temperature $T_{R}$,  or the temperature of the relativistic bath is given by:
\begin{equation}
T_{R}=\left(\frac{90}{\pi^2 g_{\ast}}\right)^{1/4}\sqrt{\Gamma_{\phi}
M_{\rm P}}=0.3\left(\frac{200}{g_{\ast}}\right)^{1/4}\sqrt{\Gamma_{\phi}
M_{\rm P}}\,,
\end{equation}
where $g_{\ast}$ denotes the effective relativistic degrees of freedom in the
plasma. However the inflaton might not decay instantaneously. In such a case
there might already exist a thermal plasma of {\it some} relativistic species
at a temperature higher than the reheat temperature already before the
end of reheating~\cite{Kolb:1988aj}. If the inflaton decays with a
rate $\Gamma_\phi$, then the instantaneous plasma temperature is found
to be~\cite{Kolb:1988aj}:
\begin{equation}
\label{instT}
T_{inst}(a)\sim \left(g_{\ast}^{-1/2}H\Gamma_{\phi}M_{\rm P}^2\right)^{1/4}\,.
\end{equation}
The temperature of the Universe reaches its maximum $T_{max}$ soon after the inflaton
field starts oscillating around the minimum. Once the maximum temperature is
reached, then $\rho_{\psi} \sim a^{-3/2}$, and $T\sim a^{-3/8}$ until
reheating and thermalization is completely over. Thermalization is achieved when both {\it kinetic} and
{\it chemical} equilibrium are reached. 

For a successful cosmology one needs to ask how the inflaton energy gets 
converted into the SM degrees of freedom. For large reheat temperatures, $T_{\rm R}\sim 10^{9}$~GeV, the Universe  could abundantly 
create thermal relics of unstable gravitinos with a mass of order $100-1000$~GeV, which could spoil the success of 
BBN~\cite{Ellis:1984eq,Moroi:1993mb,Moroi:1995fs,Bolz:2000fu,Kawasaki:2004yh,Kawasaki:2004qu} (for the effects of lighter unstable relics see \cite{Forestell:2018txr,Hufnagel:2018bjp}).
For extremely low reheat temperatures, i.e. $T_{\rm R}\sim {\cal O}(1-10)$~MeV, it becomes a great challenge to 
obtain matter-anti-matter asymmetry and the right abundance for the dark matter. Only a few particle physics
scenarios  can successfully create baryons and dark matter at such a low temperature, see for instance~\cite{Kohri:2009ka}.


\subsection{Non-thermal phase and reheating}
\label{NPID}

The Universe  after inflation could be reheated in a much more violent fashion via non-perturbative, non-thermal way. The Universe 
in this epoch prior to the attaining LTE could be completely out of equilibrium. This could lead to rapid and efficient transfer of
inflaton energy, the process is also dubbed as {\it preheating}. Indeed, preheating is model dependent, but in a wide class of inflationary 
models preheating criteria can be satisfied with ease. One of the key ingredients is that the  inflaton  couples to essentially massless field $\chi$, through
interaction term like $\phi^2\chi^2$. The quantum modes of $\chi$ can then be excited during the inflaton
oscillations via a {\it parametric resonance} \cite{Traschen:1990sw,Shtanov:1994ce,Kofman:1994rk,Kofman:1997yn,Kofman:1995fi,Yokoyama:2004pf}, for a review see~\cite{Allahverdi:2010xz}. During preheating, fermions can also be excited, but their occupation number can not grow arbitrarily large due  to Pauli blocking \cite{Dolgov:1989us,Greene:1998nh,Giudice:1999fb,Maroto:1999ch,Greene:2000ew,Peloso:2000hy,Nilles:2001fg,Nilles:2001ry}. Also, one can excite the gauge fields which may have applications for cold electroweak baryogenesis~\cite{GarciaBellido:1999sv,Cornwall:2000eu,Cornwall:2001hq}, and magnetic fields as well
~\cite{DiazGil:2008tf}. The epoch of preheating has been performed on lattice, see~\cite{Micha:2002ey,Micha:2004bv,Felder:2006cc}.


\subsubsection{\bf Parametric Resonance:}

Let us briefly discuss preheating in the simplest but most general setup.
Let us consider the relevant renormalizable couplings between the inflaton $\phi$
and a scalar field $\chi$, for which the potential will be given by:
\begin{equation} \label{nonpot}
V = {1 \over 2} m^2_{\phi} \phi^2 + \frac{1}{2}m_{\chi}^2\chi^2+\sigma \phi \chi^2 + h^2 \phi^2 \chi^2 +
\kappa \chi^4\,, 
\end{equation}
where we have considered $\phi$ and $\chi$ to be real, and the kinetic terms are all canonical. Furthermore, 
$\phi$ is a gauge singlet inflaton.
Preheating with non-canonical terms has been studied in~\cite{Mazumdar:2015pta}.
Note that $\sigma$ has a [mass] dimension. The only scalar field in
the SM is the Higgs doublet. Therefore $\chi$
denotes the real and imaginary parts of the Higgs
components. The cubic interaction term is needed for the inflaton to decay even for the preheating.  The quartic self-coupling of
$\chi$ is required to bound the potential from below along the $\chi$
direction. The dimensionless couplings $\sigma/m_{\phi}$ and $h$ (as
well as $\kappa$) are not related to each other, hence either of the
cubic or the quartic terms can dominate at the beginning of inflaton
oscillations (i.e. when the Hubble expansion rate is $H(t) \simeq
m_{\phi}$. Preheating typically occurs within a narrow window for $h$; $3 \times 10^{-4} \leq h \leq
10^{-3}$. The $h^2 \phi^2 \chi^2$ term also yields a
quartic self-coupling for the inflaton at a one-loop level which is typically 
constrained by the temperature anisotropy of the CMB 
perturbations, i.e. $\kappa\leq 10^{-12}$. 
Neglecting the self interaction for $\chi$ field, the 
equation of motion for $\chi_{k}$ quanta is given by:
\begin{equation}
\ddot\chi_{k}+3\frac{\dot a}{a}\dot\chi_{k}+\left(\frac{k^2}{a^2}+m^2_{\chi}+2(\sigma\phi+h^2\phi^2)\right)\chi_{k}=0\,.
\end{equation}
It is assumed that the inflaton oscillations are homogeneous, $\phi(t)=\hat\phi(t)\sin(m_{\phi}t)$, where 
$\hat\phi(t)\approx (M_{\rm P}/\sqrt{3\pi}m_{\phi}t)$, for the inflation mass $m_{\phi}$. 
The occupation number of the excited $\chi_k$ is given by:
\begin{equation}
n_{k}=\frac{\omega_{k}}{2}\left(\frac{|\dot\chi_k|^2}{\omega_k^2}+|\chi_k|^2\right)-\frac{1}{2}\,,
\end{equation}
There exists a possibility of a narrow resonance production of $\chi_k\propto \exp(\mu^{n}_k z)$, where  $\mu^n_k$ is set by the instability band 
$\Delta_k^{n}$ labeled by an integer $n$, and $z=m_{\phi}t$. quanta, see Refs.~\cite{Traschen:1990sw,Shtanov:1994ce,Kofman:1994rk,Kofman:1997yn},  
when expansion of the Universe and the trilinear interaction are small.
The resonance occurs for $k=0.5 m_{\phi}(1\pm q/2)$, where $\mu_k$  takes the maximum value 
$\mu_k=q/2$, where $q=g^2(\hat\phi^2/4m_{\phi}^2)$.   When the expansion rate of the Universe is taken into account, then the evolution of the oscillating
inflaton field also modifies to a damped oscillator:
\begin{equation}\label{inf-osc}
\phi (t) \simeq \frac{M_{\rm P}}{\sqrt{3\pi}}
\frac{\cos\left( m_\phi  t \right)}{m_\phi  t} \,,
\end{equation}
where $t$ is the physical time. During this period the {\it stochastic resonance } becomes important~\cite{Kofman:1997yn}, where there are resonance bands
which keep shifting from stability to instability bands.  The resonant particle production and re-scatterings of interacting quanta lead to the formation
of a plasma consisting of both $\phi$ and $\chi$ quanta with typical energies $\sim 10^{-1} \left(h m_{\phi} M_{\rm P}\right)^{1/2}$, see~\cite{Kofman:1997yn}. This plasma attains the kinetic
equilibrium first, but the full thermal equilibrium, including both kinetic and  chemical, is established over a much
longer time scale~\cite{Felder:2006cc,Micha:2004bv}. The occupation number of particles in the preheated plasma is $\gg 1$
(which is opposite to the situation after the perturbative
decay). This implies that the number density of particles is larger
than its value in full equilibrium, while the average energy of
particles is smaller than the equilibrium value.  It gives rise to
large effective masses for particles which, right after preheating, is
similar to their typical momenta~\cite{Kofman:1997yn}.  Large occupation
numbers also lead to important quantum effects due to identical
particles and significant off-shell effects in the preheat plasma~\cite{Felder:2006cc,Micha:2004bv,Podolsky:2005bw}.  In the course of
evolution towards full equilibrium, however, the occupation numbers
decrease. Therefore a proper (non-equilibrium) quantum field theory
treatment will be inevitably required at late stages when
occupation numbers are close to one~\cite{Yokoyama:2004pf}.

Preheating ends due to back reaction as well as the expansion of the Universe . Preheating
does not destroy the zero mode of the inflaton condensate completely.
The amplitude of the inflaton oscillations diminish, but the
inflaton decay is completed when the zero mode perturbatively decays
into the SM or some other degrees of freedom,
see~\cite{Traschen:1990sw,Shtanov:1994ce,Kofman:1994rk,Kofman:1997yn}.

During preheating it is possible to excite  particles which have a mass greater than
the inflaton mass $m_{\phi}$.  One of the applications is the creation of cosmologically stable 
dark matter candidate. Such processes are impossible in perturbation 
theory and in the theory of narrow parametric resonance. Superheavy $\chi$-particles with mass $M \gg m_{\phi}$ can be produced in 
the broad resonance. During the coherent oscillations of $\phi(t)$, the
adiabaticity condition is violated~\cite{Kofman:1997yn}
\begin{equation}
\label{adiab}
\frac{d\omega(t)}{d t} \geq \omega^2(t)\,.
\end{equation}
The momentum dependent frequency, $\omega_k(t)=\sqrt{k^2+m_\chi^2+2h^2\phi^2(t)}$ violates
the above condition when
\begin{equation}\label{adiabAAA}
{k^2 + m^2_{\chi}} \leq (h^2\phi m_\phi
\hat\phi)^{2/3} - h^2\hat\phi^2 \ .
\end{equation}
The maximal range of momenta for
which particle production occurs corresponds to $\phi(t) = \phi_*$, where
$\phi_* \approx {\textstyle {1 \over 2}} \sqrt {m_\phi\hat\phi\over h}$. The
maximal value of momentum for particles produced at that epoch can be
estimated by $k^2_{\rm max} + m^2_{\chi} = {h m_\phi \hat\phi \over 2}$.  The resonance becomes efficient for
$ h m_\phi \hat\phi \geq 4 m^2_{\chi} $.
Thus, the inflaton oscillations may lead to a copious production of
superheavy particles with $m_{\chi} \gg m$ if the amplitude of the field $\phi$ is
large enough, $h\hat\phi \geq 4m^2_{\chi}/m_{\phi}$. Besides narrow and broad resonances, there are other variants of preheating, 
such as {\it instant}~\cite{Felder:1998vq}, and  {\it tachyonic} preheating triggered via tachyonic instability, where at the classical level the zero mode develops exponential enhancement~\cite{Felder:2000hj, Allahverdi:2007zz}.


\subsubsection{\bf Fermionic and gauge preheating:}

The Dirac equation (in conformal time $\eta$, where $d\eta= \int dt/a(t)$) for a fermionic field is given by~\cite{Greene:1998nh,Giudice:1999fb}:
\begin{equation} \label{dir1}
\left( \frac{i}{a}\,\gamma^\mu \,\partial_\mu + i\, \frac{3}{2}
H \gamma^0 - m(\eta) \right) \psi = 0\,,
\end{equation}
where $m(\eta)=m_\psi + h \phi( \eta)$, where $m_\psi $ is the bare mass of the
fermion. $a$ is the scale factor of the Universe , $H=a'/a^2$ is the
Hubble rate and $\prime$ denotes derivative w.r.t. $\eta$.
The particle density per physical volume $V\sim a^3$ at time
$\eta$ is given by:
\begin{eqnarray}
n ( \eta ) \equiv \langle 0 | \frac{N}{V} | 0 \rangle =\frac{1}{\pi^2 \, a^3} \int\, dk \, k^2
\left| \beta_k\right|^2 ,
\end{eqnarray}
where $\alpha_k, \beta_k$ are the Bugolyubov's coefficients satisfying: $|\alpha_k|^2+|\beta_k|^2=1$.
The occupation number of fermions created is thus given by $n_k = | \beta_k |^2$, and the above
condition ensures that the Pauli limit $ n_k < 1$ is
respected.  One important physical quantity is the scaling of the total energy
\begin{equation}
\rho_\psi \propto m_\psi  N_\psi \propto q  m_\psi^{1/2}\,,
\end{equation}
which is linear in $q=h^2\hat\phi^2/m_\phi^2$~\cite{Greene:1998nh,Giudice:1999fb,Greene:2000ew,Peloso:2000hy}.
Note that $m_\psi (\eta)\propto q^{1/2}$. Note that the SM fermions are chiral, if the inflaton is a 
SM gauge singlet, then it can only couple via dimension-$5$ operators, i.e.
\begin{equation}\label{inf-ferm-coup}
\frac{\lambda}{M_{\rm P}}\phi(H\bar q_l)q_R\,,
\end{equation}
where $\lambda\sim {\cal O}(1)$, $H$ is the SM Higgs doublet and $q_l, q_R$ are the $SU(2)_l$ doublet and the
right handed SM fermions, respectively. As a result, preheating of SM fermions from a gauge singlet inflaton becomes less
important due to weak coupling. In Ref.~\cite{Giudice:1999fb}, it was shown that an inflaton coupling to the right 
handed neutrino, $h\phi \bar N N$, where $N$ is the right handed neutrino, could induce non-thermal leptogenesis, 
where the right handed neutrinos were treated as gauge singlets. Similar arguments would  hold for  the inflaton coupling to the SM gauge 
bosons, where the inflaton can only couple via non-renormalizable operator, i.e. 
\begin{equation}
\frac{\lambda}{M_{\rm P}}\phi F_{\mu\nu}F^{\mu\nu}\,, 
\end{equation}
where $\lambda\sim {\cal O}(1)$. Therefore, exciting the SM gauge bosons and the SM fermions through parametric 
resonance of a gauge singlet inflaton is a daunting task. Inflaton would rather prefer perturbative decay. The only
way one can excite SM fermions and gauge fields copiously, if they are directly excited by the oscillations of the SM Higgs boson.
This can happen in low scale electroweak baryogenesis~\cite{GarciaBellido:1999sv,Cornwall:2000eu,Cornwall:2001hq}, 
or in the context of SM Higgs inflation~\cite{GarciaBellido:2008ab}. During the Higgs oscillations the SM degrees of freedom
can be excited via parametric resonance, instant preheating and also via tachyonic preheating. All three phases of 
preheating are present. The other notable example is the MSSM inflation~\cite{Allahverdi:2011aj} where gluons and 
MSSM fermions are excited via instant preheating.


\subsubsection{\bf Fragmentation of the inflaton:}

An intriguing consequence of inflaton coupling to the fermions is the  fragmentation 
of the inflaton as a condensate~\cite{Kusenko:1997si,Enqvist:2002rj,Enqvist:2002si}. This leads to non-thermal phase where the inflaton condensate can fragment 
to form  non-topological solitons, known as Q-balls. The Q-balls can evaporate from their surface, see for a review~\cite{Enqvist:2003gh}, 
therefore suppressing the reheating and thermalization time scale.  Let us illustrate this idea by studying a simple scenario of an oscillating complex scalar field
around its minimum. Typically, the fermionic loops (assuming that the fermions live in a larger representation than bosons) yields a Logarithmic
correction to the inflaton mass. Similar corrections also arises within SUSY, in a gravity mediated
scenarios~\cite{Enqvist:2003gh}
\begin{equation}
    \label{qpotr}
    V = m^2 |\Phi|^2
    \left[ 1 - K\log\left(\frac{|\Phi|^2}{M^2}\right) \right]\,, \end{equation}
where the value of $K$ is determined by the
Yukawa coupling $h$ with $ K =-C({h^2}/{16\pi^2})$, where $C$ is
some number. If $K < 0$,
the inflaton condensate feels a negative pressure 
for field values $\phi \ll M$, we find:
\begin{equation}
    \label{pot000}
    V(\phi) \simeq \frac{1}{2}m_{3/2}^2\phi^2
    \left(\frac{\phi^2}{2M^2}\right)^K \propto \phi^{2+2K}\,.
\end{equation}
where we assume $|K|\ll 1$.  The average equation of state 
\begin{equation}
    \label{state}
    \langle p \rangle \simeq \frac{K}{2+K} \langle \rho \rangle  \simeq -\frac{|K|}{2} \langle \rho \rangle \,,
\end{equation}
where $p$ and $\rho$ is a pressure and energy density of the scalar
field, respectively.  The negative value of $K$ corresponds to the 
negative pressure, which signals the instability of the condensate. At the level of  linear perturbations~\cite{Enqvist:2002si} one can show that the field 
fluctuations grow exponentially if the following condition
is met
\begin{equation}
    \frac{k^2}{a^2}\left( \frac{k^2}{a^2}+2m_{3/2}^2K \right) < 0.
\end{equation}
The instability band exists for negative $K$, as expected from the negative pressure arguments~\cite{Enqvist:2003gh}. The 
instability band, $k$,  is in the range~\cite{Enqvist:2002si}
$  0 < \frac{k^2}{a^2} < \frac{k_{max}^2}{a^2} \equiv  2m_{3/2}^2|K|$,
where $a$ is the expansion factor of the Universe . 
The most amplified mode lies in the middle of the band, and the maximum growth rate  of the perturbations is determined by  
$\dot{\alpha} \sim ~|K|m_{3/2}/2$~\cite{Enqvist:2003gh}. When $\delta \phi/\phi_0\sim {\cal O}(1)$, the  fluctuations become nonlinear.  
This is the time when the homogeneous condensate breaks down into Q-balls and anti-Q-balls. Such a phenomena can also yield gravitational waves due to 
the anisotropic stress created during the process of fragmentation of the inflaton, and this has been studied in Refs.~\cite{Kusenko:2008zm,Kusenko:2009cv}.


\subsection{Radiation, dark matter, and dark energy}

\begin{table}[t]
    \centering
    \begin{tabular}{lc|c|c}
         & $T$ (GeV)  & $t$(s)    \\ \hline
Electroweak transition         & $\sim 20-200$ & $10^{-11}$    \\
QCD transition & $10^{-1}$ & $10^{-4}$  \\
Big bang nucleosynthesis & $\times 10^{-4}$ & $10^2$  \\
Recombination &$10^{-10}$ & $10^{12}$  \\
    \end{tabular}
    \caption{List of key times in the early Universe in terms of temperature, redshift and time. We are assuming the reheating temperature was sufficiently high and that each phase transition occurred in a single step. A large baryon chemical potential in the early Universe can also change the time of the QCD phase transition slightly and the temperature at which the electroweak phase transition occurs has some model dependence. 
    }
    \label{tab:early_Universe _summary}
\end{table}

After the end of inflation, and the end of reheating/preheating yields the most important phase of the Universe, i.e., known as the radiation domination phase.  See Table~\ref{tab:early_Universe _summary}
for known transitions in the cosmic history. The exact transition from the reheating phase, as we have seen above, depends on lots of parameters and rather model dependent on a particular nature of the BSM physics.
However, the reheating phase must come to an end before the BBN~\cite{Fields:2014uja,Cyburt:2015mya}  at temperature of $\sim 1 $~MeV, when the hadrons have already formed. After reheating,  the Universe is primarily dominated by relativistic species, assuming that they are all in LTE, the Hubble expansion rate is then determined by the ambient temperature, 
\begin{equation}
H=\sqrt{\frac{\rho}{3M_{\rm P}^2}}=1.66\times g_{\ast}^{1/2}\frac{T^2}
{M_{\rm P}}\,,
\end{equation}
where $g_{\ast}$ is the total number of relativistic
degrees of freedom and it is given by
\begin{equation}
g_{\ast}(T)=\sum_{i=\rm b}g_{i}\left(\frac{T_{i}}{T}\right)^4+\frac{7}{8}
\sum_{i=\rm f}g_{i}\left(\frac{T_{i}}{T}\right)^4\,.
\end{equation}
Here  $T_{i}$ denotes the effective temperature of the species $i$, which
has decoupled. During the radiation era when $H=(1/2t)$, one finds
\begin{equation}
\label{timetemp}
\frac{t}{1~{\rm s}}\approx 2.42 g_{\ast}^{-1/2}\left(\frac{1~{\rm MeV}}{T}
\right)^2\,.
\end{equation}
The light
elements, such as  $^2H$, $^3He$, $^4He$, and $^7Li$ are synthesized
during the first few hundred seconds~\cite{Fields:2014uja,Cyburt:2015mya}. The abundances depend on the
baryon-to-photon ratio~\cite{Steigman:2006nf,Steigman:2012ve,Cyburt:2015mya}
\begin{equation}
\eta\equiv \frac{n_B}{n_\gamma} = 273.3036\Omega_bh^2\left(1+7.6958\times 10^{-3}Y_p\right)\left(\frac{2.7255k}{T_{\gamma}^{0}}\right)^3\,,
\end{equation}
where $Y_p=2(n/p)/[1+(n/p)]\sim 0.25$, and $n/p$ is the ratio of neutron-proton abundance at the temperature of $\sim 0.1$MeV~\cite{Cyburt:2015mya}.
The latest constraint on $\Omega_b=\rho_b/\rho_c$ comes from the Planck data, where $\rho_c$ is the critical energy density of the Universe , i.e. 
$\rho_c=3H_0^2/8\pi G$, where $G$ is the Newton's constant, and $H_0=100\times h$, and $H_0=67.4\pm0.5 {\rm Km/s/Mpc}$ at $68\%$CL, while $\Omega_b h^2=0.0224\pm 0.0001$ at $68\%$CL~\cite{Aghanim:2018eyx}. Planck also gives constraint on the relativistic species which matches well with the constraints arising from the BBN, i.e.
$N_{eff}=2.99\pm0.17$.

The radiation domination ends when the non-relativistic matter starts dominating the Universe , the radiation-matter equality 
happens when
\begin{equation}
1+z_{eq}=\frac{\Omega_m}{\Omega_{r}}=2.41\Omega_m h^2\times 10^{4}\sim 3.3\times 10^{3}\,,
\end{equation}
where $\Omega_m=\Omega_c+\Omega_b$ is the total matter density.  The bound on $\Omega_c h^2=0.12\pm 0.001$ at $68\%$ CL, while
$\Omega_m=0.315\pm 0.007$ at $68\%$CL~\cite{Aghanim:2018eyx}.
The temperature of the Universe is roughly given by
$T_{eq}=T_0(1+z_{eq})=6.57\times \Omega_{m} h^2\times 10^{4}K\sim 6.2\times 10^{3}K$. The creation of non-relativistic cold 
dark matter has to happen somewhere deep inside the radiation epoch. There are number of dark matter candidates, which include 
both thermal and non-thermal candidates, see reviews~\cite{Jungman:1995df,Bertone:2004pz,Dev:2013yza}.  

Amongst thermal dark matter, the well-known candidate is hot neutrinos~\cite{Kolb:1988aj,Cowsik:1972gh}. In fact the neutrinos decouple roughly around when the temperature of the Universe falls below $1$~MeV, the interaction rate per neutrino falls below the Hubble expansion rate, after which the neutrino number density is conserved and their momentum falls as $1/a$. The neutrino has large free streaming length, which is detrimental to form large scale structures in the Universe. This is the prime reason they fall out of favour for the heavier cold dark matter candidate. The challenge remains is how to slow them down, lacking of any credible mechanism leads to postulating sterile neutrinos as a possible candidate for warm or cold dark matter, for a review see~\cite{Kusenko:2009up,Adhikari:2016bei}.  From structure formation point of view, above $3$KeV sterile neutrino, which are thermally decoupled from the plasma, can be regarded as virtually cold, see recent analysis~\cite{Dayal:2015vca,Dayal:2014nva}. As far as cold dark mater is concerned, there are plethora of models~\cite{Feng:2010gw}, but their creation mechanism remains 
predominantly thermal decoupling, such as freeze out~\cite{Kolb:1988aj,Gondolo:1990dk,Scherrer:1985zt,Griest:1990kh,Steigman:2012nb} or freeze in~\cite{Hall:2009bx}, or non-thermal processes via decay of some heavy particles, such as decay of the inflaton itself, or part of the inflaton itself~\cite{Allahverdi:2007wt}.

 
The long wavelength CMB perturbations  do not grow during the radiation epoch, but once matter domination starts, the initially induced CMB perturbations get a chance to grow and seed density perturbations in matter sector, which includes DM, baryons and photons to form first structures in the Universe , for a review see~\cite{Peebles:1994xt,2012arXiv1208.5931K,Dayal:2018hft}.

 From a particle physics perspective, the major phase occurs very late in the history of the Universe when the Universe seems to be accelerating, for a review, see~\cite{Peebles:2002gy,Frieman:2008sn}. The latest data from Placnk constraints the dark energy abundance to be $\Omega_d =0.684 \pm 0.007$~at $68\%$ CL~\cite{Aghanim:2018eyx}.
 This apparent acceleration can be explained by apparent domination of dark energy in its simplest form, i.e. the  cosmological constant in the Einstein Hilbert gravity. Indeed, the challenge lies how to protect the apparently small cosmological constant from radiative corrections, see~\cite{Weinberg:1988cp,Polchinski:2006gy}, which remains an outstanding unresolved problem. There are proposals to modify gravity in the infrared, however, without much observational or theoretical motivations in our opinion, for a review see~\cite{Joyce:2014kja}.


\section{Nature of phase transitions}

In this chapter we are mostly concerned with thermal phase transition, preheating, fragmentation were clear examples of non-thermal phase transitions, which we had briefly discussed earlier. We will begin with the nature of phase transitions below.

\subsection{First and second order transitions}
In quantum field theory a phase transition is typically thought of as a transition between one vacuum state and another. For simplicity let us consider the case where the system is in the absolute ground state at some particular high temperature and as the Universe cools a new ground state becomes energetically favorable. If the ground state evolves continuously then this is what is known as a second order phase transition (or more generally a continuous phase transition). Alternatively if there is a discontinuous change in the ground state of the quantum field theory then this is a first order phase transition. There is also a discontinuity in the entropy during a first order phase transition. As such a first order phase transition releases a large amount of latent heat.\par

To illustrate the different types of phase transition we give a graphic illustration in Fig. \ref{fig:tempevolution} where the left and right panel demonstrate a second and first order phase transition respectively. In the case of a first order phase transition there is a barrier between a local minimum and the absolute ground state. As such the phase transition occurs through quantum tunneling and initially only occurs in regions of space called bubbles. These bubbles of new phase grow and coalesce in a background of the old phase. The Universe cools until the new phase replaces the old one, completing the phase transition. \par

In the case of first order phase transitions the size of the discontinuity can be compared to the temperature and in the case where the size of discontinuity is comparable or large compared to the temperature, the transition is referred to as a strongly first order phase transition. We spell this out in more detail in later sections. In general a strongly first order phase transition is of particular interest to particle cosmology as the violent process of bubble nucleation and the subsequent collisions can result in striking primordial gravitational wave signals. Furthermore a strongly first order phase transition is of particular interest for explaining the asymmetry between matter and anti-matter. It is also possible to produce magnetic field and defects during either first and 2nd order phase transitions.

Let us conclude this section with a final note. Our statement about the ground state continuously evolving being associated with a 2nd order phase transition is some what simplified. A crossover transition also exhibits this quality. For a second order phase transition the correlation length goes to infinity and some masses as well as the specific heat becomes non-analytic at the critical temperature. Conversely a crossover transition has no non-analytic properties and all correlation lengths remain finite.

\subsection{Thermodynamical parameters}\label{sec:therm}

\begin{figure}
    \centering
    \includegraphics[width=0.48\textwidth]{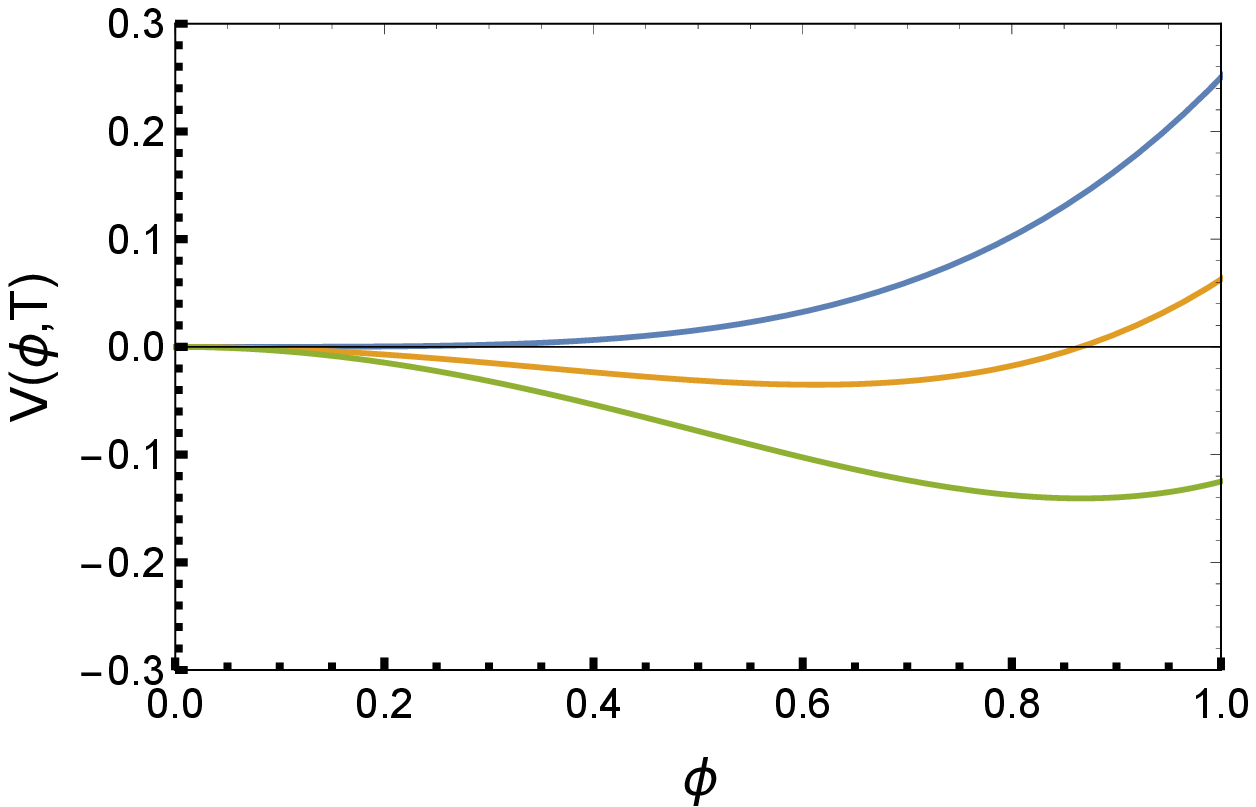}
     \includegraphics[width=0.48\textwidth]{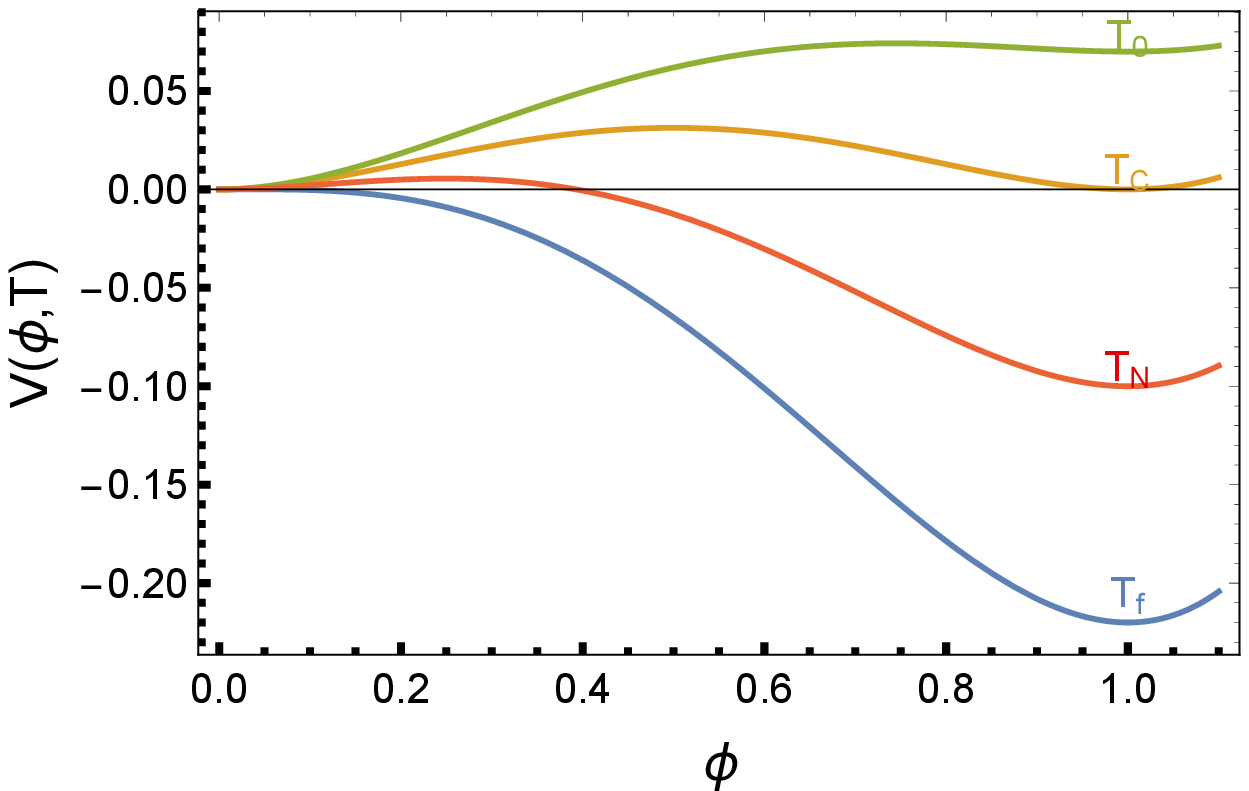}
    \caption{Evolution of an example potential with temperature with a 2nd order phase transition given in the left panel and 1st order phase transition in the right. The point where a new minimum appears away from the origin is called $T_0$. The critical temperature where the minima are degenerate is $T_C$. The nucleation temperature defined as the time where there is at least one critical bubble per Hubble volume is given by $T_N$ and the temperature at which the phase transition completes is denoted $T_f$.}
    \label{fig:tempevolution}
\end{figure}


In this section we will discuss the most important thermodynamic parameters during a first order phase transition. We focus on a strongly first order phase transition since these are the main focus of particle cosmology, be it baryogensis, or gravitational wave production. A first order phase transition proceeds through bubble nucleation~\cite{Coleman:1977py,Callan:1977pt,Coleman:1985rnk}. We will be interested in calculating the temperature at which these bubbles appear, the velocity at which they expand, the total number of bubbles, the fraction of new phase volume, the latent heat and the speed of the phase transition.

In this section our discussion will be in broad strokes with the exception of the wall velocity which we leave to its own section. To perform calculations one needs to know in more detail the effective potential and action at finite temperature which we delay to later sections. Let us begin by describing qualitatively the process of nucleation. The nucleation of bubbles doesn't occur immediately after the critical temperature even though the new phase has become energetically favourable. Instead, since the transition occurs through tunneling one must wait until the tunnelling rate is fast compared to the Hubble time. Eventually the Universe cools to the point where there is at least one critical size bubble in the Hubble volume. This temperature is denoted as the nucleation temperature. Finally when the volume fraction of the old phase is negligible the phase transition completes at a temperature $T_f$. \par

The tunnelling of the field from the false vacuum to true vacuum can be described as a solution to the classical equations of motion for the field. Assuming spherical symmetry the equation of motion for a single scalar field with a potential that is bounded from below is \cite{Kolb:1990vq}
\begin{equation}
    \frac{d^2 \phi }{d r ^2}+\frac{2}{r} \frac{d \phi }{dr} = \frac{d V}{d \phi }\,,
\end{equation}
with boundary conditions $\phi ^\prime (0) = 0$, $\phi (0) \sim \phi_{\rm true}$\footnote{This approximation will not hold if the potential is unbounded or the distance to the true vacuum is large compared to the height of the barrier.} and $\phi (\infty ) = \phi _{\rm false}$. This is the tunnelling solution to the classical equations of motion, known as the bounce, is the one where the field starts near the true vacuum and continuously evolves to the new one. It is in general non-trivial to find the bounce solution as naive attempts to find a solution tend to find the static solution - which trivially satisfies the left hand side of the above equation - where the field is in its minimum - satisfying the right hand side. As such many different approaches have been proposed to solve for the bounce \cite{Akula:2016gpl,Wainwright:2011kj,John:1998ip,Masoumi:2017trx,Kusenko:1995jv}. 
An approximate solution that is useful for illustrative purposes is the well known kink solution \cite{John:1998ip,deVries:2017ncy}
\begin{equation}
\phi(r)\approx A(r)= \frac{\phi_0}{2} \left[ 1-\tanh \left( \frac{ r - \delta}{L_w}\right) \right] - D[r,L_w,\delta] \,,\label{eq:ansatz}
\end{equation}
where $L_w$ defines the wall width, $\delta$ the offset and $D$ a regulating function to make the derivative vanish at zero. As an example consider the potential
\begin{equation}
    V=0.3 \phi ^2 - \phi ^3 +0.6 \phi ^4 \ .
\end{equation}

\begin{figure}
    \centering
    \includegraphics[width=0.42\textwidth]{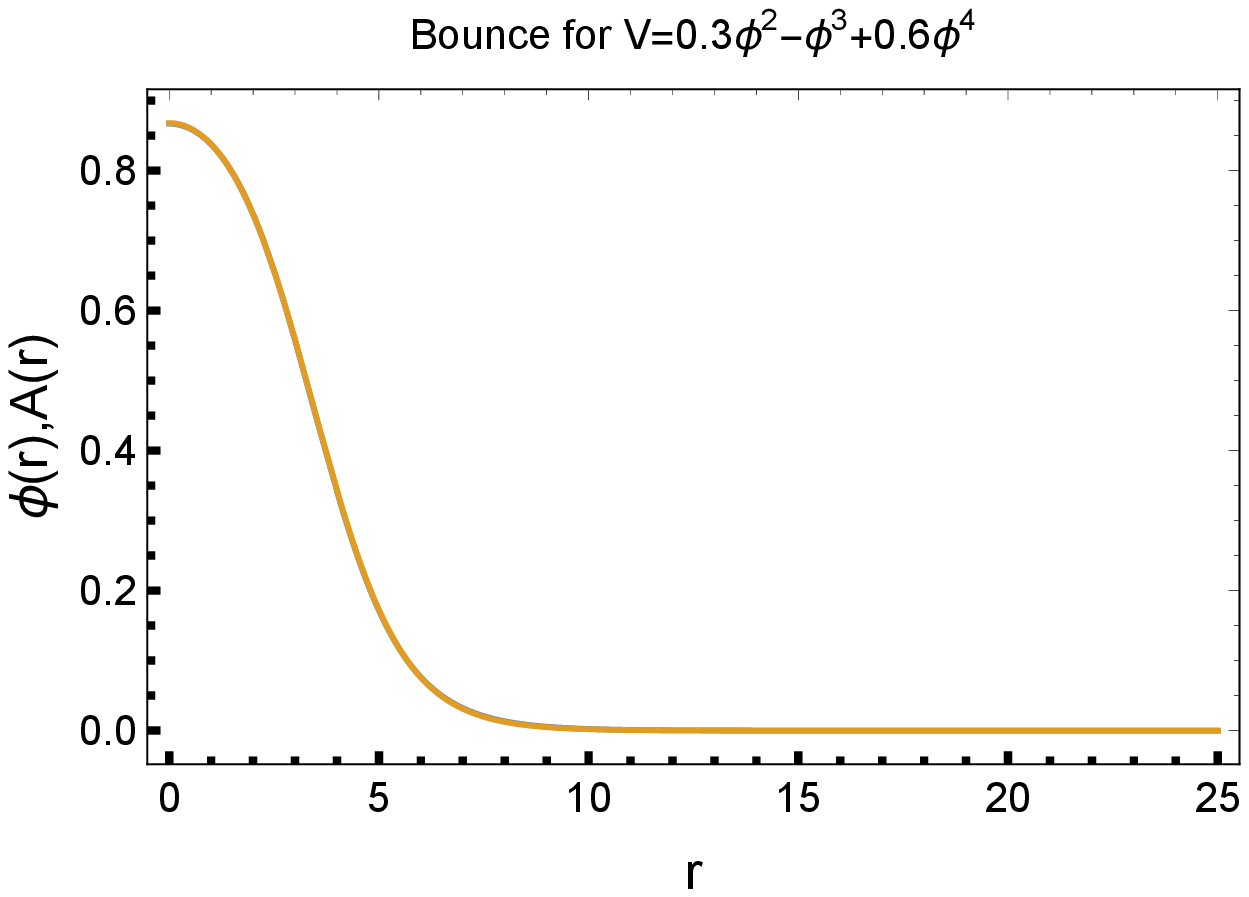}
    \includegraphics[width=0.45\textwidth]{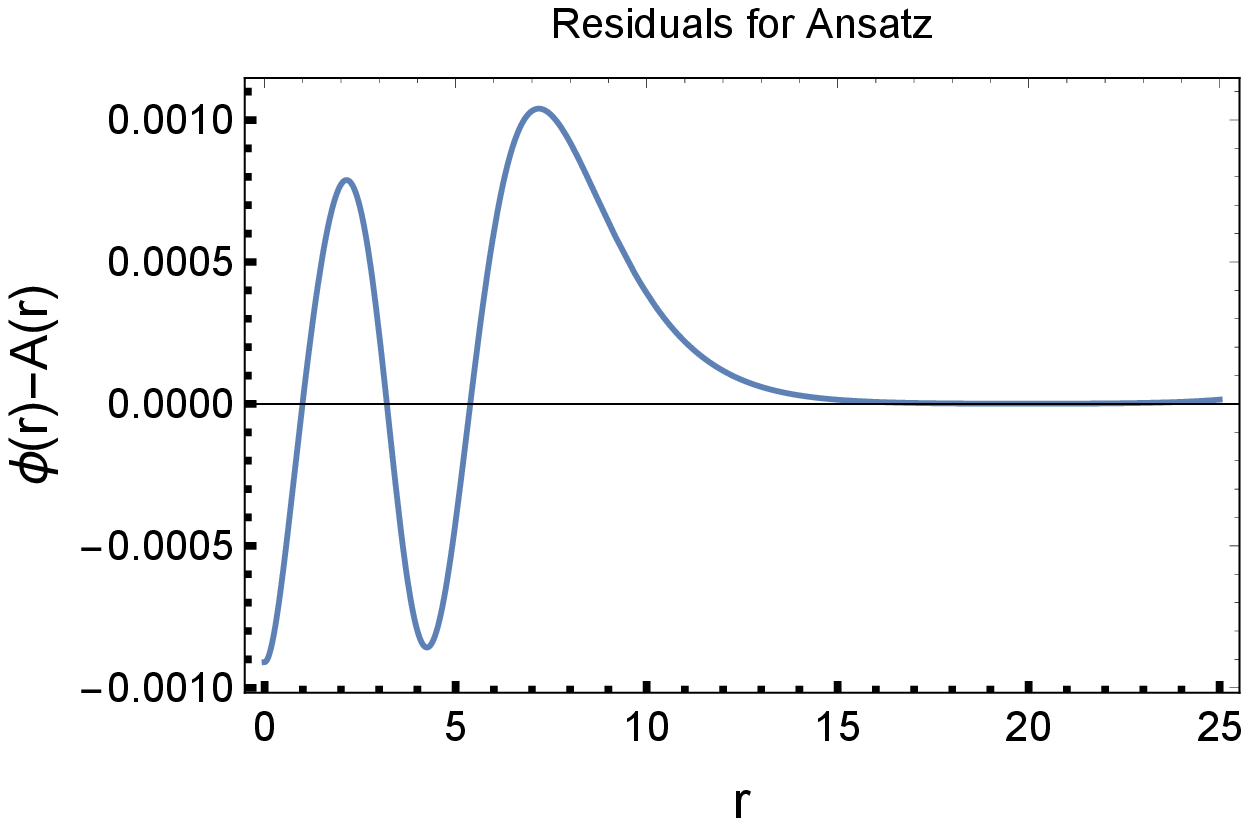}
    \caption{An example of the bounce solution for a sample potential (left panel). Above is presented the solution calculated numerically compared to the Ansatz given in Eq. \ref{eq:ansatz}. Since the difference is invisible to the naked eye we also plot the residuals (right panel).}
    \label{fig:bounce}
\end{figure}

\begin{figure}
    \centering
    \includegraphics[width=0.45\textwidth]{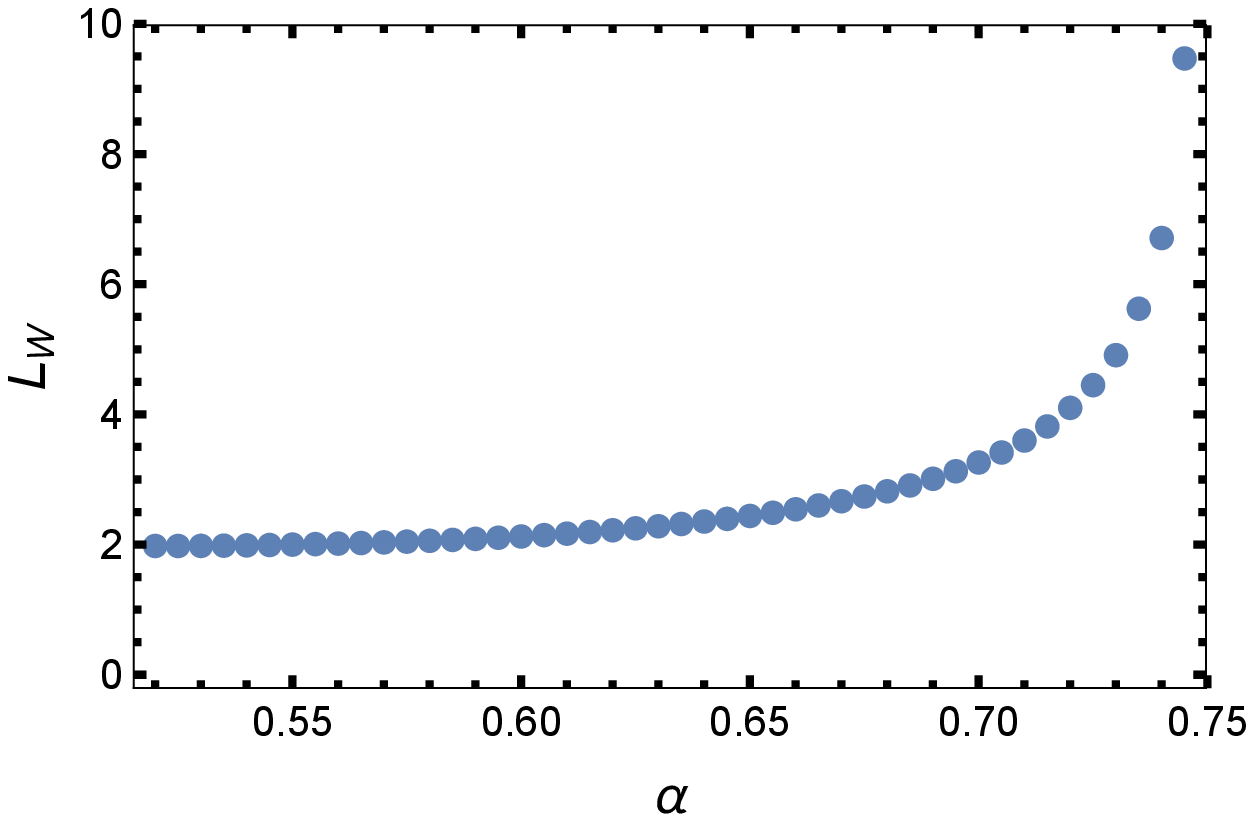}
    \includegraphics[width=0.45\textwidth]{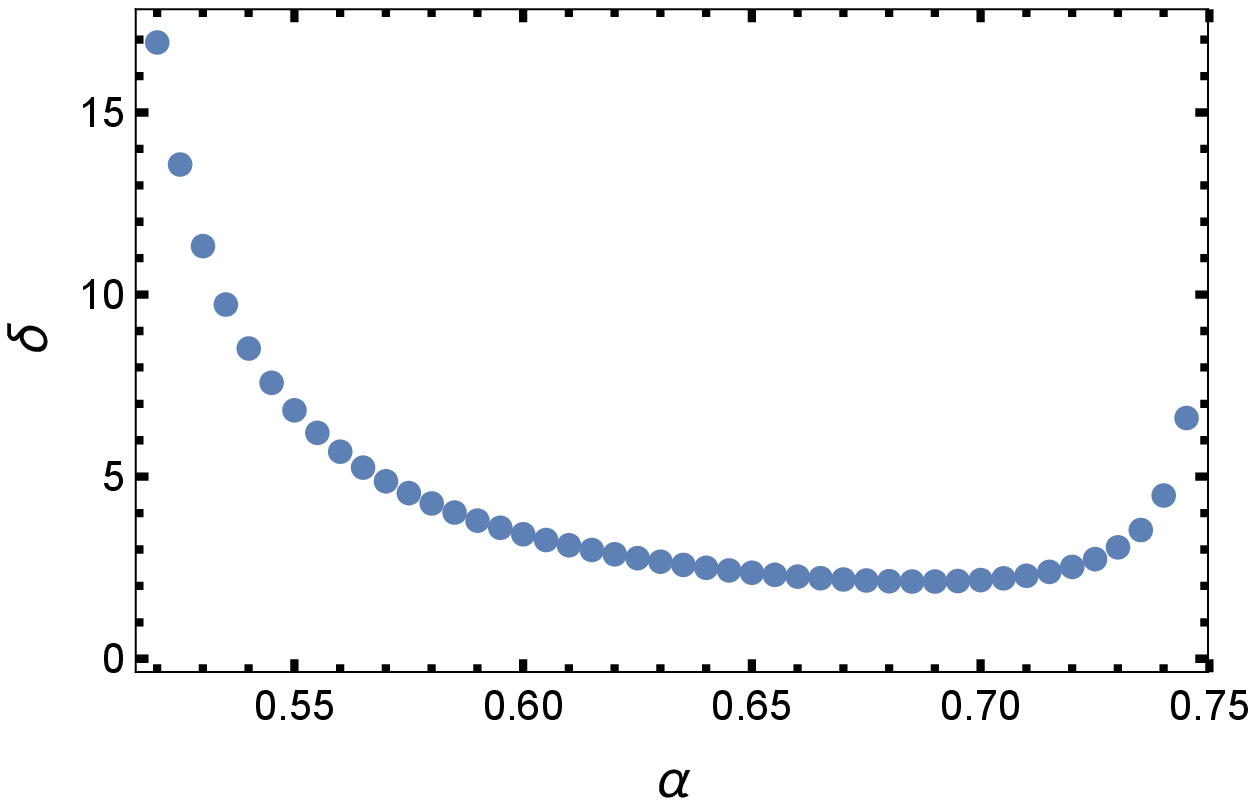}\\
    \includegraphics[width=0.7\textwidth]{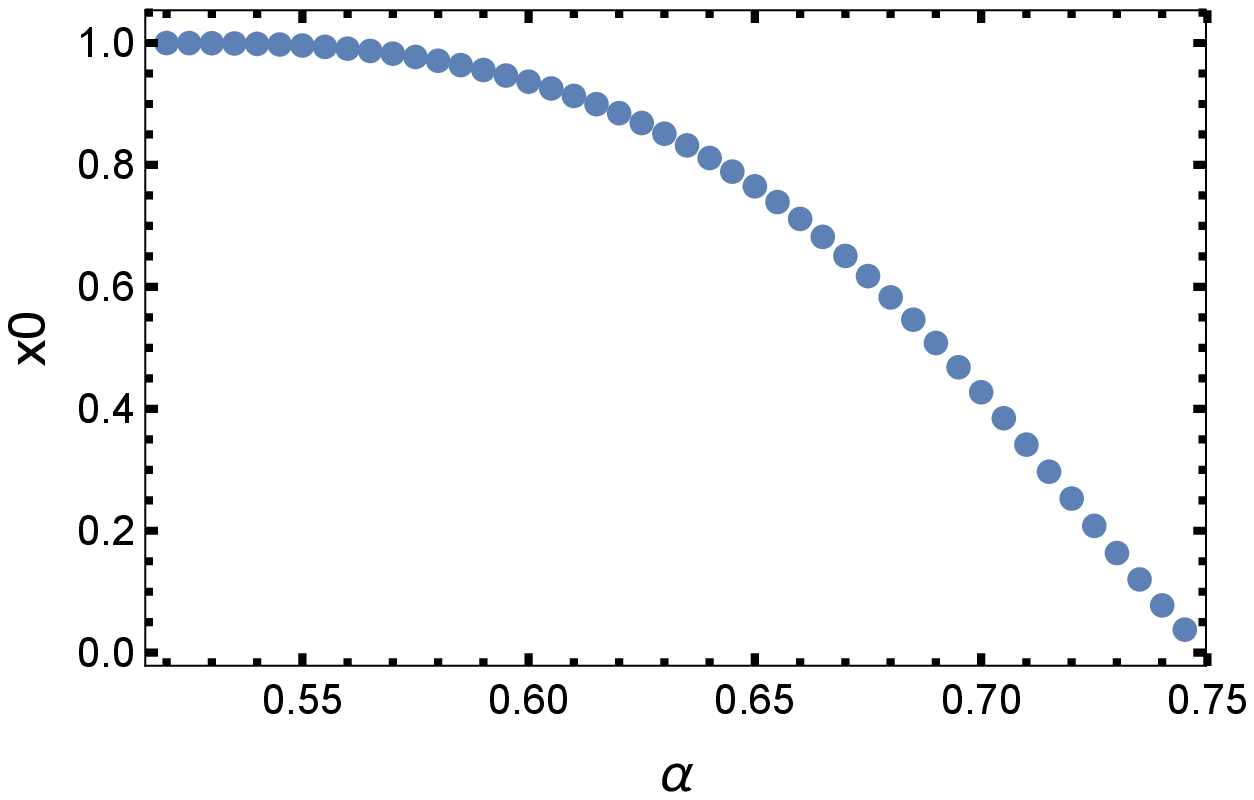}
    \caption{Numerical fitting of a tanh Ansatz given in Eq. (\ref{eq:ansatzreg}) to the bounce with the parameters as a function of $\alpha$. The top panels are the bubble wall and offset respectively whereas the bottom panel is the prefactor.}
    \label{fig:alphaplots}
\end{figure}

The kink solution that approximates the true solution is given by \cite{Akula:2016gpl}
\begin{equation}
    A(r)= \frac{0.937}{2} \left[1-\tanh\left( \frac{r-3.42}{2.12} \right) \right] - \frac{1}{2} \left| \frac{0.937 {\rm Sech}  \left( \frac{3.42}{2.12}\right)^2}{2.12} \right| e^{-r} \ .
\end{equation}
The above compares remarkably well to the true bounce solution as can be shown in Fig. \ref{fig:bounce}.
An alternative Ansatz was proposed in \cite{Sarid:1998sn} and takes the form
\begin{equation}
A(r)=\frac{A_0}{1+r^\gamma e^{\sqrt{2}(x-R}}
\end{equation}
with the parameters given in the thick and thin wall limits in the above reference.
 Since all potentials with two minima separated by a maximum can be approximated by a quartic function over the region of interest, it is useful to generically solve general quartic potentials. Indeed since one can always make rescalings and shifts of the form $\phi \to \phi + a$, $\phi \to b \phi$, $V \to V+c$ and $V \to d V$ it turns out to be sufficient to solve the class of potentials of the form
\begin{equation}
    V(\alpha ) = \frac{-4 \alpha +3}{2} \phi ^2 - \phi ^3 + \alpha \phi ^4 \,,
\end{equation}
for the parameter range $\alpha \in [0.5,0.75]$.
Using the Ansatz \cite{John:1998ip,Akula:2016gpl,deVries:2017ncy}
\begin{equation}
    A(r,\alpha) = \frac{x _0 (\alpha)}{2} \left[1- \tanh \left( \frac{r- \delta (\alpha)}{L_w (\alpha)} \right) \right] - \frac{1}{2} \left|\frac{\phi_0(\alpha ) {\rm Sech}\left( \frac{\delta(\alpha)}{L_w (\alpha )}\right)^2 }{L_w (\alpha ) } \right| e^{-r} \,, \label{eq:ansatzreg}
\end{equation}
where we plot the dependencies of the Ansatz parameters in Fig. \ref{fig:alphaplots}. In principle if the path between the true and false vacuum can be approximated by a quartic one can use such an ansatz to approximate the bounce solution. The main source of error in this will be the fact that the true bounce solution follows a curved path in field space when multiple fields are involved.

Not that for $\alpha \sim 0.5$ one has degenerate minima and this is where the bubble wall is the thinnest. In practice the bubble wall width  tends to be in the range $ 1/T$ to $\sim 20/T$ for phenomenologically viable phase transitions. 
Denoting the bounce solution as $\phi _{\rm B}$, the decay rate of the old phase to the new phase is controlled by the Euclidean action 
\begin{equation}
   S_{\rm E}(\phi _{\rm B}) = 4 \pi \int _0 ^\infty dr r^2 \left[ \frac{1}{2} \left( \frac{d \phi _{\rm B} }{ d r} \right) ^2 + V(\phi _{\rm B} ) \right] \,,
\end{equation}
where $V$ is the effective potential at finite temperature which for now we leave unspecified. The lack of angular variables in the integration reflects the fact that we are considering a spherically symmetric solution. The decay rate per unit volume of the effective potential is given by \cite{Linde:1978px}
\begin{equation}
    \Gamma (T) \sim T^4  \left\{ \frac{S_{\rm E}(\phi _{\rm B})}{2 \pi T} \right\}^{3/2} e^{ -\frac{S_{\rm E}(\phi _{\rm B})}{ T}} \ .
\end{equation}
From the decay rate one can write the differential decay probability for a given temperature $T$ as \cite{Enqvist:1991xw}
\begin{equation}
    \frac{d P}{d \ln T} \sim \Gamma (T) \frac{M_{\rm p} }{T^2} \left( \frac{t_U T_0}{T} \right) ^3\,,
\end{equation}
where $t_U$ is the age of the Universe and $T_0$ is the temperature today. \par 
Using the relationship between temperature and time
\begin{equation}
    T^2 t = \sqrt{\frac{45}{16 \pi ^3}} \frac{M_{\rm p}}{\sqrt{g_{\ast}}}\,,
\end{equation}
one can derive an approximate condition for when $1/e$ volume fraction is in the new phase 
\begin{equation}
    \frac{S_{\rm E}(\phi _{\rm B})}{ T} = 170- \log[\frac{T}{\rm GeV}]-2 \log[g _{\ast}] \ .
\end{equation}
The temperature which satisfies this equation is known as the nucleation temperature. A more precise way of calculating the nucleation temperature is through calculating the total number of bubbles in a Hubble volume at a given temperature, $T$, \cite{Enqvist:1991xw},
\begin{equation}
    N = \int _T ^{T_C} \frac{H^{-1}(T ^\prime )}{T^\prime } V(T^\prime ) \Gamma (T^\prime ) d T^\prime \ .
\end{equation}
where $T_c$ is defined in Fig.~\ref{fig:tempevolution}.
The nucleation temperature is defined when the above expression is equal to 1. At a given time the total volume fraction of space in the false vacuum is \cite{Enqvist:1991xw}
\begin{equation}
    f_{\rm false}(t)=e^{-I(t)},~~~ \quad I(t) = \int _{t_c}^tdt ^\prime  p(T(t^\prime )) V(t^\prime, t) \ .
\end{equation}
Here $V(t^\prime, t) $ is the volume of a bubble formed at time $t^\prime $ evaluated at time $t$. If one assumes spherical symmetry one can write $V(t^\prime ,t)=4 \pi [r(t)-r(t^\prime)]^3/3$. Taking $r(t)-r(t^\prime) \sim v_w(t-t^\prime ) $ for a constant wall velocity, $v_w$, one can write the fraction of volume in the false vacuum as \cite{Enqvist:1991xw},
\begin{equation}
    f_{\rm false} = \exp\left[ - \frac{4\pi}{3} v_w^3 \int _{t_c}^t dt ^\prime p(T(t^\prime)) (t-t^\prime )^3 \right] \ .
\end{equation}
Dropping the time dependence of the temperature, the nucleation probability per unit volume in the above equation is given by \cite{Callan:1977pt,Enqvist:1991xw} 
\begin{equation}
    p(T) = \frac{\omega _-}{2 \pi } \left( \frac{S_{\rm E}(\phi _{\rm B})}{2 \pi T} \right)^{3/2} A(T) e^{-\frac{S_{\rm E}(\phi _{\rm B})}{ T}} \ .
\end{equation}
In the above $\phi _B$ is the bounce solution to the classical equations of motion, $\omega _-$ is the angular frequency of the unstable mode and $A(T)$ is the fluctuation determinant. The phase transition completes when $f_{\rm false}(t)$ becomes negligible. The duration of the phase transition can then determined by taking the difference between this time and the nucleation time. The speed of the phase transition is often parametrized in terms of the time rate of change of the effective action
\begin{equation}
   \beta = - \frac{d S}{dt } \approx \left. H T \frac{d(S_E/T)}{dT} \right|_{T_N} \ . 
\end{equation}
The speed of the phase transition is a parameter which controls the frequency and amplitude of relic gravitational wave backgrounds left by cosmic phase transitions.
Other parameters that control the amplitude and frequency of relic gravitational waves are the Latent heat and the bubble wall velocity. The latent heat divided by the radiation energy density is given by
\begin{equation}
    \alpha = \frac{\Delta \rho }{\rho _N}
\end{equation}
with
\begin{equation}
    \rho _N = \frac{\pi ^2 g^* T_N^4}{30} , \quad \Delta \rho = \left[ V- \frac{dV}{dT} T_N \right] _{\rm False}-\left[ V- \frac{dV}{dT} T_N \right] _{\rm True} \ .
\end{equation}

\subsection{bubble wall velocity}
After a bubble nucleates it expands creating an ever larger region of the new vacuum. The act of expansion leads to more particles in the plasma either acquiring or receiving a mass. Furthermore the equilibrium distributions of particles in the plasma gets perturbed near the bubble wall. These processes costs energy and results in resistance to the bubble expansion. If such friction is large, the bubble may not necessarily go ultra relativistic. \par 

Let us begin with the classical equations of motion for a scalar field interacting with fermions and gauge bosons \cite{Moore:1995si}
\begin{equation}
    \Box h - \frac{\partial V_0}{\partial h}- \frac{d m_h^2}{d h} \langle \delta h ^2 \rangle - \sum \frac{d m_A^2}{d h} \langle A^2 \rangle - \sum  \frac{d m_\psi}{d h} \langle \bar{\psi}_R \psi _L \rangle =0\,,
\end{equation}
where $A$ and $psi$ are gauge bosons and fermions that acquire a mass when $h$ acquires a vacuum expectation value.
Also the angular brackets denotes the vacuum distribution at finite temperature but not necessarily in equilibrium.
If the particle distributions are their equilibrium functions then the above are the same equations of motion one solves when finding the bounce solution that describes the profile of a bubble wall during nucleation. We will derive this fact in more detail when we review the fate of the effective potential at finite temperature in a later section. The expansion of the bubble wall however perturbs the plasma away from equilibrium and the energy required to perform such a perturbation resists the expansion of the bubble wall. The box operator contains a curvature friction term, $(2/r)dh/dr$, which can be neglected when the bubble expands to a sufficiently large size that we can neglect the curvature of the bubble wall. In this case the box operator becomes $\Box = (1-v_w^2)\partial _z^2$ with the bubble moving along in the $z$ direction. Such an approximation underestimates the total friction and therefore will provide a slight overestimate of the bubble wall velocity.
The 2nd moments of the fields can be written in terms of their vacuum expectation values (which will be zero) and their distributions
\begin{eqnarray}
   \langle \delta h ^2 \rangle &=& \langle \delta h^2 \rangle _{\rm vac} +  \int \frac{d^3k}{(2 \pi )^3 E_k}f_{\delta h}(k,x) \,, \nonumber \\
    \langle A ^2 \rangle &=& \langle A^2  \rangle _{\rm vac} +  \int \frac{d^3k}{(2 \pi )^3 E_k}f_A(k,x)\,, \nonumber \\       
     \langle \bar{\psi }_R \psi _L \rangle &=& \langle \bar{\psi }_R \psi _L \rangle _{\rm vac} +  \int \frac{d^3k}{(2 \pi )^3 E_k}f_\psi (k,x) \ . 
    \end{eqnarray}
Writing the distributions as $f_X=f^0_X+\delta f_X$, where the $\delta f_X$ piece corresponds to the departure from equilibrium, and noting that 
\begin{equation} 
\frac{\partial V(h,T)}{\partial h}=\frac{\partial V_0}{\partial h}+\sum _j \frac{\partial m_j}{\partial h} \int \frac{d^3 p}{(2 \pi )^3(2 E_p)} f_j^0(p,x) \,,
\end{equation} 
we can then rewrite the classical equations of motion such they explicitly contain the wall velocity
\begin{equation}
    -(1-v_w^2)h ^{\prime \prime } +\frac{\partial V(h,T)}{\partial h }+\sum _j \frac{\partial m_j }{\partial \phi} \int \frac{d^3 p }{(2 \pi )^3 2 E_j} \delta f_j (p,x)=0 \ .
\end{equation}
The sum in the above equation represents the resistance to bubble expansion due to the new phase causing particles in the plasma to acquire a mass and depart from their equilibrium distributions. When the WKB condition of $p_j>>1/L_w$ is satisfied~\cite{Moore:1995si}, the particle distributions satisfy Boltzmann equations~\cite{Bernstein:1988bw}
\begin{equation}
    \left( \partial _t  + \dot{x} \partial _x  +\dot{p}_x \partial _x \right) f_j = -C[f_j]\,.
\end{equation}
We can parametrize the distribution function as
\begin{equation}
    f_x= \frac{1}{e^{(E+\delta)/T} + x }\,,
\end{equation}
where the value of $x=\pm 1$ denotes fermions (+) or bosons (-). The various terms in the Boltzmann equation can be written as
\begin{eqnarray}
    \partial _t f = f_0' (\partial _t E + \partial _t \delta ) = f_0' (\frac{1}{2E} \partial _t m^2 + \partial _t \delta ) \,,\nonumber \\
    \dot{x} \partial _x f = f_0^\prime \frac{p_x}{E} \partial _x \delta \,, \nonumber \\
    \dot{p}_x \partial _{p_x}f \sim 0 \ .
\end{eqnarray}
The perturbations can be written as a sum of a perturbation in the chemical potential, the fluctuation in the temperature and the fluctuation in the velocity of each species. That is \cite{Moore:1995si}
\begin{equation}
    \delta _i = -\left[ \delta \mu _i  + \frac{E}{T} \left( \delta T_i + \delta T _{bg} \right) + p_x(\delta v + \delta v _{bg}) \right] \ .
\end{equation}
To solve the Boltzmann equations one takes the moments - $\int d^3p/(2\pi )^3$, $\int (E/T) d^3p/(2\pi )^3$ and $\int (p_x/T) d^3p/(2\pi )^3$- of the Boltzmann equations to acquire a set of linear equations
\begin{eqnarray}
&&    c_{2x}^i \partial _t \mu _i +c_{3x} ^i \partial _t (\delta T_i + \delta T_{bg}) +(c^i_{3x}/3) T \partial _x (\delta v_i +v_{bg}) \nonumber \\  &&+ \int \frac{d^3 p}{(2\pi)^3 T^2} C(f)_i = \frac{c^i_{1x}}{2T }\frac{\partial m_i^2}{\partial t}\,, \nonumber \\  
&&        c_{2x}^i \partial _t \mu _i +c_{3x} ^i \partial _t (\delta T_i + \delta T_{bg}) +(c^i_{3x}/3) T \partial _x (\delta v_i +v_{bg}) \nonumber \\ && + \int \frac{E}{T} \frac{d^3 p}{(2\pi)^3 T^2} C(f)_i  = \frac{c^i_{2x}}{2T }\frac{\partial m_i^2}{\partial t}\,, \nonumber \\
&&            c_{2x}^i \partial _t \mu _i +c_{3x} ^i \partial _t (\delta T_i + \delta T_{bg}) +(c^i_{3x}/3) T \partial _x (\delta v_i +v_{bg}) \nonumber \\ && + \int \frac{p_z}{T} \frac{d^3 p}{(2\pi)^3 T^2} C(f)_i  = 0 \ . 
\end{eqnarray}
where we have defined 
\begin{equation}
    c_{nx} = \int \frac{d^3p}{(2\pi)^3} \frac{E^{n-2}}{T^{n+1}} f_x(p) \ .
\end{equation}

One can write the above set of equations in a matrix equation \cite{John:2000zq,Moore:1995si,Kozaczuk:2014kva}
\begin{equation}
    {\bf M} \vec{\delta } ' + {\bf \Gamma } \vec{ \delta } = \partial _t m^2 \vec{F} \ . 
\end{equation}
In the above the matrix $\bf \Gamma $ is from the set of relevant collision integrals.
If we assume that the derivatives of the fluctuations $\delta$ are small then one can invert the above matrix equation to obtain an expression for the fluctuations $\vec{\delta } \sim  {\bf \Gamma }^{-1}  \partial _t m^2 \vec{F}$. 
For a single field undergoing a phase transition the equations of motion including these perturbations can be written \cite{Moore:1995si}
\begin{equation}
    - h^{\prime \prime} + \frac{\partial V}{\partial h } + \frac{1}{2}T n_i \frac{\partial m_j^2}{\partial h} \left( c_{1x} \delta \mu _j + c_{2x} \delta T_j + c_{2x} T \delta v_j \right)=0\,.
\end{equation}
Substituting the approximation for $\delta$ into our equations of motions and taking the bubble wall velocity to be small gives
\begin{equation}
    h ^{\prime \prime } - \partial _h  V = \eta v_w \gamma \frac{h^2}{T} h ^\prime\,,
\end{equation}
where $\eta$ is given by
\begin{equation}
    \eta = \frac{T}{4} G \Gamma ^{-1} \vec{F}\,,
\end{equation}
and isolates the part of the friction that is independent of the wall velocity.
Here the term $G$ is a vector that for a particle species $i$ has the form $2N_i m_i/v \vec{c}_\pm $, with $\vec{c}_\pm = (c_{1\pm},c_{2\pm},c_{3\pm})$, and we give the standard model friction terms in Table \ref{tab:friction} from Ref.\cite{Kozaczuk:2015owa}. 

A rule of thumb for whether bubble walls can run away was developed by Bodecker and Moore \cite{Bodeker:2009qy}. Assuming the bubble wall reaches ultra relativistic speeds, the pressure that resists expansion due to particles crossing the wall and changing its mass reaches an asymptotic value independent of the Lorentz factor $\gamma$. In the ultra relativistic regime, one needs to only consider particles crossing the wall from the false vacuum to the true. Since no particles enter the false vacuum phase either through reflecting off the wall, or through exiting the expanding bubble, the particle distributions can be assumed in the equilibrium distribution.  Specifically the pressure reaches the value \cite{Bodeker:2009qy}
\begin{equation}
    P_{1 \to 1} = \sum _i n_i \int \frac{d^3 p}{2E(2 \pi )^3} f_{0,n} (p) (m_i^2(h_T) - m_i ^2(h_F))\,,
\end{equation}
where $h_{(t,f)}$ are the true and false vacuum respectively and the subscript $1 \to 1$ refers to the process where particles which cross the wall and acquire a contribution to their mass. 
If the minimum of the mean field potential
\begin{equation}
     \tilde{V} = \left. V \right|_{T=0} +\frac{1}{2}(h-h_t)^2 T^2\,, \label{eq:bmcriteria}
\end{equation}
 either does not exist or is higher than the false vacuum of the full thermal potential, the bubble wall cannot run away. This criteria implies that phase transitions involving scalar singlets tend to runaway as they introduce additional expansion pressure without introducing too much additional friction. Even still Ref. \cite{Kozaczuk:2015owa} found there are in fact some cases where the bubble wall in a singlet catalyzed electroweak phase transition can expand subsonically, that is the velocity of the bubble wall is slower than the speed of sound within the plasma.
\begin{table}[]
    \centering
\begin{tabular}{c|c}
Coefficient &  \\ \hline 
    $ \Gamma _{m_1,h}^h$ & $\left(  1.1 \times 10^{-3} g_3^2 y_t^2 1.4\times 10^{-3} y_t^4 \right)  T$   \\
    $ \Gamma _{T_1,h}^h = \Gamma ^h _{\mu _2,h}$  & $\left( 2.5 \times 10^{-3} g_3^2 y_t^2 +1.4 \times 10^{-3} y_t^4 \right) T$ \\ 
    $ \Gamma _{T_2 ,h}^h$ & $\left( 8.6 \times 10^{-3} g_3^2 y_t^2+1.8 \times 10^{-3} y_t^4 \right) T$ \\
    $\Gamma _{v,h}^h$ & $\left(3.5 \times 10^{-3} g_3^2 y_t^2 + 1.8 \times 10^{-3} y_t^4 \right) T$ \\ \hline 
    $-\Gamma _{\mu _1 t}^h$ & $\left( 10^{-3} g_3^2 y_t^2 +5.8 \times 10^{-4} y_t^4 \right) T$ \\
    $-\Gamma _{T_1,t}^h =- \Gamma ^h _{\mu _2 ,t}$ & $\left( 2.5\times 10^{-3} g_3 ^2 y_t^2 + 1.5 \times 10^{-3}y_t^4 \right) T$ \\
    $-\Gamma _{T_2,t}^h$ &$ \left( 8.5 \times 10^{-3} g_3^2 y_t^2 + 4.8 \times 10^{-3} y_t^4 \right) T$ \\ 
    $-\Gamma _{v,t}^h $& $\left( 2.8 \times 10^{-3} g_3 ^2 y_t^2 + 1.4 \times 10^{-3} y_t^4  \right) T$ \\ \hline 
    $\Gamma _{\mu _1 ,t } ^t$ & $\left(5.0\times 10^{-4} g_3^4+5.8 \times 10^{-4} g_3^2 y_t^2 + 1.5 \times 10^{-4} y_t^4 \right)T$ \\ 
    $\Gamma _{t_1,t}^t$ & $\left( 1.2 \times 10^{-3} g_3^4 + 1.4 \times 10^{-3} g_3^2 y_t ^2 + 3.6 \times 10^{-4} y_t^4 \right) T$ \\
    $ \Gamma _{T_2,t} ^t$ & $\left( 1.1 \times 10^{-2} g_3^4 + 4.6 \times 10^{-3} g_3^2 y_t^2 +1.1 \times 10^{-3} y_t^4 \right) T$ \\ 
    $\Gamma _{v,t} ^t$ & $\left( 2.0 \times 10^{-2} g_3^4 +1.7 \times 10^{-3} g_3^2 y_t^2 +4.3 \times 10^{-4} y_t^4 \right)T$ \\ \hline 
   $-\Gamma _{\mu _1 h}^t$ & $\left( 9.3 \times 10^{-5} g_3^2 y_t^2 +1.3 \times 10^{-4} y_t^4 \right) T$ \\
    $-\Gamma _{T_1,h}^t =- \Gamma ^t _{\mu _2 ,h}$ & $\left( 2.2\times 10^{-4} g_3 ^2 y_t^2 + 1.3 \times 10^{-4}y_t^4 \right) T$ \\
    $-\Gamma _{T_2,h}^t$ &$ \left( 7.2 \times 10^{-4} g_3^2 y_t^2 + 4.0 \times 10^{-4} y_t^4 \right) T$ \\ 
    $-\Gamma _{v,h}^t $& $\left( 2.4 \times 10^{-4} g_3 ^2 y_t^2 + 1.2 \times 10^{-4} y_t^4  \right) T$ \\ \hline 
    $-\Gamma _{\mu _2 t}$ & $\left( 1.4 \times 10^{-2} g_3^4  +1.3 \times 10^{-2}g_3^2 y_t^2 +2.6\times 10^{-3} y_t^4 \right) T$ \\
    $-\Gamma _{T_2,t} $ & $\left( 1.4\times 10^{-3} g_3 ^4 +4.6 \times 10^{-2} g_3^2 y_t^2 + 8.7 \times 10^{-3}y_t^4 \right) T$ \\
    $-\Gamma _{v,t}$ &$ \left( 2.4 \times 10^{-1}g_3^4+ 1.7 \times 10^{-2} g_3^2 y_t^2 + 3.4 \times 10^{-3} y_t^4 \right) T$ \\ 
    $-\Gamma _{T_2,h} $& $\left( 1.0 \times 10^{-3} g_3 ^2 y_t^2 + 9.8 \times 10^{-5} y_t^4  \right) T$ \\
    $-\Gamma _{v,h} $& $\left( 1.6 \times 10^{-3} g_3 ^2 y_t^2 + 4.6 \times 10^{-3} y_t^4  \right) T$ \\
    
\end{tabular}
    \caption{Table of friction coefficients for relevant SM processes. Table contents taken from Ref. \cite{Kozaczuk:2015owa}.}
    \label{tab:friction}
\end{table}

\subsection{Multistep phase transitions}
\begin{figure}
    \centering
    \includegraphics[width=0.7 \textwidth]{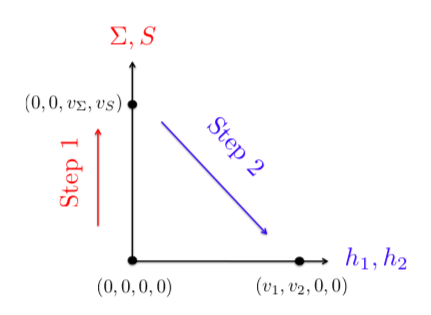}
    \caption{An example of a two step phase transition for a triplet Higgs extension of the standard model. In the first transition, denoted ``step 1'', a singlet along with an SU(2)$_L$ triplet denoted $\Sigma$ acquire a vacuum expectation value. In the second step these fields loose their vacuum expectation value while the Higgs fields acquire one. Figure taken from \cite{Inoue:2015pza}}
    \label{fig:2step}
\end{figure}
Thus far we have been considering the case where the phase history of a system is the simple case where one has one phase at high temperature and a different one at low temperature. Condensed matter systems teach us that things may not be so simple. Indeed there has been much recent interest in the case of multistep phase transitions and their application to baryogenesis \cite{Kuzmin:1985mm,Shaposhnikov:1987tw,McLerran:1990zh,Farrar:1993sp,Rubakov:1996vz,Inoue:2015pza,Ramsey-Musolf:2017tgh} and gravitational waves \cite{Croon:2018new}. In general there are four distinct cases of interest. The first where a symmetry is not broken in one of the phase transitions (for example a gauge singlet field may tunnel from one vacuum to another). The second case is where the same symmetry is broken in both phase transitions such as a two step electroweak phase transition. One can also break two different symmetries in subsequent transitions such as the case where the standard model is extended by a singlet field with a discrete $Z_2$ symmetry  where $V(\phi) =V(-\phi )$. Such a model was considered in Ref. \cite{Matsui:2017ggm} in the context of gravitational waves. Finally, there is the case where one has a a symmetry at zero temperature is broken at an intermediate temperature (and then possibly restored at high temperature).

For the first case consider the example of a real singlet where the first phase transition proceeds as $(0,\epsilon) \to(0 ,v_s)$ where the first transition does not begin exactly at the origin as finite temperature effects generate a linear term that shifts the minimum away from $v=0$.\footnote{Note that for this to work one needs some fields to couple to the scalar in such a way that the effective quadratic temperature correction can prevent a linear thermal correction from lifting the potential at the origin too quickly for a phase transition to be strongly first order.} Second, the electroweak phase transition proceeds from $(0 ,v_s) \to (v_h, v_s^\prime)$ which can have a tree level barrier between the true and false vacuum catalyzing a strongly first order phase transition. In principle such a scenario could lead to exotic gravitational wave effects. In the context of  NMSSM, a phase transition where the singlet changes sharply during the electroweak phase transition can boost the efficiency of baryon production \cite{Akula:2017yfr}. Detailed phenomenological scans of a real singlet extension to the standard model have shown that a scalar as heavy as 800 GeV can still catalyze a SFOEWPT \cite{Profumo:2007wc,Profumo:2014opa}. Such a scenario would take a 100TeV collider to fully probe \cite{Kotwal:2016tex}. The requirement that a singlet must be no heavier than a 800 GeV sounds model specific leading to a question as to whether a more complicated scalar sector could push the scale of new physics even higher and still catalyze a SFOEWPT. However the result for the singlet extension of the standard model seems to agree with the effective field theory result which also sets the scale of new physics at $800$ GeV \cite{Grojean:2004xa}. Recent work has stressed the difficulties of using effective field theory during a phase transition so this result should be taken with a grain of salt \cite{deVries:2017ncy}.

Next let us consider the second possibility raised above. The idea of a symmetry being broken through multiple transitions has been mainly of interest in the case of electroweak symmetry breaking. One can break electroweak symmetry in a multistep transition when the standard model is extended by at least an additional scalar SU(2)$_L$ multiplet. The simplest case is when one adds a scalar SU(2)$_L$ triplet to the standard model along with a scalar gauge singlet to act as a catalyst. A component of the SU(2)$_L$ triplet can acquire a vacuum expectation value at high temperature before the deepest minima is in the direction of the SU(2)$_L$ doublet fields. This situation was considered in Ref. \cite{Inoue:2015pza} to catalyze electroweak baryogenesis at a higher scale and is illustrated in Fig. \ref{fig:2step}. A key attraction of this scenario is the physics that leads to the electroweak phase transition being strongly first order, a requirement for electroweak baryogenesis, can be above the electroweak scale avoiding current bounds. 

Finally the last possibility raised above is arguably the most exotic. Weinberg demonstrated by example that counter to our intuition, symmetries can be broken at high temperature \cite{Weinberg:1974hy}. Indeed this is the case in some condensed matter systems such as Rochelle salt which prompts us to consider this scenario. It was shown in ref \cite{Patel:2013zla} that if one extends the scalar sector of the standard model by a colored scalar field, one can indeed generate a scenario where the colored scalar acquires a vacuum expectation before losing it during the electroweak phase transition. If one also includes a gauge singlet the scale of both the color breaking and electroweak phase transitions can both be multi-TeV. This scenario was recently considered as a mechanism for producing the baryon asymmetry of the Universe \cite{Ramsey-Musolf:2017tgh}. Note that the colored scalar cannot be a stop \cite{Cline:1999wi} as if one tunnels into a phase where the stop has a vacuum expectation value, one cannot efficiently tunnel back into the SU(3)$_C$ symmetric phase when the Universe cools. In addition to adding extra scalars to the effective potential, one can achieve symmetry breaking at high temeprature through modifying the effective potential with large chemical potentials \cite{Senjanovic:1998xc,Bajc:1999cn,McDonald:1999he}.


\section{Effective potentials at finite temperature}

It is possible for a scalar field to acquire a vacuum expectation value. This vacuum expectation value can be space-time dependent so we can treat it like a field. It is however, a classical field rather than a quantum field as it does not have excitations that correspond to particle states. To derive the effective potential the process is to shift the scalar fields by the expectation value such that the expectation value of the scalar fields are zero.  The part of the shifted Lagrangian that is purely made up of the classical field we call the effective potential. The global minimum of the effective potential is the vacuum expectation value of the unshifted field. It is common to refer to local minima as false vacua as they can decay to the true vacuum through tunnelling. \par
Let us consider some examples. In the case of the standard model, the Higgs is complex SU(2)$_L$ doublet so it formally contains four scalar quantum fields. All four fields can acquire a vacuum expectation value so we can shift by a SU(2)$_L$ doublet of four classical fields which are the vacuum expectation values for each field. However, gauge invariance then allows us to perform a rotation such that only a single classical field is necessary. It is customary to choose the shift to be in the following direction in the internal space
\begin{equation}
    H \to \left( \begin{array}{c} \phi _1 +i \phi _2 \\ \frac{ \phi _3 +i \phi _4}{\sqrt{2}}  \end{array} \right) + \frac{1}{\sqrt{2}} \left( \begin{array}{c}
         0  \\
          v
    \end{array}
    \right)\,,
\end{equation}
where $v$ is the vacuum expectation value. For now we have omitted any possible space time dependence of $v$. The vacuum expectation value spontaneously breaks SU(2)$_L$ symmetry and its associated Noether current (as well as the associated charge) is no longer conserved. Instead the conserved charge is the one that returns zero when acting on the shifted classical fields. It is easy to see that if the Higgs has a hypercharge of $Y=1/2$, the linear combination $Q_{\rm EM} =\tau _3 +Y$ satisfies this criteria and is then the conserved charge of the theory with broken symmetry. \par
Next let us consider the case where there are two Higgs doublets with the same hypercharge. In this case we have a total of 8 scalar fields between the two SU(2)$_L$ doublets. We can again shift all 8 scalar fields with a classical field corresponding to their vacuum expectation values. Once again we can use gauge invariance to render some of the classical fields redundant. However, this time we are still left with 5 classical fields. If we use our freedom to rotate away all but one classical field on the first Higgs doublet one has
\begin{eqnarray}
    H_1 \to \left( \begin{array}{c} \phi ^1 _1 +i \phi ^1 _2 \\ \frac{ \phi ^1 _3 +i \phi^1 _4}{\sqrt{2}}  \end{array} \right) + \frac{1}{\sqrt{2}} \left( \begin{array}{c}
         0  \\
          v^1
    \end{array} \right)\,, \\ H_2 \to \frac{1 }{\sqrt{2}} \left( \begin{array}{c} \phi ^2 _1 +i \phi ^2 _2 \\  \phi ^2 _3 +i \phi^2 _4  \end{array} \right) + \frac{1}{\sqrt{2}} \left( \begin{array}{c}
         v^2 _1 +i v^2_2  \\
          v^2 _3 + i v^2 _4 
    \end{array} \right) \ .
\end{eqnarray}
In the above, the vacuum expectation value $v^2_3$ violates the same charge as the standard model case. In contrast, $v_1^2$ and $v_2^2$ violate $Q_{\rm EM}$ and $v_4^2$ is a $CP$ odd vacuum. In zero temperature equilibrium QFT, the  vacuum expectation values are fixed and it is unnecessary to consider their space time dependent behaviour. During a cosmic phase transition however, the vacuum expectation value can evolve with space and time.


\subsection{Coleman-Weinberg potential}

\begin{figure}
    \centering
    \includegraphics[width=\textwidth]{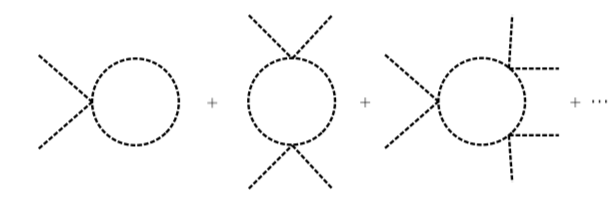}\\
    \includegraphics[width=\textwidth]{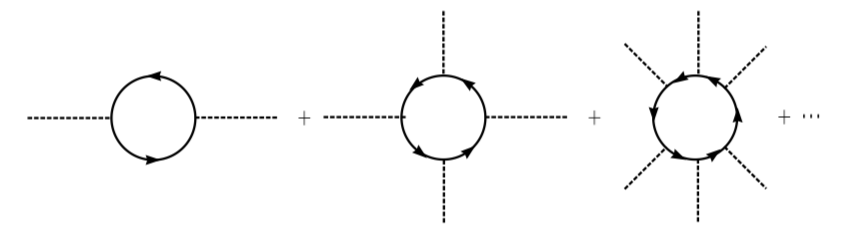}\\
    \includegraphics[width=\textwidth]{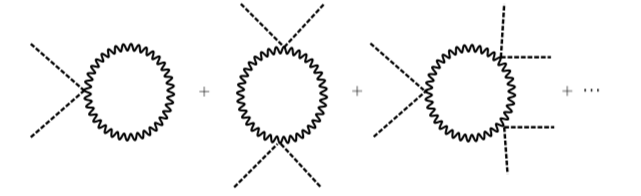}
 \caption{1PI diagrams contributing to one loop corrections to the effective potential including scalar contributions (top panel), fermion contributions (central panel) and gauge contributions (bottom panel). Figures taken from \cite{Quiros:1999jp}}
    \label{fig:1pI}
\end{figure}

The process for calculating temperature effects on the evolution of the vacuum follows the same recipe as the process of calculating loop effects albeit with finite temperature ingredients. Specifically, in the real time formalism the propagators are replaced by their finite temperature counterparts, the masses are corrected by a Debye term and the time contour changes (as will be discussed in the forth coming subsection). Therefore we first summarize the effects of zero temperature loop interactions on the effective potential and, by extension, the vacuum expectation value. \par
The shifted Lagrangian contains interactions between the classical field and quantum fields. One can therefore consider diagrams such as shown in Fig. \ref{fig:1pI}. Calculating the case of Higgs self interactions, $\lambda h^4/4!$, with the physical Higgs one has \cite{Quiros:1999jp} 
\begin{equation}
    V_{\rm self, h}= i \sum _{n=0} ^\infty \int \frac{d^4p}{(2 \pi )^4} \frac{1}{2n}
 \left[ \frac{\lambda v^2/2}{p^2-m_H^2 + i \epsilon } \right]^n\,,
 \end{equation}
whereas interactions between the classical field and the Goldstone modes at one loop gives a contribution to the effective potential
\begin{eqnarray}
    V_{\rm self, GM} =  i \sum _{n=0} ^\infty \int \frac{d^4p}{(2 \pi )^4} && \frac{2}{2n}
 \left[ \frac{\lambda v^2/12}{p^2-(m_{GB}^2+ \xi m_W^2) + i \epsilon } \right]^n \nonumber\,, \\ 
&&+  \frac{1}{2n}
 \left[ \frac{\lambda v^2/12}{p^2-(m_{GB}^2+ \xi m_Z^2) + i \epsilon } \right]^n\,,
\end{eqnarray}
where GB refers to the Goldstone mode and $\xi$ is the gauge fixing parameter.
One can see explicitly in the above that the one loop correction to the effective potential acquires both gauge and renormalization dependence. Indeed, the one loop corrections even depend on the renormalization scheme. These issues we return to later. For now we can work in the Landau gauge ($\xi=0$) as this conveniently hides the gauge dependence. We also use dimensional regularization in the 
$\overline{MS}$ renormalization scheme. We can categorize the one loop corrections to the effective potential by the virtual state in the interaction the correction corresponds to. The one loop corrections due to scalar (including goldstone boson), gauge boson and scalar interactions are respectively, with respect to the running energy $\mu$

\begin{eqnarray}
V^{CW} _S &=& \frac{m^4}{64 \pi ^2} \left( \log \left[ \frac{m^2}{\mu ^2} \right] - \frac{3}{2} \right)\,, \\
V^{CW} _{GB} &=& \frac{m^4}{64 \pi ^2} \left( \log \left[ \frac{m^2}{\mu ^2} \right] - \frac{5}{6} \right)\,,\\
V^{CW} _{F} &=& - \frac{m^4}{64 \pi ^2} \left( \log \left[ \frac{m^2}{\mu ^2} \right] - \frac{3}{2} \right) \ ,
\end{eqnarray}
where $CW$ stands for the Coleman-Weinberg potential and $S,GB$ and $f$ refer to scalars, gauge bosons and fermions respectively. One then has the total one loop correction to the effective potential as
\begin{equation}
    V_1 (T=0) =\sum _{b} n_b V^{CW} _b  -\sum _{f} n_f V^{CW}_F \,,
\end{equation}
where $n_{b/f}$ is the number of bosonic/fermionic multiplicity factors $n_t=12$, $n_W=6$ etc.


\subsection{Thermal corrections from scalars, fermions and gauge bosons}

\begin{figure}[hbtp]
\centering
\includegraphics[height=5cm]{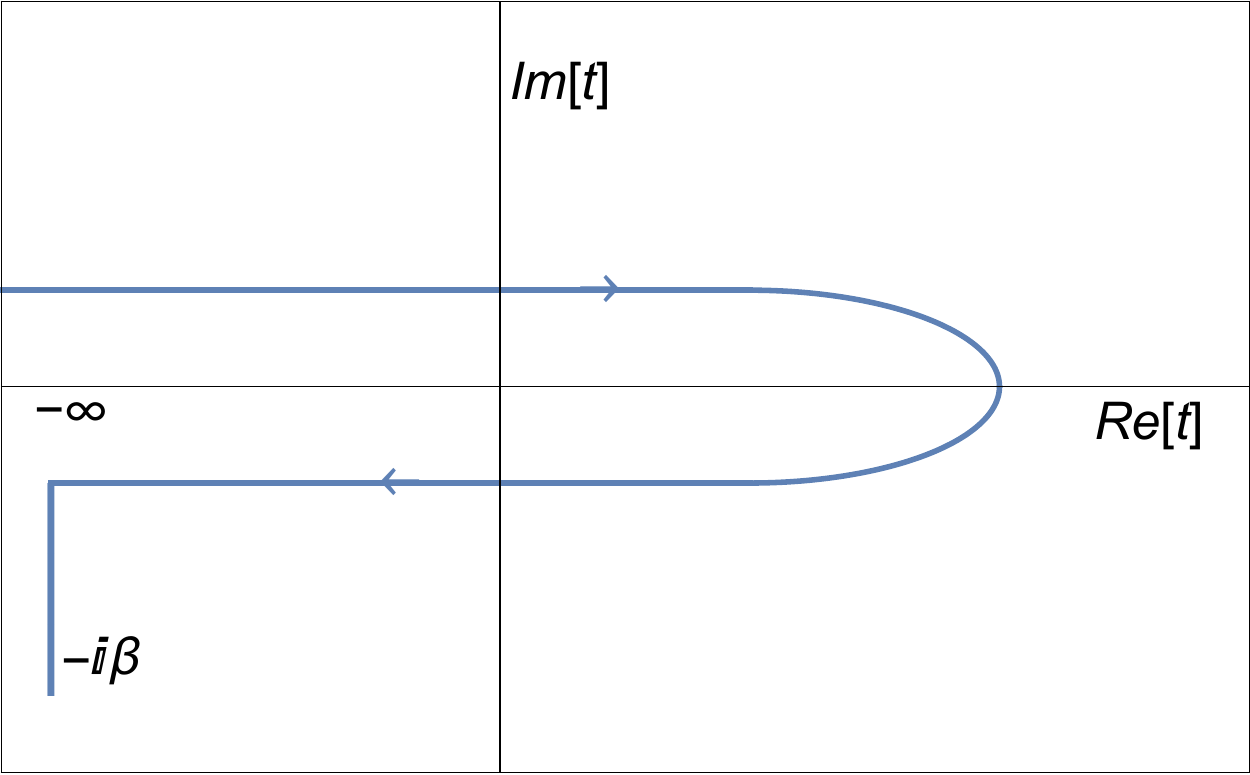}
\caption{The closed time path contour. The contour begins above the real line far in the past, evolves to the present before dropping below the real line and evolving back into the past. At $T=-\infty$ the contour evolves perpendicularly to the real line.}\label{fig:CTP}
\end{figure}

Calculating the finite temperature corrections to the effective potential essentially means repeating the previous analysis using finite temperature propagators, modifying the masses by temperature dependent ``Debye'' corrections and modifying how we treat time. There are two commonly used treatments of time at finite temperature that give the same result for the corrections to the effective potential. One involves performing calculations in imaginary time; in this case the time domain is compactified to an interval $p_0 \in [0, \beta ]$. The other approach remains in real time formalism where the time contour is modified as shown in Fig. \ref{fig:CTP}. This approach is slightly more complicated in its formalism, however a few features are more transparent. Therefore we will briefly summarize the closed time path formalism here. \par

The reason behind the different contour can be understood as follows. Suppose some system at a time $t=0$ is in equilibrium. The density matrix then has the familiar form \cite{Das:1997gg,White:2016nbo}
\begin{equation}
    \rho _0 = \frac{e^{-\beta H}}{{\rm Tr} e^{-\beta H}} \ .
\end{equation}
The time dependent density matrix can then be derived by evolving the equilibrium density matrix with time evolution operators,
\begin{equation}
    \rho (t) = U(t,0) \rho (0) U ^\dagger (t,0) \ .
\end{equation}
Note the explicit form of the time evolution operators
\begin{equation}
    U(t^\prime , t) = T \left( e^{-i \int _{t ^\prime } ^t dt ^{\prime \prime} H(t ^{\prime \prime})} \right) \ .
\end{equation}
 Here $T$ refers to time. Both the equilibrium and time dependent density matrices can be expressed in terms of the equilibrium operators
\begin{eqnarray}
    \rho (0) &=& \frac{U(T- i \beta ,T) }{{\rm Tr} U(T-i\beta ,T)} \,,\\
    \rho (t) &=& \frac{U(t,0)U(T- i \beta ,T)U(0,t)}{{\rm Tr} U(T-i\beta ,T)} \ .
\end{eqnarray}
We can then calculate the time dependent expectation value of any operator, $\langle A(t) \rangle $ and express it purely in terms of time evolution operators and $A$  \cite{White:2016nbo,Das:1997gg}
\begin{eqnarray}
    \langle A(t) \rangle &=& Tr[  \rho (t) A \rangle ] \,, \\
    &=&  \frac{{\rm Tr}U(T- i\beta , T)U( T,T^\prime )U(T^\prime ,t)AU(t, T)}{{\rm Tr} U(T- i \beta ,T)U( T , T^\prime )U(T^\prime ,T)} \ .
\end{eqnarray}
Taking $T \to - \infty$ and considering some $n$-point correlator for some source, ${\cal J}$, we see that we can interpret the above as taking the time contour given in Fig. \ref{fig:CTP}. 
As a result we have four distinct types of propagators depending on where the components of the bilinears are on the time contour. It is convenient to represent these in matrix form. We will restrict ourselves to the scalar case \cite{White:2016nbo,Das:1997gg}
\begin{eqnarray}
\tilde{\Delta }(x,y) &=& \left( \begin{array}{cc}  \Delta ^{++} & - \Delta ^{+-} \\ \Delta ^{-+} & - \Delta ^{--}  \end{array} \right)  \\ &=& \left( \begin{array}{cc}
\langle T\left[\phi (x) \phi ^\dagger (y) \right] \rangle     & -\langle  \phi ^\dagger (x) \phi (y) \rangle \\
\langle \phi (x) \phi ^\dagger (y)  \rangle     & - \langle \bar{T} \left[ \phi (x) \phi ^\dagger (y) \right] \rangle 
\end{array} \right) \ .
\end{eqnarray}
The four propagators in momentum space we give explicitly
\begin{eqnarray}
    i \Delta ^{+-} (p) &=&  2 \pi \delta \left(p^2-m(T)^2 \right) \left[ \Theta (p_0) f(\vec{p} + \Theta (-p_0)\left( 1+\bar{f} (-\vec{p}) \right) \right] \,,\\
    i \Delta ^{-+} (p) &=& 2 \pi \delta \left(p^2-m(T)^2 \right) \left[ \Theta (p_0) \left( 1+ f(\vec{p}) \right) + \Theta (-p_0) \bar{f}(-\vec{p}) \right] \,,\\
    i \Delta ^{++} (p) &=& \frac{1}{p^2-m(T) ^2+ i \epsilon } \nonumber \\ && + 2 \pi \delta \left( p^2-m(T)^2 \right) \left[ \Theta (p_0) f (\vec{p}) + \Theta (-p_0) \bar{f} (-\vec{p}) \right]  \,, \\
        i \Delta ^{--} (p) &=& \frac{-1}{p^2-m(T) ^2- i \epsilon } \nonumber \\ && + 2 \pi \delta \left( p^2-m(T)^2 \right) \left[ \Theta (p_0) f (\vec{p}) + \Theta (-p_0) \bar{f} (-\vec{p}) \right] \,.
\end{eqnarray}
These propagators can be essentially derived from unitarity and causality  \cite{Millington:2013ema}. Note that the first two propagators vanish at zero temperature. This is expected as they have no zero temperature counterpart. The last two operators (referred to as the time and anti time order propagators) contain a sum of zero temperature and finite temperature pieces. The finite temperature pieces are Boltzmann suppressed when the temperature drops well below the masses so the time and anti-time ordered propagators reduce to their zero temperature counterparts as the temperature goes to zero. \par
The finite temperature corrections to the effective potential (apart from the Debye corrections to the mass) then are produced by recalculating the one loop corrections to the effective potential, but this time with the finite temperature versions of the propagator. As an example let us consider just the one loop corrections due to the interactions between the physical Higgs and the classical field. It is actually easier to calculate the derivative of this term with respect to the mass. In this case we just need to calculate a single bubble diagram \cite{Quiros:1999jp,White:2016nbo} 
\begin{equation}
    \frac{\partial V_1}{\partial m^2(v) } = \frac{1}{2} \int \frac{d^4 p}{(2 \pi )^4 }\Delta ^{++} (p) \ .
\end{equation}
Note the appearance of the time orders propagator which, as we have stated, is a sum of the zero temperature and finite temperature pieces. Therefore we recover the zero temperature loop correction but now have an additional finite temperature piece given by \cite{Quiros:1999jp,White:2016nbo} 
\begin{eqnarray}
    \Delta V _B &=& \frac{T^4}{2 \pi ^2} n_B J_B \left(\frac{m^2}{T^2} \right) \,,
    \\
    J_B(z^2) &=&  \int _0 ^\infty dx x^2 \log \left[1-e^{\sqrt{x^2+z^2}} \right] \,,
\end{eqnarray}
where $z=m/T$ has been implicitly defined.
Note that the above function is complex for negative arguments. The imaginary parts of the effective potential corresponds to some decay which reflects an instability in the system. This issue we return to later in this section. Performing the same analysis with fermions as virtual states one then can derive the finite temperature contributions to the effective potential due to fermions \cite{Quiros:1999jp,White:2016nbo} 
\begin{eqnarray}
    \Delta V _F &=& \frac{T^4}{2 \pi ^2} n_F J_F\left(\frac{m^2}{T^2} \right)\,,
    \\
    J_F(z^2) &=&  \int _0 ^\infty dx x^2 \log \left[1+e^{\sqrt{x^2+z^2}} \right] \,.
\end{eqnarray}
The total contribution at one loop including finite temperature corrections we can then write as \cite{Quiros:1999jp,White:2016nbo} 
\begin{equation}
    V_1 (T) =\sum _{b} n_B( V^{CW} _B + \Delta V_B) -\sum _{f} n_B( V^{CW}_F + \Delta V_F) \,.
\end{equation}
For the standard model one usually includes the thermal corrections due to the top quark, the physical Higgs and the Goldstone bosons as well as the massive gauge bosons  with $n_f=12$ and $n_B\equiv \{n_H,n_{GB},n_W,n_Z \} = \{1,1,6,3 \}$.

\subsubsection{\bf Debye Masses and daisy diagrams:}

\begin{figure}
    \centering
    \includegraphics[width=0.6\textwidth]{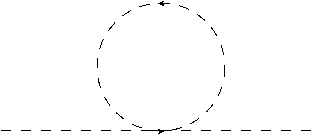}
    \caption{An example of a self energy contributing to Debye mass through a scalar loop.}
    \label{fig:scalardebye}
\end{figure}

\begin{figure}
    \centering
    \includegraphics[width=\textwidth]{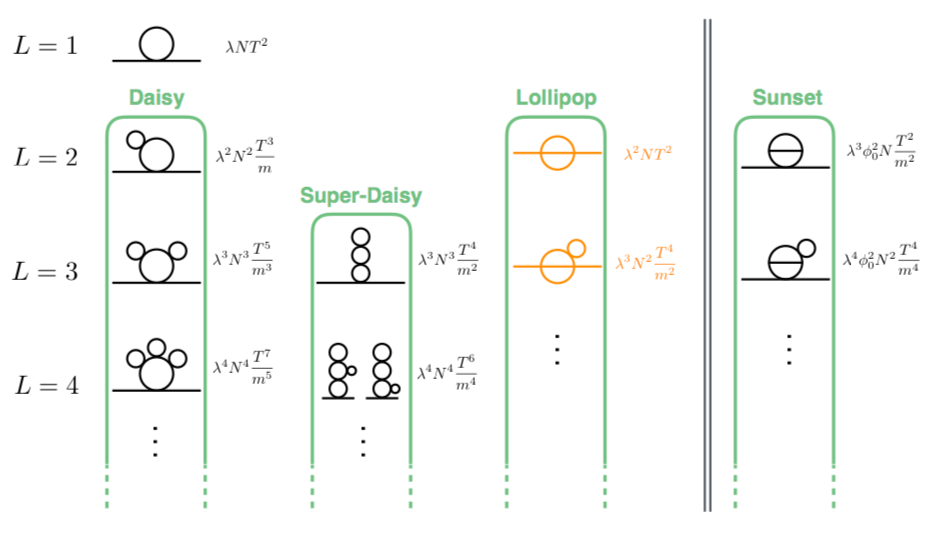}
    \caption{Various contributions to the thermal mass for a general scalar potential including super daisy and daisy contributions which we cover in detail here. Note that the lollipop and sunset diagrams require a trilinear coupling. Figure taken from \cite{Curtin:2016urg}.}
    \label{fig:daisy}
\end{figure}

\begin{figure}
    \centering
    \includegraphics[width=0.8\textwidth]{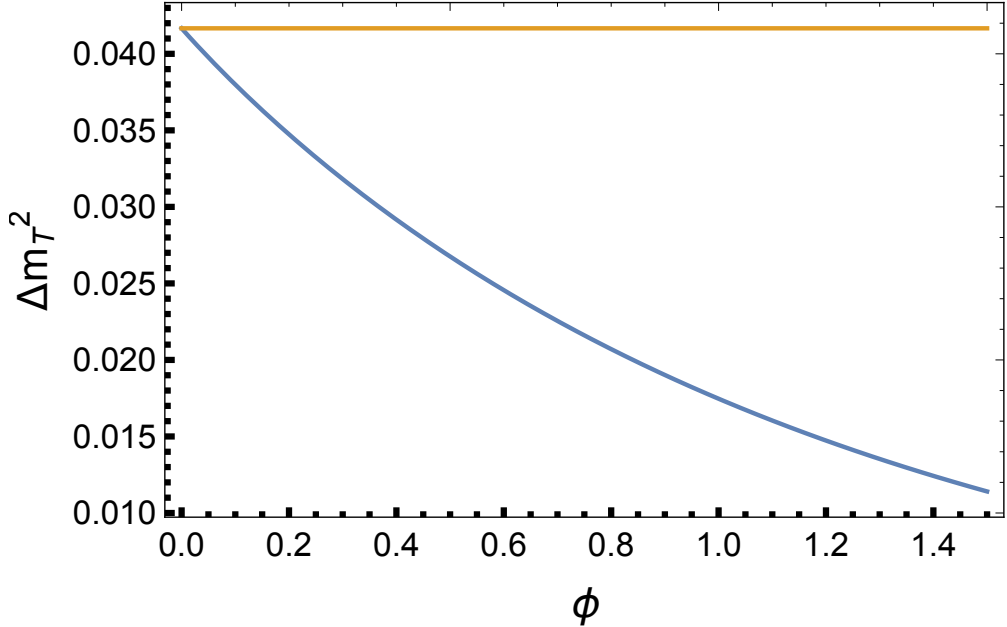}
    \caption{Thermal mass for a particle with mass $\lambda \phi$ and units scaled such that $T=1$. The gold line represents the high temperature expansion whereas the blue line represents the numerical calculation of the integral. The discrepancy is remarkably large even for $T \sim \phi$. Furthermore the numerical result shows the expected decoupling behaviour.}
    \label{fig:daisys}
\end{figure}

Hard thermal loops, $p\sim T$,  can cause perturbation theory to break down at finite temperature. One can delay the break down of perturbation theory by performing a resummation  in figure \ref{fig:daisy} which results in the shift of the pole mass by a temperature dependent Debye term. We will discuss different derivation schemes in this section as well as what limits different corrections become important.  The dangerous diagrams can be categorized as in Fig. \ref{fig:daisy}: daisy diagrams, super daisy diagrams, lollipops and Sunsets. Note that the last two types of diagrams only exist in the case where you have a dimensionful trilinear coupling. We will focus therefore on the first two. Daisy contributions of the form given in Fig. \ref{fig:scalardebye} become important when the mass of a particle is small compared to the temperature \cite{Parwani:1991gq}. Super-daisy diagrams such as the form given in Fig. \ref{fig:daisy} are important when the couplings are large and the masses are small compared to the temperature \cite{Curtin:2016urg}.
Consider the simplest possible model, a $\phi ^4$ model.
\begin{equation}
    {\cal L} = \frac{1}{2} \partial _\mu \partial ^\mu - \frac{m^2}{2} \phi ^2 - \frac{\lambda}{4!} \phi ^4\,.
\end{equation}
Diagrams such as the one given in Fig. \ref{fig:scalardebye} contribute to a thermal correction to the mass. Let us explicitly calculate such a diagram in the imaginary time regime \cite{Das:1997gg}
\begin{eqnarray}
    \Delta m^2 &=& \frac{\lambda}{2 \beta} \sum _n \int \frac{d^3k}{(2 \pi)^3} \frac{1}{\left(\frac{2n \pi}{\beta}\right)^2+k^2+m^2} \\
    &=&\frac{\lambda}{4} \int \frac{d^3 k}{(2 \pi)^3} \frac{1}{\sqrt{k^2+m^2}}+\frac{\lambda}{2} \int \frac{d^3 k}{(2 \pi)^3} \frac{1}{\sqrt{k^2+m^2}} \frac{1}{e^{\beta \sqrt{k^2+m^2}}-1} \ . \nonumber \\
\end{eqnarray}
In the limit where the temperature is large compared to the mass one can write the high temperature expansion~\cite{Kapusta:1989tk,Das:1997gg}.
\begin{equation}
    \Delta m^2 _T = \frac{\lambda T^2}{24}\,.
\end{equation}
This indeed is the typical form used for the Debye mass. If one takes the temperature dependent part of the potential evaluated at the thermally corrected mass, $V_T(m^2+\Delta m^2 )$, and performs a high temperature expansion one encounters a common approximation for the daisy contributions \cite{Arnold:1992rz}
\begin{equation}
    \frac{T^4}{2 \pi ^2}J_{B} \left( \frac{m^2+\Delta m^2}{T^2} \right)  \sim  \frac{T^4}{2 \pi ^2}J_{B} \left( \frac{m^2}{T^2} \right) + V_{\rm Daisy}\,,
\end{equation}
with
\begin{equation}
    V_{\rm Daisy} = \frac{T^4}{12 \pi} \sum \left( \left[ \frac{m^2 _i}{T^2}\right] - \left[ \frac{m^2_i+\Delta m^2_i }{T^2} \right] ^{3/2} \right) \ .
\end{equation}
The  terms $m_i^2 +\Delta m^2_i$ are the eigenvalues of $m^{ij}_{s,G} +\Pi^{ij}_{s,G}$ where $m^{ij}_{s,G}$ are the scalar and gauge boson mass matrices respectively and \cite{Basler:2018cwe}
\begin{eqnarray}
\Pi ^{ij} &=& \frac{T^2}{24} n_k (-1)^{s+1} \partial ^{kkij}  \tilde{L} - \frac{T^2}{48} n_k \partial ^{kkij}L_Y \\
\Pi ^{ab}_G &=& \frac{2}{3} T^2 \left( \frac{\tilde{n}_H}{8} + 5 \right) \frac{1}{\tilde{n}_H} \sum _{m}^{n_{\rm Higgs}} \delta _{ab} \partial _a^2 \partial _m^2 L
\end{eqnarray}
with $n_x$ the appropriate multiplicity factors, $s=(0,1)$ the spin of the boson and $\tilde{n}_H$ is the number of Higgs that couple to a guage boson. Finally the derivatives $\partial _x$ are derivatives with respect to field $x$, $L_Y$ is the part of the Lagrangian that contain Yukawa interactions and $\tilde{L}=L-L_Y$.  Note that only longitudinal gauge bosons acquire thermal mass corrections. 
It was recently shown by \cite{Curtin:2016urg} that the high temperature estimation of the daisy diagrams can become a poor approximation. As can be seen in Fig. \ref{fig:daisys} the high temperature approximation is quite poor for large values of the quartic coupling. More seriously, the high temperature expansion does not show how decoupling occurs when the mass is large compared to the temperature. A more accurate way of calculating the Debye mass is through the self consistency relation \cite{Curtin:2016urg}
\begin{eqnarray}
m_0^2(\phi )+    &\Delta m_T^2  &= \frac{\partial ^2 }{\partial \phi ^2} \left[ V_0(\phi)+V_{\rm CW}(\phi ) \right] + \frac{\partial ^2}{\partial \phi ^2} V_T(m_0^2(\phi) + \Delta m_T^2 )\,, \nonumber \\
&\Delta m_T^2  &=  \frac{\partial ^2}{\partial \phi ^2} V_T(m_0^2(\phi) + \Delta m_T^2 )\,.
\end{eqnarray}

Furthermore Ref.\cite{Curtin:2016urg} developed a scheme for calculating the superdaisy contributions, by solving the equation  \cite{Curtin:2016urg}
\begin{equation}
    V_{\rm super daisy} = \int d \phi \frac{d V_T(m^2(\phi ) +\Delta m_T ^2 )}{d m^2(\phi )} \frac{d m^2(\phi )}{d \phi } \ .
\end{equation}


\subsubsection{\bf Gravitational corrections:}
Gravitational corrections to false vacuum decay suppress the decay rate slightly. Consider the action of a single scalar field coupled to gravity defined by the Euclidean action \cite{Coleman:1980aw,Salvio:2016mvj}
\begin{equation}
    S= \int d^4x \sqrt{g} \left[ \frac{1}{2} \partial _\mu h \partial ^\mu h + V(h) - \frac{R}{2 \kappa } - \frac{R}{2} f(h)  \right]\,,
\end{equation}
where $\kappa = 8 \pi G = 8 \pi /M_{\rm P}^2$ and the metric is given by $ds^2 = dr^2 + \rho (r) ^2 d\Omega ^2$, where $d\Omega^2$ contains the angular part of the metric. In this metric the Ricci scalar has the simple form
\begin{equation}
    R= \frac{-6 (\rho ^2 \rho ^{\prime \prime} + \rho \rho ^\prime {}^2 -\rho )}{\rho ^3}\,,
\end{equation}
where the prime is a derivative with respect to the radial coordinate, $r$. The equations of motion are
\begin{eqnarray}
h^{\prime \prime } + 3\frac{\rho ^\prime}{\rho} h^\prime &=& \frac{dV}{dh} -\frac{1}{2} \frac{df}{dh}R \\
\rho^\prime {}^2 &=& 1 + \frac{\kappa \rho ^2}{2(1+\kappa f(h))} \left(\frac{h^\prime {}^2}{2} - V - 3 \frac{\rho ^\prime }{\rho} \frac{df}{dh} h^\prime \right) \ .
\end{eqnarray}
One can find approximate solutions to the above equations of motion by expanding in the Planck mass, 
\begin{eqnarray}
h(r) &\sim & h_0 (r) + \kappa h_1 (r)\,, \\
\rho (r) & \sim & r+\kappa \rho _1 (r)\,,
\end{eqnarray}
where $\kappa=\sqrt{8\pi G}$ and $h_0$ is the bounce solution without gravitational corrections. The change in the action, $S=S_0+\Delta S$, where $S_0$ corresonds to the action of the bounce has a formidable form \cite{Salvio:2016mvj}
\begin{eqnarray}
\Delta S &=& \frac{48 \pi^2}{M_{\rm P}^2} \int dr \left[ r^2 \rho _1 \left( \frac{1}{2} h_0^\prime {}^2   + V(h_0) \right)  \right. \nonumber \\ && \left. + (r \rho _1 ^\prime {}^2 +2 \rho _1 \rho _1 ^\prime +2 \rho _1 r \rho _1 ^{\prime \prime } + r f(h_0) (r \rho _1 ^{\prime \prime } + 2 \rho _1 ^\prime ) \right]  \ .
\end{eqnarray}
One can greatly simplify the above expression by making the rescaling $\rho _1 \to \rho _1 \times s$ where the value of s is chosen by requiring the derivative with respect to s to vanish at $s=1$. The change in the effective action then has the remarkably simple form
\begin{equation}
    \Delta S = \frac{48 \pi ^2}{M_{\rm P } ^2} \int dr r \rho _1 ^\prime {}^2 \geq 0 \ . 
\end{equation}
Note that, as we indicate, this correction to the effective action from gravitational effects is always positive indicating that the tunneling rate in turn is suppressed by gravitational corrections. Also it is useful to note that the first order correction to $\rho$ is independent of the gravitational corrections to the bounce action \cite{Salvio:2016mvj}
\begin{equation}
    \rho _1 ^\prime = \frac{r^2}{6} \left[ \frac{1}{2} h_0^\prime {}^2 -V- \frac{3}{r} f^\prime (h_0) h_0 ^\prime  \right] \ .
\end{equation}
The correction tends to be quite small for weak scale phase transitions, scaling as $\sim v^6/(\Lambda^4 M_{\rm P} ^2)$ where $v$ is the vacuum expectation value of the non trivial minimum (assuming the false vacuum is at the origin in field space) and $\Lambda$ is the scale of the potential.


\subsubsection{\bf Finite number density contributions:}
Consider a scalar field, $\phi$, with a Global $U(1)$ symmetry which corresponds to a Noether charge $Q$. For a charge density $n=Q/V$ the effect of a non-zero charge density for $T>m$ is 
\cite{Senjanovic:1998xc,Haber:1981ts,Benson:1991nj}
\begin{eqnarray}
    V(n,T,\phi) &\sim & V(\phi ,T)- \frac{1}{2} n^{a} (M^{-1})_{ab} n^b \nonumber \\ 
    &\sim &\frac{n^2}{2\lambda (3 |\phi|^2+T^2/2)}+\frac{\lambda}{4} T^2 |\phi |^2 +\frac{\lambda }{4} |\phi |^4\,,
\end{eqnarray}
where we have indicated how to generalize to more complicated models in the first line of the above equation. If the number density scales with the temperature cubed then the potential actually grows a minimum at high temperature. In this case the symmetry breaking isn't caused by the microphysics of the field's couplings and mass, but is instead caused by the macroscopic conditions.


\subsection{Topical theoretical issues} 

\subsubsection{\bf Gauge invariance:} 
The gauge dependence of the effective potential has been the subject of much debate. Some have approached the issue by arguing for the benefits of a particular gauge (usually the Landau gauge because of the simplifications that there is no mixing between longitudinal vector boson and goldstone modes and that the gauge fixing parameter is not renormalized). Others have proposed novel solutions \cite{Buchmuller:1994vy,Nielsen:2014spa}. A couple particularly creative ones involve coupling the source to a composite field \cite{Buchmuller:1994vy}. Unfortunately \cite{Hu:1996qa} argued that one cannot calculate a finite effective potential this way if there are more than 3 space time dimensions. Another approach was to demonstrate that a gauge independent effective potential can be derived via a field redefinition \cite{Nielsen:2014spa}. While ingenious, this approach invites the criticism that an effective potential that is sensitive to field redefinitions isn't an improvement on one that is sensitive to gauge transformations. 
Recent work by Schwarz et al \cite{Andreassen:2014gha} argued that the effort to produce a gauge invariant version of the effective action was misguided. They argue that the effective action itself is unphysical as its construction involves a test of how the system responds to an external source, $J$. If $J\neq 0$ this is a non-dynamical background charge density. This charge density does not couple to the gauge bosons which means Ward identities are violated \cite{Andreassen:2014gha}. The $J=0$ case means that $\phi _0$ is extremal. They then argue that all physical quantities are gauge invariant, demonstrating that one can write the minimum of the potential to two loops if one expresses things in a gauge invariant scale, $\mu _X$. The trick is then to do an expansion in $\hbar$ rather than the usual loop expansion as such an expansion manifestly conserves gauge invariance order by order whereas the usual perturbative expansion fails in this regard. Explicitly one has in a $\hbar$ expansion \cite{Patel:2011th,Andreassen:2014gha}
\begin{eqnarray}
    v &=& v_0 + \hbar v_1 +\hbar ^2 v_2 + \cdots \\
    V &=& V^{\rm LO}+\hbar V^{\rm NLO} +\hbar ^2 V^{\rm NNLO}
\end{eqnarray}
One can then express the minimum of the potential, itself a physical quantity, in terms of a physically meaningful scale, $\mu _X$ by grouping terms together in a $\hbar$ expansion as follows  \cite{Patel:2011th,Andreassen:2014gha}
\begin{eqnarray}
   && V_{\rm min} = V^{\rm LO}(\mu _X)+\hbar \left( V^{\rm NLO}(\mu _X) +v_1 V^\prime {}^{\rm NLO} (\mu _X) \right) \nonumber \\ &+& \hbar ^2 \left( V^{\rm NNLO}(\mu_X) + v_1 V^{\prime} {}^{NLO} (\mu _X) + v_2 V^{\prime }{}^{\rm LO}( \mu _X)+\frac{v_1^2}{2} V^{\prime \prime} {}^{\rm LO} (\mu _X) \right) \ . \nonumber \\
\end{eqnarray}
At finite temperature, one is often interested in calculating the order parameter as a measure of the strength of the phase transition. Explicitly it is the ratio of the vev at the non-trivial minima to the temperature at which the potential has degenerate minima, $T_C$. One approach was to expand $T_C$ in a $\hbar$ expansion. Such an approach suffers from infrared divergences. Another approach, is to expand both the minima and the  potential evaluated at the minima in a $\hbar$ expansion. Such an approach seems sufficient for calculating the physical sphaleron energy which is the true quantity controlling whether the yield of any particle produced during a phase transition is washed out. 


\subsubsection{\bf Model dependence of the order parameter:}

A popular measure of the strength of the phase transition is through the order parameter $\phi _C/T_C$ where the critical temperature, $T_C$, is the temperature at which the minimum is degenerate and $\phi_C$ is the value of the non-trivial minimum at the critical temperature. The order parameter is not gauge invariant and a gauge invariant treatment is given in \cite{Patel:2011th}. Generally many just use the Landau gauge to calculate the order parameter. In baryogenesis one needs the phase transition to be sufficiently strong such that the phase with broken electroweak symmetry has electroweak sphalerons sufficiently suppressed such that a sufficient percentage of any baryon asymmetry produced through CP violating interactions with the bubble wall is preserved and not washed out \cite{Patel:2011th}. A rule of thumb that gets used in the literature is that the phase transition is sufficiently strong if $\phi _C/T_C \geq 1$~\cite{Quiros:1999jp}. 

A more precise condition involves calculating how much initial baryon asymmetry is preserved during the transition. 
 However, the sphaleron rate depends on the sphaleron energy and the fluctuation determinant. Both of these are very model dependent and even in the standard model with a variable higgs mass the true condition can range from $\phi_c/T_c > [0.7,1.5]$ \cite{Patel:2011th,Fuyuto:2015jha}. 


\subsubsection{\bf Imaginary part of effective potential:}

At both zero and finite temperature the loop corrections to the effective potential is not real everywhere. This occurs when the mass squared for some values of the classical field the mass squared of the physical Higgs and goldstone bosons can be negative leading to complex logarithms in the Coleman Weinberg potential as well as complex contributions from the thermal functions. Furthermore daisy contributions also can be responsible for imaginary contributions. This leads to two related theoretical issues: first the effective potential is convex by construction and yet a negative mass squared appears to contradict this, second what is the physical interpretation of the imaginary components of the effective potential. \par 
The first theoretical issue is known as the convexity problem where the effective potential is convex by construction and yet we frequently encounter effective potentials which are definitely not convex everywhere. The solution to this problem is found in merely bringing clarity to what it means to say the effective potential is convex by construction \cite{Dannenberg:1987fw,Plascencia:2015pga}.
The effective action is derived as the functional Legendre transform of $W[J(x)]=-\ln \langle 0 ^+ | 0^- \rangle _J $. This implies that the effective action is concave and the effective potential is convex. This follows from the definition of a Legendre transform, $L(p)={\rm min}[xp-f(x)]$, which can be written as,  $x_0 p - f(x_0(p))$, where $x_0(p)$ satisfies the conditions that $\partial _x f(x_0) = p $ and $\partial ^2_x f(x_0) \leq 0$. In other words the  Legendre transform of $f$ is concave by definition.
Now the effective action is the Legendre transform, $\Gamma [\bar{\phi }] = {\rm min} [ J \phi - W(J)]$, which is concave.
When one calculates the effective action by summing 1PI diagrams one finds a non-convex effective potential. $\Gamma [\phi]$.
However one does not require, $\Gamma [ \phi ] = \Gamma [\bar{\phi}]$, the latter is the concave envelope of the former (and $V(\bar{\phi } )$ is the convex envelope of $V(\phi)$). The two are equivalent for a constant background evaluated at the absolute minimum.

The second issue is a little more subtle. It was shown by Weinberg and Wu \cite{Weinberg:1987vp} that the imaginary contributions to the effective potential have the interpretation of decay processes. The decay in question is not the scalar fields decaying into other particles. This can be demonstrated from the fact that even a theory with a single scalar field that has no decay modes still obtains imaginary contributions. The decay also is not the non-perturbative decay of the false vacuum as these imaginary contributions are perturbatively derived. In fact, the decay corresponds to an instability in the system where fluctuations around the classical field become large in a set of uncorrelated domains of size $(V^{\prime \prime}) ^{-1/2}$. Within each domain the fluctuations grow exponentially with time and the system becomes unstable. This instability becomes important when the false vacuum is decaying and one needs to expand around a space time varying background that includes regions of negative field curvature. If the imaginary part of the effective potential is large compared to the real part then the system is unstable and the usual process of calculating bubble nucleation may be invalid.

This is best understood in direct analogy with the Schwinger effect in electromagnetism~\footnote{We thank Hiren Patel for clarifying this issue}. If the electric field in some volume, $V$, is strong enough, electron anti-electron pairs will be spontaneously produced via interactions between the vacuum and the strong electric field. These electron anti-electron pairs will split, aligning themselves with the background electric field which lowers both the background and the energy of the system. The Schwinger effect can be formally understood in terms of effective actions. Defining the one loop correction to the Lagrangian density as ${\cal L}_1$ one can write the pair creation rate per unit volume and time as \cite{Dunne:2008kc}
\begin{equation}
    \Gamma = -2 {\cal L}_1 = \frac{1}{4 \pi ^3} \left(e E \right)^2 \sum _{n=1}^\infty \frac{1}{n^2} {\rm exp}\left[ -n \frac{\pi m_e^2}{|e| E} \right] \ . 
\end{equation}
Similarly the imaginary components of the effective potential correspond to spontaneous production of scalar quanta. This percolation serves to drive the field to the inflection point lowering the total energy to a point that is higher than the minimum. Note that this lowering of the energy is a purely quantum mechanical effect that is different from a classical roll. Thermal corrections are also responsible for an imaginary component to the finite temperature version of the effective potential. These components arise from the fact that we have calculated the effective potential under the assumption of equilibrium and if there is an imaginary component it means that you have a thermal instability to your equilibrium state. The thermal imaginary components will then proceed to take the system out of equilibrium.

It was recently shown that for the standard model case the imaginary parts of the ring sum term effectively cancel the imaginary parts of the one loop corrections guaranteeing the stability in this case. Explicitly one has in the high temperature limit \cite{Delaunay:2007wb}
\begin{eqnarray}
    {\rm Im}\left[ V_1 (T) \right]  &\to & \sum _{i \in \{h,GB \}} \Theta \left( -m_i^2 \right) n_i \left[ -\frac{|m_i|^4}{64 \pi} + \frac{|m_i|^3T}{12 \pi} \right] \nonumber \\
    && + \sum _{i \in \{h,GB \}} \Theta \left( -m_i^2 \right) n_i \left[ \frac{|m_i|^4}{64 \pi}  \right] \,,\nonumber \\
     {\rm Im}\left[ V_{\rm Ring} (T) \right]  &\to & \sum _{i \in \{h,GB \}} \Theta \left( -m_i^2 \right) n_i \left[ - \frac{|m_i|^3T}{12 \pi} \right] \,. \\
\end{eqnarray}
At high temperature these contributions can cancel for the standard model case (apparently also do at $m \sim T$ which isn't much of a surprise since the high temperature limit holds very well until $m \sim 2 T$). So for the standard model case at least this presents no issue at least in terms of the stability of the system during a phase transition. However, one should note that the cancellation occurs only in the finite temperature expansion and small imaginary component remains even at the origin when electroweak symmetry is restored.


\subsubsection{\bf Back reaction of the soliton:}
To self consistently calculate the tunneling rate which is relevant for calculating various thermodynamic parameters, one needs to include the correction that is due to the fact that one has a space time varying field configuration. Recent works \cite{Garbrecht:2015cla,Garbrecht:2015yza,Garbrecht:2017idb} have addressed this issue and suggest the following recipe
\begin{itemize}
    \item[1] find the approximate bounce solution to the classical equations of motion
    \item[2] Insert the bounce solution into the equation for the Greens function and find the new greens function that solves for the case of a $\phi^4$ theory with quartic term $(\lambda /4! ) $
    \begin{equation}
        (-\Box + \frac{\lambda}{2} \phi^2) G(x,y) = \delta (x-y) \ ,
    \end{equation}
    where as before $G$ is a propagator.
\item[3] Calculate the tadpole corrections $\Sigma ^R$ renormalizing in the homogeneous false vacuum
\item[4] Insert tadpole into the equations of motion
\begin{equation}
    -\Box \phi +\Sigma ^R \phi = 0\,,
\end{equation}
where $\Sigma ^R= \lambda  S(\phi) +\delta \Sigma$ and $\delta \Sigma $ is contains all the relevant counter terms. Solve the bounce which now solves this corrected equation of motion. 
\item[5] Repeat steps 2-4 until one has convergence.
\end{itemize}
The corrections were found to be very small in the thin wall regime \cite{Garbrecht:2015yza,Garbrecht:2017idb} but are expected to be more relevant when one is beyond the thin wall limit.

\section{Examples of phase transitions}

\subsection{Electroweak phase transition}

\begin{figure}
    \centering
    \includegraphics[width=0.8\textwidth]{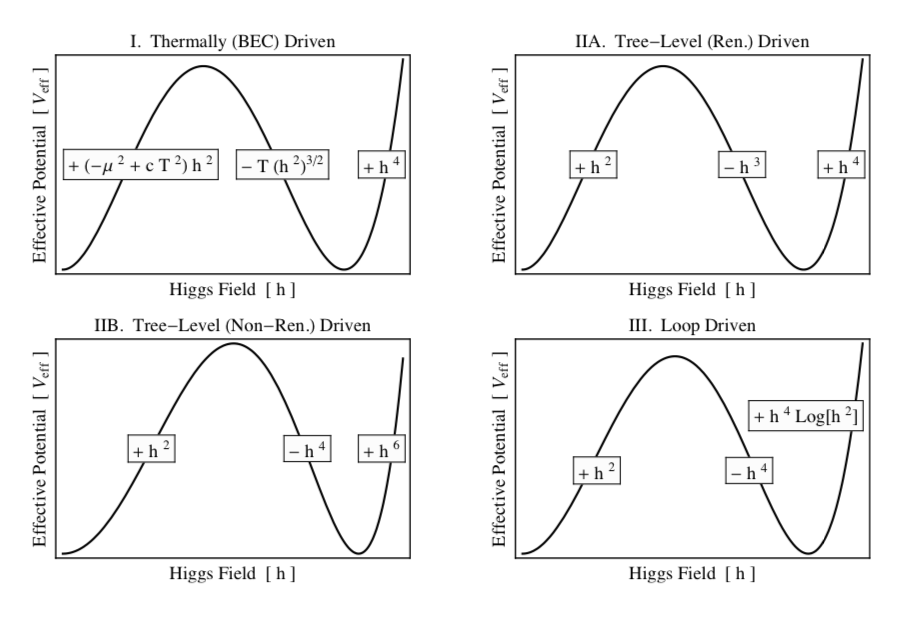}
    \caption{Model classes for catalyzing a strongly first order electroweak phase transition. From top left to bottom right the barrier is caused by thermal loops, tree level triscalar interactions, non-renormalizable operators and Coleman Weinberg corrections respectively.  Figure taken from \cite{Chung:2012vg}}
    \label{fig:Chung:2012vg}
\end{figure}

\begin{figure}
    \centering
    \includegraphics[width=0.8\textwidth]{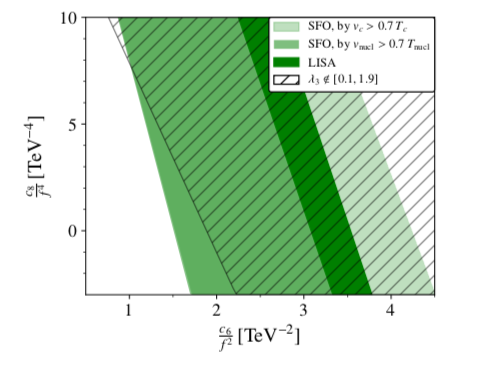}
    \caption{Values of Wilson coefficients for $c_6/(8f^2)h^6$ and $c_8/(16f^4)h^8$ operators with LISA and future $e^+e^-$ circular collider (FCC-ee) constraints given. Note $\lambda _3$ refers to the usual Higgs coupling and FCC-ee can probe values outside the interval $\lambda _3 \in [0.1,1.9]$. Figure taken from \cite{Chala:2018ari}.}
    \label{fig:Chala:2018ari}
\end{figure}

No other phase transition gains as much attention as the electroweak phase transition \cite{Kuzmin:1985mm,Shaposhnikov:1987tw,McLerran:1990zh,Farrar:1993sp,Rubakov:1996vz,Cline:2008hr,Kozaczuk:2014kva,Basler:2018cwe,Profumo:2007wc,Enqvist:1991xw,Patel:2011th,Cline:2009sn,Cline:1996cr,Chao:2014ina,Noble:2007kk,Huang:2014ifa,Ham:2007wu}. Reheating models generically tend to predict a reheating temperature high enough to restore electroweak symmetry.\footnote{One caveat is that it is possible through the introduction of many singlets for electroweak symmetry to not restore at high temperature \cite{Meade:2018saz}} Furthermore, if electroweak symmetry is broken during the cooling of the Universe after reheating via a strongly first order electroweak phase transition, the baryon asymmetry of the Universe can be generated during the phase transition (at the small cost of diluting thermal relics \cite{Wainwright:2009mq}). 
Let us begin with the standard model to understand why it does not accommodate a strongly first order phase transition and how to extend the SM to catalyze such a scenario. We will follow the conceptual organization of \cite{Chung:2012vg} which categorized the different classes of extensions to the Standard model that can successfully accommodate a strongly first order electroweak phase transition. The Higgs potential in the standard model in the high temperature expansion expressed in the Landau gauge is given by \cite{Quiros:1999jp}
\begin{equation}
    V = D(T^2-T_0^2) h^2 -ET h^3 + \frac{\lambda}{4} h^4 \,,\label{eqn:ewpt}
\end{equation}
where \cite{Quiros:1999jp}
\begin{eqnarray}
    D&=& \frac{g_1^2+3g_2^2 +4y_t^2+8\lambda }{32}\,, \nonumber \\
    T_0 &=& \frac{m_H^2}{4 D} \,,\nonumber \\
    E&=& \frac{3}{96 \pi} \left(2 g_1^3 + (g_1^2+g_2^2)^{3/2} \right) \ ,
\end{eqnarray}
where in the above $g_1$ and $g_2$ are the standard model gauge boson couplings. The strength of the phase transition is given by \cite{Quiros:1999jp}
\begin{equation}
    \frac{\phi _c}{T_c} = 2 \frac{E}{\lambda } = 4 E \frac{v^2}{m_H^2} \sim \frac{2000({\rm GeV})^2}{m_h^2} \ .
\end{equation}
So for a Higgs mass of $125$ GeV one has a very weak first order phase transition with an order parameter $\sim 0.1$. In reality, lattice simulations indicate that the electroweak transition is a smooth crossover. To boost the strength of the electroweak phase transition there are four model classes to achieve this which are depicted in figure \ref{fig:Chung:2012vg}. These are 
\begin{itemize}
    \item[I] Boost the effective $E$ parameter in  Eq.\ref{eqn:ewpt} by a factor of at least $5$. This can only be achieved through the introduction of new bosonic degrees of freedom that acquire a part of their mass through electroweak symmetry breaking. Also the total mass of the new boson cannot be too heavy as the cubic term is only manifest when the high temperature expansion is valid - that is when the temperature is large or comparable to the mass. Above such a mass the thermal contribution from such a boson is heavily Boltzmann suppressed.  The most celebrated example of such an approach is the light stop scenario \cite{Carena:1996wj}. Such a scenario is very efficient as the contribution to $E$ has a multiplicity factor of $12$. However, the light stop scenario is highly constrained as it requires a stop lighter than the SM top quark and it is difficult, but not impossible, for such a light stop to evade detection \cite{Liebler:2015ddv}. Another possibility that is equally efficient and can evade detection is that of folded supersymmetry \cite{Katz:2014bha} where the SU(3)$_C$ quantum numbers of the stop is not the standard model colour. It is also in principle possible to boost the value of $E$ through light scalars fields. 
    \item[IIA] The second scenario attempts to introduce a tree level effective cubic term to provide a barrier between the true and false vacuum during the phase transition (a barrier that can persist at zero temperature). Such an operator is forbidden due to gauge invariance unless there are additional scalar fields \cite{OConnell:2006rsp}. Such a scalar field must have their vev also substantially change during the phase transition. 
    \item[IIB] In this scenario the barrier between the true and false vacuum is created by the effective quartic changing signs and the vacuum is stabilized by the non-renormalizable sextet term. Such a theory can be an effective theory that is valid up to the cutoff scale $\Lambda $. The scale of new physics needs to be relatively low compared to the standard model - between about $500$ and $800$ GeV - in order to catalyze a strongly first order electroweak phase transition \cite{Grojean:2004xa}. If the cutoff is too low then the tunneling probability becomes large compared to the age of the Universe . If the cutoff is too high then the effect of the new physics is too feeble to catalyze a strongly first order electroweak phase transition. Recent work has demonstrated that the dimension 8 operators are also important for the electroweak phase transition \cite{Chala:2018ari}. The dependence on the dimension 6 and dimension 8 Wilson Coefficients we show in Fig. \ref{fig:Chala:2018ari}. 
    \item[III] Perhaps the least explored option of the four is to induce a large contribution from the Coleman Weinberg potential to catalyze a strongly first order electroweak phase transition. For instance in the case where one has a large number of inert scalar singlets (say 12 or more) the contribution to the Coleman Weinberg potential can be large enough to catalyze a strongly first order electroweak phase transition. A more recent paper achieved this with the addition of two fermion fields \cite{Egana-Ugrinovic:2017jib}.
\end{itemize}
On top of these possibilities some more exotic possibilities include having cosmologically varying Yukawa couplings \cite{Baldes:2016rqn} or cosmologically varying the gauge coupling such that a strongly first order EWPT is catalyzed by a QCD transition at a higher scale \cite{Ipek:2018lhm}.
\begin{table}[]
    \centering
    \begin{tabular}{c|c|c}
    Model    & Couplings & Wilson coefficient of $H^6$ \\ \hline
    $\mathbb{R}$ Singlet     &$-\frac{1}{2}\lambda _{HS} |H|^2 S^2-g_{HS} H^\dagger H S $ & $-\frac{\lambda_{HS}}{2} \frac{g^2_{HS}}{M^4}$ \\
    $\mathbb{C}$ Singlet & $-g_{HS}|H|^2 \Phi -\frac{\lambda_{H \Phi}}{2} |H|^2 \Phi^2-\frac{\lambda_{H \Phi}^\prime }{2} H^\dagger H |\Phi |^2 +h.c.  $ & $-\frac{|g_{HS}|^2\lambda ^\prime _{H\Phi}}{2M^4} - \frac{{\rm Re}[g_{HS}^2\lambda _{H\Phi }] }{M^4}$ \\
    2HDM &$ -Z_6 |H_1|^2 H_1^\dagger H_2-Z_6^* |H_1|^2 H_2^\dagger H_1 $ &  $\frac{|Z_6|^2}{M^2}$ \\
    $\mathbb{R}$ triplet & $g H^\dagger \tau ^a H \Phi ^a-\frac{\lambda _{H\Phi }}{2} |H|^2 |\Phi ^a|^2 $ &$-\frac{g^2}{M^4}\left(\frac{\lambda _{H\Phi} }{8} -\lambda \right)$ \\
    $\mathbb{C}$ triplet & $ g H^T i\sigma _2 \tau ^a H \Phi ^a-\frac{\lambda_{H \Phi }}{2}|H|^2|\Phi ^a|^2$  & $-\frac{g^2}{M^4}\left(\frac{\lambda _{H\Phi} }{4} + \frac{\lambda ^\prime}{8}-2\lambda \right)$ \\ & $-\frac{\lambda^\prime}{4}H^\dagger \tau ^a\tau ^b H \Phi^a(\Phi^b)^\dagger +h.c. $ & \\
    $\mathbb{C}$ $4-$plet & $-\lambda _{H3 \Phi} H^* _i H^*_j H^*_k \Phi ^{ijk} +h.c. $ & $\frac{|\lambda _{H3 \Phi}|^2}{M^2} $
    \end{tabular}
    \caption{List of operators in scalar extensions that lead to a non-zero Wilson coefficient for the $H^6$ operator necessary to catalyze a strongly first order electroweak phase transition through mechanism $IIB$. Note the Wilson coefficient for $H^6$ must be positive to catalyze the electroweak phase transition. Notation and results taken from \cite{Corbett:2017ieo}.}
    \label{tab:H6}
\end{table}

With the light stop scenario becoming more constrained, most phenomenology research focuses on IIA and IIB scenarios. Both of these types of phase transition often causes substantial supercooling which also implies a larger gravitational wave signal as we will see in forth coming sections. For type IIB scenarios the Higgs quartic being negative is a generic consequence of fixing the mass and vacuum expectation value to agree with experiment in the presence of a sizable positive Wilson coefficient for a $H^6$ operator.  Some examples of extended scalar sectors that generate this operator are given in Table \ref{tab:H6}. Note from the form of the Wilson coefficients given in this table it is of course not guaranteed it has the needed positive sign and some models more easily accommodate this than others. 

For scenarios of type IIA one requires an effective trilinear coupling to provide a barrier between the true and false vacuum. For a second field, $\phi$, the electroweak phase transition proceeds along the field space path $(0,v_\phi) \to (v_h,v_\phi ^\prime)$ where in principle $v_\phi$ can be negligible. Rotating and shifting to field space coordinates $\varphi = a h +b( \phi + v_\phi)$ and $\vartheta = -b h +a ( \phi + v_\phi)$ where $a$ and $b$ are chosen so that in the rotated and shifted coordinates the phase transition proceeds as $(0,0) \to (0,v_\varphi)$. Even though a term of the form $h^3$ is forbidden by gauge invariance, if a trilinear coupling between $h$ and $\phi$ exists then in the rotated coordinates this leads to a term of the form $\varphi ^3$. As an example consider the real singlet, $S$, with potential \cite{Profumo:2014opa}
\begin{eqnarray}
V&=&-\mu^2 |H|^2 +\lambda |H|^4+ \frac{a_1}{2}|H|^2 S+\frac{a_2}{2}|H|^2S^2\nonumber \\ &&+ \frac{b_2}{2}S^2 + \frac{b_3}{3} S^3+\frac{b_4}{4}S^4 \ .
\end{eqnarray}
After rotating to the coordinates $v_s = \varphi (T) \sin \alpha (T) , \ v_h/\sqrt{2} =\varphi (T) \cos \alpha (T) $ and ignoring the resulting linear term (which means ignoring the existence of a high temperature singlet vev) one has \cite{Profumo:2014opa}
\begin{equation}
    V(T,\varphi) =  D(T^2-T_0^2) \varphi^2 +E \varphi ^3  + \lambda ^\prime \varphi ^4\,,
\end{equation}
with
\begin{eqnarray}
    D&=& \frac{g_1^2+3g_2^2 +4y_t^2+8\lambda }{32}\,, \nonumber \\
    T_0 &=& \frac{-\mu ^2 \cos ^2 \alpha-\frac{b_2}{2} \sin ^2 \alpha }{ D}\,, \nonumber \\
    E&\sim& \frac{a_1}{2} \cos^2  \alpha \sin \alpha +\frac{b_3}{3} \sin ^3 \alpha \,,  \nonumber \\ 
    \lambda ^\prime &=& \lambda \sin ^4 \alpha + \frac{a^2}{2} \sin^2 \alpha \cos ^2 \alpha +\frac{b_4}{4} \sin ^4 \alpha \ ,
\end{eqnarray}
where we remind the reader that $\alpha$ is the angle of the phase transition in field space.
In the above we have ignored the small corrections to the effective cubic term due to the gauge bosons. Note that the trilinear couplings $a_1$ and $b_3$ enter directly into the effective cubic term. To generate a large enough effective cubic to catalyze a strongly first order electroweak phase transition one usually has $a_1$ as quite sizeable, $-1000GeV\leq a_1\leq -100GeV$. In order to comply with LHC constraints on the zero temperature mixing angle between the singlet and Higgs, one requires that the other portal coupling $a_2$ be large and anti correlated with $a_1$ to supress the mixing angle. For a sub $TeV$ singlet mass, current constraints on the mixing angle are $|\sin \theta| \leq 0.2$ \cite{Chalons:2016jeu,Ilnicka:2018def} with this bound expected to tighten with future colliders \cite{Chang:2018pjp,Kotwal:2016tex}.

For both IIA and IIB type phase transitions, one can have a barrier between the true and false vacuum that is so large at zero temperature that the false vacuum decay rate is never fast enough compared to Hubble for the phase transition to proceed. A recent proposal demonstrates that one can have the QCD transition reduce the barrier between the true and false vacuum \cite{vonHarling:2017yew}. Specifically in a Randall Sundrum model the radion potential acquires a contribution from gluon condensates. The contribution is negative and becomes important near the origin thus it removes some of the barrier between true and false vacuua. Therefore when the gluons form a condensate electroweak symmetry breaking can occur. Thus the electroweak phase transition could occur at a much lower scale than usual. Alternatively, it was recently shown that if the electroweak phase transition occurs in two steps, the scale of the electroweak phase transition can be multi-TeV \cite{Patel:2012pi}.


\subsubsection{\bf QCD phase transition: an example of fermion condensation:}
\begin{figure}
    \centering
    \includegraphics{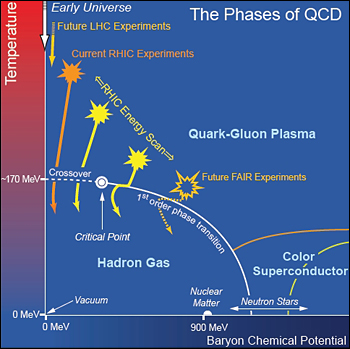}
    \caption{Phase diagram of QCD as a function of temperature and baryon chemical potential. Note that in the absense of a large chemical potential the QCD is expected to have a crossover transition. Image taken from \cite{Mohanty:2013yca}}
    \label{fig:QCDPT}
\end{figure}

The QCD phase transition generally occurs when the temperature of the Universe is $170$ MeV assuming no significant baryon chemical potential in the early Universe . The transition is caused by the temperature evolution of the strong coupling constant $g_s$. At temperatures above the transition temperature the coupling constant is small enough to treat the system perturbatively and the system is an a phase of quark-gluon plasma. As the Universe cools the strong coupling constant grows and quarks and gluons hadronize into colour neutral objects. All colour multiplets are confined to exist then within colour singlet objects such as baryons. Since perturbation theory breaks down during the QCD phase transition, they are best analyzed through lattice simulations. The phase diagram of QCD is shown in terms of temperature and baryon chemical potential in Fig. \ref{fig:QCDPT} which is taken from Ref. \cite{Mohanty:2013yca}. Some intuition can be obtained through the bag model~\cite{Chodos:1974je}, for a review see \cite{DeTar:1983rw}. \par 
Although vanilla cosmology would predict that QCD underwent a crossover transition, there have been some recent proposals to catalyze a strongly first order phase transition. One approach is to delay the electroweak phase transition until after the QCD phase transition such that the number of light quarks is large enough for the transition to be strongly first order \cite{Bai:2018vik}. The quark nuggets that form during such a transition are a dark matter candidate \cite{Bai:2018vik}. Another approach is to take advantage of the fact that the lepton asymmetry is relatively unconstrained. A large enough lepton asymmetry could catalyze the QCD transition \cite{Schwarz:2009ii}.\footnote{BBN bounds constrain the lepton asymmetry from being this large if a large amount of the lepton asymmetry is first generation \cite{Serpico:2005bc}. However, this can be avoided if the initial lepton asymmetry is second or third generation \cite{Barenboim:2016shh,Barenboim:2016lxv}. This is true even when one continues oscillations.}  Such a phase transition could leave the signature of observable low frequency gravitational waves \cite{Caprini:2010xv,Anand:2017kar,Chen:2017cyc}. 

\subsubsection{\bf An example of a multistep phase transition:}

\begin{figure}
    \centering
    \includegraphics[width=0.9\textwidth]{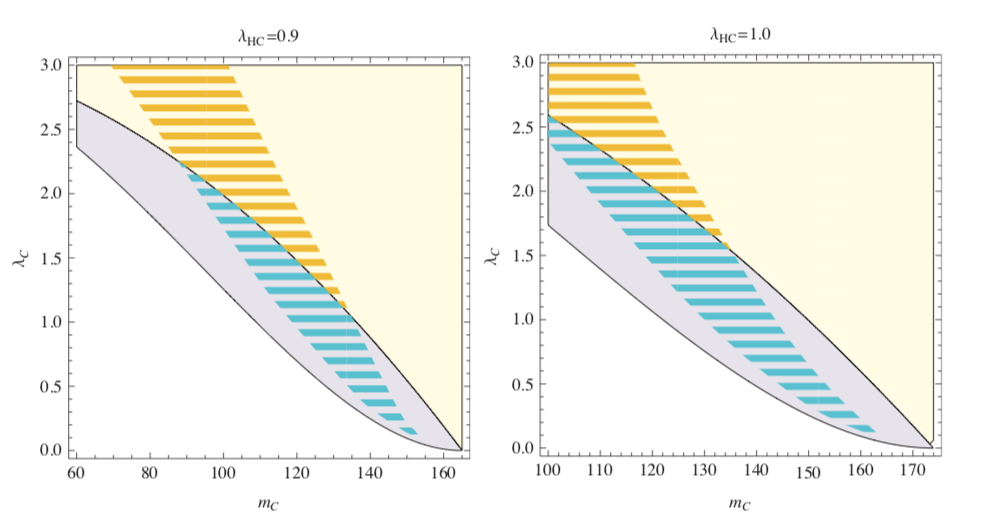}\\
    \includegraphics[width=0.7\textwidth]{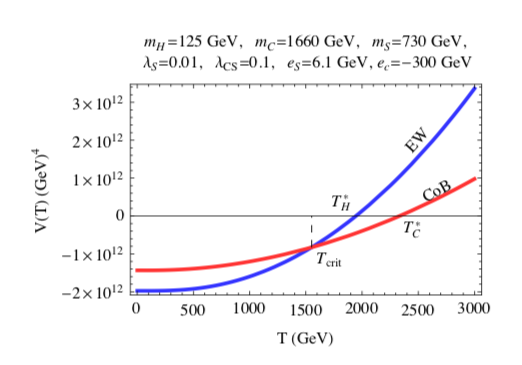}
    \caption{Top panel: parameter range that permits an intermediate colour breaking phase transition. Yellow hatched region are regions with electroweak symmetry at zero temperature and have an intermediate color breaking phase. Bottom pane: Tempearture evolution of the potential evaluated at the electroweak and color breaking vacuum for a benchmark scenario that includes the addition of a gauge singlet. The electroweak minimum is deeper at zero temperature evolves faster with temperature. The addition of a gauge singlet allows the phase transition to occur at multi-TeV scale.  Figure taken from \cite{Patel:2013zla}.}
    \label{fig:cobPT}
\end{figure}

In this section we briefly give an example of a multi-step phase transition. We will focus on the case where a zero temperature symmetry, SU(3)$_C$ in particular, is broken in an intermediate phase before being restored. This can be achieved either by having a large number density or introducing new colored scalars which acquire a vacuum expectation value at an intermediate temperature. Let us consider the latter case. Consider an effective potential that includes the standard model Higgs coupled to a colored scalar field $C$, which is an SU(2)$_L$ and U(1)$_Y$ singlet but a triplet under SU(3)$_C$
\begin{equation}
    V=- \mu^2 H |H|^2-\mu _C |C|^2 + \lambda _{HC} |H|^2 |C|^2 + \frac{\lambda _H }{2}|H|^4+ \frac{\lambda _C}{2} |C|^4 \,.
\end{equation}
For simplicity let us only include thermal corrections which are quadratic in temperature and fields. Symmetry is restored in the $H$ and $C$ directions above a temperature $T_X^f$ where \cite{Patel:2012pi}
\begin{eqnarray}
(T_H^f)^2&=&\frac{\lambda _H v_h^2}{ \left( \frac{\lambda _H}{4}+ \frac{\lambda _{HC}}{4}+\frac{3g^2_2}{16} +\frac{y_t^2}{4} \right)} \,,\\
(T_C^f)^2&=& \frac{\mu _C^2}{ \left( \frac{\lambda _C}{3}+ \frac{\lambda _{HC}}{6}+\frac{g^2_3}{3} \right)} \ .
\end{eqnarray}
If one has $T_H^f<T_C^f$ then one has a range of temperatures where color can be spontaneously broken but electroweak symmetry is restored. In figure \ref{fig:cobPT} we show in the top panel from \cite{Patel:2012pi} the possibility of color breaking and restoration. Note that this scenario isn't particularly fine tuned however the mass range of the colored scalars is quite light. The addition of a gauge singlet allows the mass of the colored scalar to be multi-TeV.


\subsection{Topological/non-topological defects and solitons}\label{sec:partphys:symmbreak}

\subsubsection{\bf Topological defects:}

The phase transition can also give rise to topological defects~\cite{Kibble:1976sj,Vilenkin:1984ib,Vachaspati:1993pj,Vilenkin:1994,Achucarro:1999it,Durrer:2001cg,Sakellariadou:2007bv}, for a review see~\cite{Hindmarsh:1994re}. Let us discuss briefly the microscopic origin of 
the formation of topological defects. Let us suppose that there is a non-trivial charge for  $\psi$ field under 
some gauge symmetry $G$, and then $\psi$ field obtains a non-vanishing VEV due to phase transition,  the 
symmetry group is broken now; $G\rightarrow H$. The manifold of all the vacua 
accessible to $\psi$ is then given by the quotient group  after breaking, i.e. $\mathcal{M}=G/H$. As an
example, in the case of an abelian Higgs model, the symmetry breaking pattern is very simple
$U(1)\rightarrow I$, and the manifold of vacua is $\mathcal{M}=U(1)$, 
corresponding to the circle of constant radius in the complex plane 
$|\psi|={\rm constant}$. Therefore, the formation and the type of topological defects 
depend on  the topological properties of 
$\mathcal{M}$~\cite{Kibble:1976sj,Vilenkin:1994}, which is classified by the homotopy groups 
$\pi_n$ of order $n$. Each group $\pi_n(\mathcal{M})$ is composed of all classes of 
hyper surfaces of dimension $n$. If any hyper surface can 
shrink to a point inside  $\mathcal{M}$, then the homotopy group contains only one element and becomes trivial
and is simply connected~\cite{Nakahara:2003nw}. In the opposite case, 
if $\mathcal{M}$ is not simply connected (for example during the breaking of a 
discrete group $Z_n \rightarrow I$), uncorrelated regions of the Universe 
would have different vacua separated by the domain 
walls~\cite{Kibble:1976sj,Vilenkin:1984ib}. The domain walls with tiny energy scale may yield some 
interesting cosmological consequences~\cite{Zeldovich:1974uw,Frieman:1991tu}, including mild acceleration of cosmic expansion~\cite{Brandenberger:2003ge}.
If their energy scale is high, and if they persist in the late Universe , they would simply cause cosmological disasters over dominating the energy density of the 
Universe ~\cite{Zeldovich:1974uw}. There are ways to tackle the problem if we change the nature of phase transition from $G\rightarrow H$ to a smooth adiabatic transition~\cite{Mazumdar:2015dwd}. Note that defects can also be formed in a slow first order phase transitions~\cite{Borrill:1995gu}.

The formation of topological defects also depends on the space-time dimensions, a $d$-dimensional defect is governed 
by the non-triviality of the homotopy group~\cite{Nakahara:2003nw}:
\begin{equation}
\pi_{3-d}\neq I~. 
\end{equation}
Any symmetry breaking of the form $G\rightarrow H\times U(1)$ gives rise to 
monopole (point-like defects). Since the Standard Model group contains the 
$U(1)$ factor, any GUT group breaking down to the Standard Model gauge group leads to this monopole problem. 
This formation of unwanted defects was one of the original 
motivation to introduce a phase of primordial inflation. 

There is also a class of unstable topological defects which can form even when the topology is 
trivial~\cite{Vachaspati:1993pj,Achucarro:1999it}. 
The electro-weak strings can be formed during the 
electroweak symmetry breaking which are perturbatively stable
for a range of parameters which are not realized in nature.
In general, the defects are a priori unstable due to plasma 
effects.


\subsubsection{\bf Non-topological solitons:}

The phase transition can also yield non-topological solitons, such as $Q$-balls. The $Q$-ball becomes a 
generic ground state in interacting scalar fields carrying some conserved global charge~\cite{Lee:1974ma,Coleman:1985ki,Lee:1991ax},
whose boundary condition at infinity is the same as that for the vacuum state, unlike in the case
of topological solitons such as magnetic monopoles \cite{tHooft:1974kcl,Polyakov:1974ek}. A detailed review of non-topological solitons can be found in, e.g.
\cite{Lee:1991ax}.  Formation of $Q$-balls can be extended to many scalar fields with various
$U(1)$ charges \cite{Kusenko:1997ad,Kusenko:1997zq},  with a non-Abelian symmetries
\cite{Axenides:1998fc}, and also with local gauge symmetries
\cite{Anagnostopoulos:2001dh}. The main
difference between global and local $Q$-balls  is that
in the latter case the charge of the stable $Q$-ball is bounded from above.

There is a theorem \cite{Coleman:1985ki,Lee:1991ax}, which 
states that if there exists a range for a field $\phi$, in a  potential $U(\phi^2)$, which
contains an attractive interaction, then a non-topological soliton should exist for
\begin{equation}
\label{theorem}
\nu^2 \leq \omega^2 < m_{\phi}^2\,,
\end{equation}
where $U(\phi^2) \rightarrow m_{\phi}^2\phi^2$ when $\phi \rightarrow 0$.
The value of $\omega =\sqrt{k^2+m_{\phi}^2}$ determines the frequency of
the $\phi$ quanta in the field space. A necessary condition for the existence of a solitonic solution is 
$\omega^2<m_{\phi}^2$, which means that there exists a parabola
$\nu^2\phi^2$ tangent to $U(\varphi^2)$ at $\phi =\pm \phi_{0}$, with
$\nu^2 <m^2_{\phi}$. For a sufficiently large $Q$, the energy
of a soliton is then given by
\begin{equation}
\label{minen1}
E=|\nu Q|< m_{\phi}|Q|\,,
\end{equation}
which ensures its stability against decay into plane wave solutions
with $\phi \simeq \phi_{0}$ inside the $Q$-ball, and $\phi \simeq 0$ outside.
The global $U(1)$ symmetry is broken inside and remains unbroken outside.

The $Q$-balls can be formed after inflation, as we had discussed earlier, but can be formed at later stages
by the dynamics of a scalar field, such as present in supersymmetric theories due to plenty of supersymmetric  flat directions, 
made up of squarks and sleptons, for a review see~\cite{Enqvist:2003gh,Allahverdi:2012ju}. The stability of the $Q$-balls can contribute to 
the dark matter abundance~\cite{Kusenko:1997si}, see review~\cite{Enqvist:2003gh}. During the formation, gravitational waves can also be generated~\cite{Kusenko:2008zm,Kusenko:2009cv}.


\section{Phase transitions and Cosmic signatures}

\subsection{Gravitational waves}

\begin{figure}
    \centering
    \includegraphics[width=\textwidth]{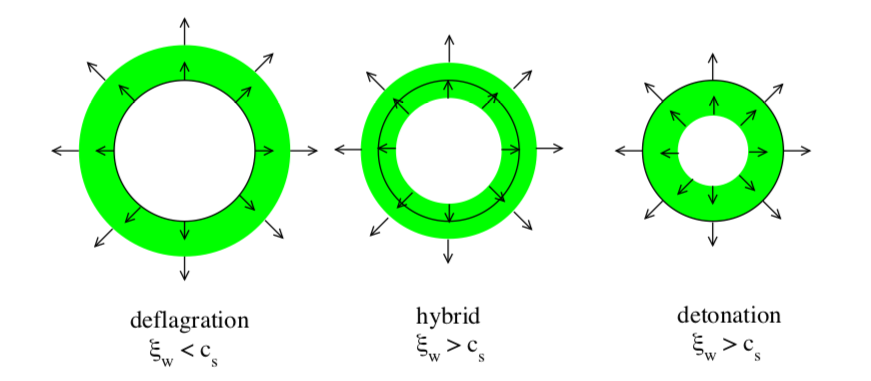}
    \caption{From left to right: depiction of deflagration, detonation and hybrid phase transitions. The green bands denote the sound shell whereas the solid black line denotes the scalar shell. Figure taken from \cite{Espinosa:2010hh}}
    \label{fig:my_label}
\end{figure}

First order cosmological phase transitions proceed via bubble nucleation. While an isolated spherical bubble produces no gravitational waves as such an event has no quadupole moment, the violent process of bubble collision does \cite{Coleman:1980aw,Kamionkowski:1993fg,Kosowsky:1992vn,Nicolis:2003tg,Kosowsky:1992vn,Katz:2016adq,Jinno:2016vai,Huber:2008hg,Jinno:2017fby,Caprini:2018mtu,Kosowsky:1992rz,Kosowsky:1991ua,Caprini:2007xq,Watanabe:2006qe,Saito:2012bb,Dev:2016feu,Ashoorioon:2015hya}, for a review of gravitational waves, see~\cite{Maggiore:2018sht}. Upon collisions of such bubbles, the latent heat will be converted to bulk flow of the plasma, as well as to kinetic energy of the scalar fields. The fraction of energy converted to gravitational waves per decade is
\begin{equation}
    \Omega _{\rm GW} = \omega \frac{d E_{\rm GW}}{d\omega } \frac{1}{E_{\rm tot}}\,,
\end{equation}
where
\begin{equation}
    \frac{dE_{\rm GW}}{d\Omega d\omega } = 2 G \omega ^2 \Lambda _{ij,lm} (\hat{k}) \tilde{T} _{ij}^* (\hat{k}, \omega ) \tilde{T} _{l,m}(\hat{k}, \omega )\,,
\end{equation}
where $\tilde{T} _{ij}$ is the fourier transform of the stress energy tensor and 
\begin{equation}
    \Lambda _{ij, lm} = \delta _{il} \delta _{jm}
 - 2 \hat{k} _j \hat{k} _m \delta _{ij} + \frac{1}{2}\hat{k}_l \hat{k}_j \hat{k}_l \hat{k}_m -\frac{1}{2} \delta _{ij} \delta _{lm} + \frac{1}{2} \delta _{ij} \hat{k}_l \hat{k}_m +\frac{1}{2} \delta _{lm} \hat{k}_i \hat{k}_j \,,\end{equation}
is a projection operator. 
The contributions to the gravitational wave spectrum can be modeled as a sum of three contributions characterized by a contribution to the stress energy tensor and an efficiency parameter $\kappa _x$ which denotes the efficiency that the latent heat can be converted into a particular source of gravitational waves. The three contributions to the stress energy tensor are as follows
\begin{itemize}
    \item[1] A contribution from the initial collision of scalar field shells. The stress energy tensor contribution is
    \begin{equation}
       T_{\mu \nu} = \partial _\mu \phi \partial _\nu \phi -g_{\mu \nu} \left[ \frac{1}{2} \partial _\rho \phi \partial ^\rho \phi - V_0 \right]\,.
    \end{equation}
    \item[2] The interaction between kinetic shells going at the speed of sound \cite{Hindmarsh:2013xza}. The stress energy contribution is \cite{Weir:2017wfa}
    \begin{equation}
      T_{\mu \nu} = \sum _i  \int \frac{d^3k}{(2 \pi)^3 2E_i}k_\mu k_\nu f_i(k)=w u_ \mu u _ \nu -g_{\mu \nu } p\,,
    \end{equation}
    where $w$ is the enthalpy, $u=(\gamma ,\gamma v)$ is the four velocity,  $v$ is the 3 velocity of the relativistic fluid, $\gamma$ is the Lorentz factor, and $p$ is the pressure.
    \item[3] A contribution due to magnetohydrodynamic turbulence \cite{Caprini:2006jb} which again is prominent after the collision of the scalar shells. This contribution is usually subdominant. The spatial components of this contribution to the stress energy are
    \begin{equation}
         T_{ij} (x, \eta ) = \frac{4 \pi}{3} \rho (\eta)  v_i(x , \eta ) v_j (x,\eta) \,,
     \end{equation}
     where $v_i$ is the turbulent velocity and $\eta $ is the conformal time.
\end{itemize}
Recent work has suggested the existence of a fourth contribution from quantum fluctuations in bubble wall collisions \cite{Braden:2014cra,Braden:2015vza,Bond:2015zfa}. They considered a double well potential and demonstrated that quantum fluctuations break the SO(2,1) symmetry of bubble wall collisions. One has a parametric instability and wiggles on the wall from the collision grow and break SO(2,1). The size of this contribution relative to LISA sensitivity is an open problem and we therefore focus on the contributions to the gravitational wave spectrum that are better understood.
The total gravitational wave spectrum can then be modeled as a sum of the three contributions \cite{Weir:2017wfa}
\begin{equation}
    \Omega (f) h^2 = \Omega _{\rm col} (f) h^2 + \Omega _{\rm sw} (f) h^2 + \Omega _{\rm turb} (f) h^2 \ . 
\end{equation}
The change in free energy during the phase transition gives a limit to how much vacuum energy can be converted into gravitational waves. The efficiency of converting vacuum energy into scalar field gradient energy is denoted by $\kappa _{\rm col}$ (the first contribution in the above list) controls the efficiency of producing contributions to $\Omega _{\rm col}$. The efficiency parameter, $\kappa _{\rm col}$, is typically small, making this contribution sub-dominant. Specifically it be found by calculating the gradient density $\rho _D = \frac{1}{2}(\nabla \phi )^2$, and potential energy density $\Delta V(\phi)$ for a bounce solution \cite{Cutting:2018tjt}
\begin{equation}
    \kappa _{\rm col} = \frac{ 2 \rho _D}{\rho _{\rm vac} -\rho _V}\,.
\end{equation}
Ignoring the turbulence contribution, the conservation of energy and momentum gives
\begin{equation} 
\partial _\mu T^{\mu \nu} _{\rm plasma} +\partial _\mu T^{\mu \nu} _{\rm field}=0 \ . 
\end{equation}
One can parametrize the plasma contribution as follows \cite{Espinosa:2010hh}
\begin{equation} 
T_{\mu \nu }^{\rm plasma} = w u_ \mu u _ \nu -g_{\mu \nu } p\,,
\end{equation}  
where $u=(\gamma , \gamma v)$ is the four velocity field of the plasma. If we ignore the field contribution we can calculate the fluid velocity from the equations $\partial _z T^{zz}=\partial _z T^{0z}=0$ from which one obtains \cite{Espinosa:2010hh}
\begin{equation} 
w_+v_+^2 \gamma_+^2=w_-v_-^2 \gamma _-^2 \,,
\end{equation} 
and  
\begin{equation} 
w_+v_+ \gamma_+^2=w_-v_- \gamma _-^2 \,,
\end{equation}  
where $\pm$ denotes the symmetric and broken phases respectively. Defining \cite{Espinosa:2010hh}
\begin{equation} a_+\sim \frac{\pi^2}{30} \sum _i (N_i^b + \frac{7}{8} N_i^f) \end{equation}
one can define an expression for the fluid velocity \cite{Espinosa:2010hh}
\begin{equation}
        v_+=\frac{1}{1+a_+}\left[ \frac{v_-}{2}+\frac{1}{6 v_-} \pm \sqrt{\left(\frac{v_-}{2}+\frac{1}{6 v_-} \right)^2+a_+^2+\frac{2}{3}a_+ -\frac{1}{3}} \right] \,,
\end{equation}
    a detonation has $v_+>v_-$ and deflagration is $v_-<v_+$. The latter only exists only when $a_+<1/3$. In a detonation the wall moves at supersonic speed and the plasma it expands into is at rest. In contrast, a deflagration has the wall expanding into the perturbed plasma. Simulations show that the efficiency coefficient for a deflagration (wall velocity smaller than the speed of sound) is \cite{Espinosa:2010hh}
    \begin{equation}
        \kappa _{sw} = \frac{6.9 v_w \alpha }{1.36-0.037 \sqrt{\alpha}+\alpha }\,,
    \end{equation}
    whereas for detonations (runaway walls) one has
    \begin{equation}
        \kappa _{sw} = \frac{\alpha }{0.73+0.083 \sqrt{\alpha} +\alpha }\,.
    \end{equation}
Here $\alpha $ is the ratio of Latent heat to vacuum energy and $v_w$ is the wall velocity. These thermodynamic quantities are defined in section \ref{sec:therm}. 
Alternatively if one knows the fluid radial velocity profile, $V_r(\xi )$ one can explicitly calculate the efficiency as \cite{Weir:2017wfa}
\begin{equation}
    \kappa _f = \frac{3}{\epsilon v_w^2} \int d \xi \omega (\xi ) V_r ^2(\xi) \gamma ^2 \xi ^2 \ .
\end{equation}
Here $\xi =r/t$ and $\omega$ is the enthalpy. If the bubble wall does not runaway, the sound wave and turbulence terms are expected to dominate. If the bubbles runaway, the collision term becomes more important and in fact dominates for very large $\alpha$. 

Apart from the efficiency parameters that define the efficiency of converting energy available to gravitational wave energy, the gravitational wave power spectrum is controlled by the ratio of Latent heat to vacuum energy, the bubble wall velocity and the speed of the phase transition compared to the Hubble rate $\beta/H^*$ as well as parameters that are numerically derived from analytical fits to numerical simulations.


\subsubsection{\bf Collision term:}

The interaction of the bubbles can be well approximated by the ``envelope approximation'' \cite{Huber:2008hg}  which is the combination of two approximations - first that the stress energy tensor is non-zero only in an infinitesimal region at the bubble wall and second that the stress energy tensor vanishes when the bubble overlaps. This contribution becomes most significant when the bubble runs away $\gamma \to \infty$. This contribution can be derived analytically through a calculation the correlation of the stress energy tensor $\langle T(x) T(y) \rangle$ \cite{Jinno:2016vai}. Under the envelope approximation the stress energy tensor due to a bubble nucleated at $x_N=(t_N,\vec{x}_N)$ is given by
\begin{equation}
    T_{ij} = \rho (x) \widehat{(x-x_N)}_i \widehat{(x-x_N)}_j\,,
\end{equation}
with the energy density localized around he bubble wall in accordance with the envelope approximation \cite{Jinno:2016vai}
\begin{equation}
    \rho (x) = \left\{ \begin{array}{cc}
    \frac{4\pi}{3} r_B(t)^3 \frac{\kappa _{\rm col} \rho _0}{4 \pi r_B^\prime (t) ^2 l_B}    & r_B(t) < |x-x_N| < r^\prime _B (t)   \\
    0     & {\rm otherwise} .
    \end{array} \right.
\end{equation}
Here $r(t)=v(t-t_N)$ and $r_B^\prime (t) = r_B(t) +l_B$ are the interior and exterior edge of the bubble wall respectively and $\rho _0$ is the latent heat released by the transition. The nucleation rate is controlled by the time rate in change of the effective action $\beta$. If the phase transition is sufficiently quick, $\beta/H>>1$ one can ignore the expansion of the Universe and write the metric as
\begin{equation}
    ds^2 = -dt^2 + (\delta _{ij} +2 h_{ij} ) dx^i dx^j \ .
\end{equation}
From the equations of motion the tensor perturbations satisfy the following
\begin{equation}
    \ddot{h}_{ij}(t,k) + k^2 h_{ij} (t,k) = 8 \pi G \Pi _{ij}(t,k) \,,
\end{equation}
where $\Pi _{ij}$ is related to the fourier transform of the stress energy tensor via a projection operator 
\begin{equation}
    \Pi _{ij}(t,k) = \Lambda_{ij,kl} T_{kl }(t,k) \ .
\end{equation}
The tensor perturbations can be solved in terms of a Greens function. 

The total energy of the gravitational waves is given by the oscillation and ensemble average of the correlator \cite{Jinno:2016vai}
\begin{equation}
    \rho _{\rm GW} = \frac{\langle \dot{h}_{ij} (t,x)\dot{h}_{ij} (t,x)  \rangle _T}{8 \pi G} \ .
\end{equation}
From which we can derive the gravitational wave spectrum \cite{Jinno:2016vai}
\begin{eqnarray}
\Omega _{\rm GW} &=& \frac{1}{\rho _t} \frac{d \rho _{\rm GW}}{d \ln k} \\ 
&=& \frac{2G k^3}{\pi \rho _{\rm tot} } \int _{t_i}^{t_f} dt_x \int _{t_i}^{t_f}dt_y \cos (k (t_x-t_y)) \Pi (t_x,t_y,k) \,,
\end{eqnarray}
where $\rho _t$ is the total energy $\rho _0 +\rho _{\rm rad}$ and $\Pi(t_x,t_y,k)$ is the fourier transform of the stress energy correlation function contracted with projection operators \cite{Jinno:2016vai}
\begin{equation}
    \Pi (t_x,t_y,k) = \int d^3 r e^{i k \dot r} \Lambda_{ij ,kl}\Lambda_{ij,mn} \langle T_{kl}(t_x,x) T_{mn} (t_y,y) \rangle \,,
\end{equation}
with $r=x-y$. Defining the quantity $\alpha = \rho _0 / \rho _{\rm rad} $ and using the fact that $H_*^2 = \frac{8\pi }{3G} \rho _t$ we can write
\begin{eqnarray}
    \Omega _{GW} &=& \kappa ^2  \left( \frac{\beta}{H_*} \right) ^{-2} \left( \frac{\alpha }{1+\alpha} \right) ^2 \Delta (k/\beta , v_w)\,, \nonumber \\
    \Delta (k/\beta , v_w) &=& \frac{3 \beta ^2 k ^3}{4 \pi ^2 \kappa ^2 \rho _0^2 }\int _{t_i}^{t_f} dt_x \int _{t_i}^{t_f}dt_y \cos (k (t_x-t_y)) \Pi (t_x,t_y,k) \ . \nonumber \\
\end{eqnarray}
The ratio of the scale factor at the time of transition to the scale factor today is $a_0/a^* = 8\times 10^{-16}(100/g^*)(100{\rm GeV}/100)$ which can be used to relate the gravitational wave spectrum at transition to its spectrum today \cite{Jinno:2016vai}
\begin{eqnarray}
  f&=& 1.65 \times 10^{-5} {\rm Hz} \left( \frac{f_*}{\beta} \right) \left( \frac{\beta }{H_* } \right) \left( \frac{T}{100 {\rm GeV}} \right) \left( \frac{g_*}{100} \right)^{1/6}  \nonumber \\ && \times \frac{0.35 }{1+0.069 v_w +0.69 v_w^4} \,,\nonumber  \\
  \Omega _{GW} h^2 &=& 1.67 \times 10^{-5} \left( \frac{g_*}{100} \right)^{-1/3} \kappa ^2 \left( \frac{\beta}{H_*} \right) \left( \frac{\alpha }{1+\alpha} \right)^2  \nonumber \\ && \times \frac{0.48 v_w^3}{1+5.3 v_w^2 +5.0 v_w^4} \Delta \ . \nonumber \\
\end{eqnarray}
The dependence on $v_w$ unfortunately comes from numerically fitting. All that remains is an analytical calculation of $\Delta$. Such a calculation is difficult in practice however one can acquire a closed form solution in terms of integrals of spherical Bessel functions. The asymptotic form can be derived from the asymptotic expansions of the spherical bessel functions and one finds that $\Delta \sim k^3$ for $k/\beta < 1$ and $k^{-1}$ for $k/\beta >1$. Numerically fitting to the integral over Bessel functions for $v_w$ close to unity one has for the frequency spectrum, one finds that $\Delta$ is well approximated by \cite{Weir:2017wfa}
\begin{eqnarray}
\Delta &=& \frac{\Delta _{\rm peak}}{c_l \left(\frac{f}{f_{\rm peak}}  \right) ^{-3} +(1-c_l-c_h) \left( \frac{f}{f_{\rm peak}} \right)^{-1} +c_h \left( \frac{f}{f_{\rm peak}} \right)} \,,
\end{eqnarray}
where fitting yields $c_l=0.064$ and $c_h=0.48$. Note that recent work analyzing a vacuum transition (that is, a case where the plasma is ignored) \cite{Cutting:2018tjt} demonstrated that the envelope approximation breaks down right when it starts to become visible and the true spectrum is dampened. This seems to imply that the collision term is always sub-dominant.

\subsubsection{\bf Sound waves:}
The contributions from the plasma flow are much harder to capture in a model. Moreover, recent studies indicate \cite{Hindmarsh:2017gnf} that the plasma flow contributions dominate over the scalar field contributions, since the plasma flow continues to source GWs long after the collisions of the bubbles. 

Progress in this area has been largely dominated by large-scale hydrodynamic simulations. Nevertheless, well-motivated simplified models have been developed recently, such as the recent bulk flow model \cite{Jinno:2017fby} and sound shell model \cite{Hindmarsh:2016lnk}. Such models may describe the physics in regimes in where simulations have limitations \cite{Konstandin:2017sat}.

The sound wave contribution is typically larger than the other contributions. Its power spectrum is \cite{Hindmarsh:2017gnf}
\begin{equation}
h^2\Omega _{\rm sw}    = 8.5 \times 10^{-6} \left( \frac{100}{g_*} \right)^{-1/3} \Gamma ^2 \bar{U}_f^4  \left( \frac{\beta}{H} \right)^{-1}  v_w S_{\rm sw}(f) \,,
\end{equation}
and the spectral shape is given by \cite{Hindmarsh:2017gnf}
\begin{equation}
    S_{\rm sw} =  \left( \frac{f}{f_{\rm sw}} \right) ^3 \left( \frac{7}{4+3\left( \frac{f}{f_{\rm sw}}\right) ^2} \right)^{7/2}\,,
\end{equation}
with \cite{Hindmarsh:2017gnf}
\begin{equation}
    f_{\rm sw} = 8.9 \times 10^{-8} {\rm Hz} \frac{1}{v_w} \left( \frac{\beta}{H} \right) \left( \frac{T_N}{{\rm Gev}} \right) \left( \frac{g_* }{100} \right)^{1/6} \ .
\end{equation}
where $\Gamma \sim 4/3$ is the adiabatic index, and $\bar{U}_f^2\sim (3/4) \kappa _f \alpha _T$ is the rms fluid velocity.
Note that the above fits for SW are not valid for all possible values of $(\alpha,v_w)$. The fit instead was chosen to work for typical thermal parameters, namely cases where $v_w$ is within 10 percent of either the speed of sound or the speed of light and $\alpha <0.1$. A feature of the soundwave source is that it is only supressed by one power of $(\beta/H_*)^{-1}$ in contrast to the collision of scalar shells. This $\beta/H$ enhancement captures the fact that this source is longer lasting as the dissipation of kinectic energy in the sound shell takes several Hubble times \cite{Caprini:2015zlo}. If the phase transition involves a large amount of super cooling the strength of the gravitational wave background will grow. However, in the limit of high supercooling, the expansion of the Universe can be vacuum dominated which can prevent the phase transition from completing \cite{Ellis:2018mja}. This implies the strength of the gravitational signal from sound waves can't be arbitrarily large.


\subsubsection{\bf Turbulence:}

 Kolmogorov turbulence \cite{frisch1995turbulence} can be modeled by considering a flow made up of eddy's of different length scales. Large eddies break up into smaller eddies and so on. For rate of energy dissipation $\epsilon$ and viscosity $\nu$ one has the  Kolmogorov length scale, or the dissipation scale, which defines the length scale at which the dissipation of kinetic energy occurs \cite{frisch1995turbulence},
 \begin{equation} L_K=(\nu ^3/\epsilon)^{1/4} \ . \end{equation}
This is compared to the largest scale of the flow, $L_B$. Eddies exist in the range $L_K<<r<<L_B$ and KE is not dissipated in this range but merely transferred to smaller scales. We would need some characteristic vector field and its correlation. The turbulent KE of the flow is $(1/2) \langle v_i v_i \rangle $
for a phase transition the size of the largest eddies, $L_B<<H^{-1}$ is the comoving size of the largest bubbles when they collide.
Energy dissipation is \cite{frisch1995turbulence} 
    \begin{equation}
        \epsilon = - \frac{d}{d\eta} \frac{\langle v^2 \rangle}{2}\ .
    \end{equation}
The power spectrum is given by Fourier transform of 2 point correlator
    \begin{equation}
        P(k) = \frac{1}{2} \int d^3 x e^{i \vec{k} \dot \vec{x}} \langle v_i(x) v_i (x) \rangle\,,
    \end{equation}
    where $v$ is the turbulent velocity which is a random variable, 
    \begin{equation}
        v_i(k,\eta) = \begin{array}{cc}
            v_i (k)  & {\rm for} \ k < L_B^{-1} \,, \\
            V_i(k) e^{i \omega _k \eta} &  {\rm for} \  L_B^{-1} < k < L_K ^{-1}   \,,
        \end{array} 
    \end{equation}
    and $\omega _k$ is the frequency associated with an eddy of size $l=1/k$. The needed fourier transform of the stress energy tensor is as follows \cite{Caprini:2006jb,Kahniashvili:2008pf,Kahniashvili:2008pe,Kahniashvili:2009mf,Caprini:2009yp,Kisslinger:2015hua}
     \begin{equation}
         T_{ij} (k, \eta ) = \frac{4 \pi}{3} \rho (\eta) \int \frac{d^3q }{(2 \pi )^3} v_i(q) v_j (k-q) e^{i \omega _q  \eta} e^{i \omega _{|k-q|}\eta } \,,
     \end{equation}
and $\rho (\eta)$ is the energy density at conformal time $\eta$. This contribution can only be modeled numerically. Caprini et al \cite{Kosowsky:2001xp,Caprini:2009yp}, noted that when modeling the contribution from turbulence, one needs to take into account that the turbulence continues long after after the phase transition is complete. If the source is long lasting one needs to take expansion into account. For example for $T_*=100$ and $\beta/H_*=100$ one finds the turbulence is not complete until $T\sim 120MeV$. This causes some amplification. The effect is rather modest however, as the decay time of source (controlled by eddy turnover time) is much smaller than Hubble time. Indeed they found an amplification of a factor of about 2. Taking this into account, simulations show that one can achieve a reasonable fit with a power spectrum governed by our usual thermal parameters.  \cite{Binetruy:2012ze,Caprini:2009yp}
\begin{equation}
    h^2 \Omega _{\rm turb}(f) = 3.35 \times 10^{-4} \left( \frac{\beta }{H} \right) ^{-1} \kappa _{\rm turb}^{3/2}  \left( \frac{\alpha }{1+\alpha} \right)^{3/2} \left( \frac{100}{g_*} \right) ^{1/3} v_w S_{\rm turb} (f) \ . \end{equation}
    There is as yet no known method for directly calculating the efficiency parameter, however, this contribution is expected to be sub-dominant. The spectrum is  \cite{Binetruy:2012ze,Caprini:2009yp}
    \begin{equation}
        S_{\rm turb} = \frac{(f/f_{\rm turb})^3}{\left[ 1+(f/f_{\rm turb})\right]^{11/3}(1+8 \pi f/h*)}\,,
    \end{equation}
    with
    \begin{eqnarray}
    h_* &=& 1.65 \times 10^{-7} {\rm Hz} \left(\frac{T_N}{{\rm GeV}} \right) \left( \frac{g_*}{100} \right)^{1/6} \,,\\
    f_{\rm turb} &=& 27 \times 10^{-8} {\rm Hz} \frac{1}{v_w} \left( \frac{\beta}{H} \right)\left( \frac{T_N}{\rm GeV} \right) \left( \frac{g_*}{100} \right) ^{1/6}\,.
    \end{eqnarray}

\subsubsection{\bf Detection of gravitational waves from cosmic phase transitions:}
 
\begin{figure}
    \centering
    \includegraphics[width=0.95\textwidth]{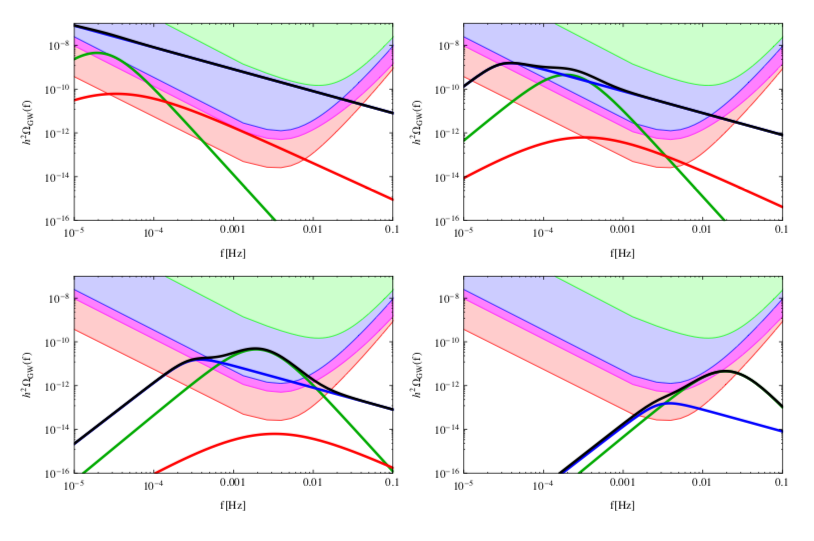}
    \caption{Gravitational wave spectrum against Lisa 1 year sensitivity curves. Thermal parameters are $T_{*}=100$, $\alpha = 1$ and $v_w=1$. From left to right the top panels have $\beta/H=(1,10)$ respectively and the bottom panels are $\beta/H=(10,100)$ respectively. The black line is the total spectrum whereas the blue, green and red lines are the collision, sound wave and turbulence terms respectively. Figure taken from \cite{Caprini:2015zlo}}
    \label{Fig:3peakGW}
\end{figure}

\begin{figure}
    \centering
    \includegraphics[width=0.7\textwidth]{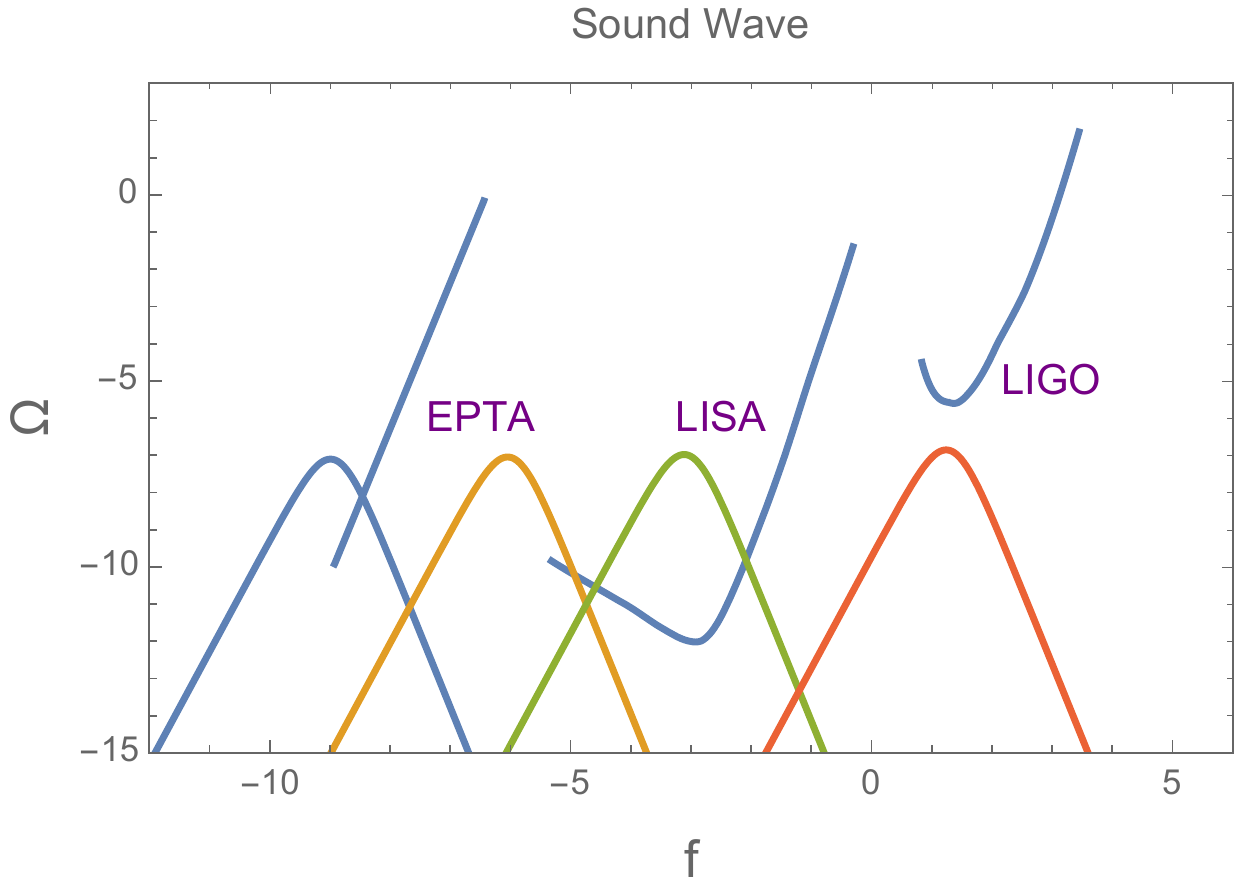}\\
    \includegraphics[width=0.45\textwidth]{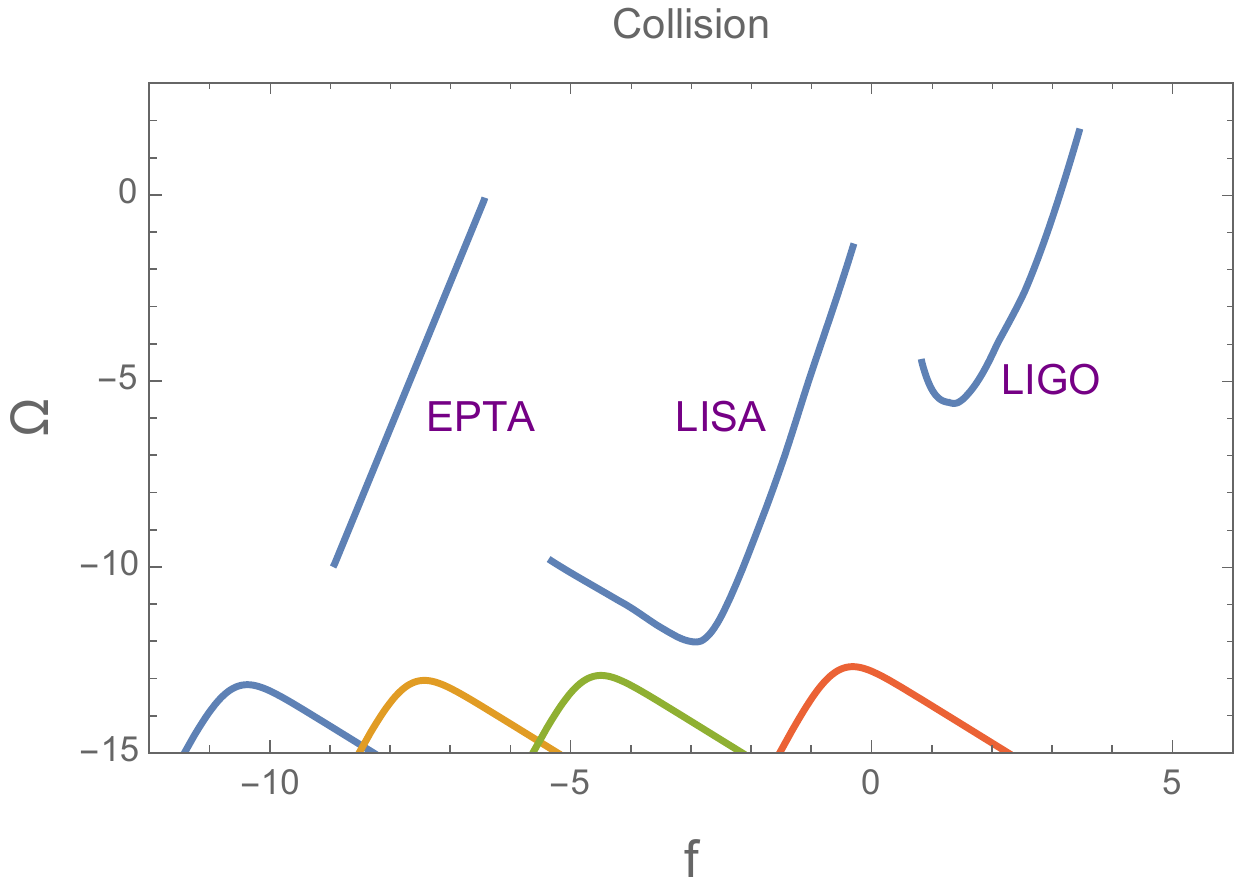}
    \includegraphics[width=0.45\textwidth]{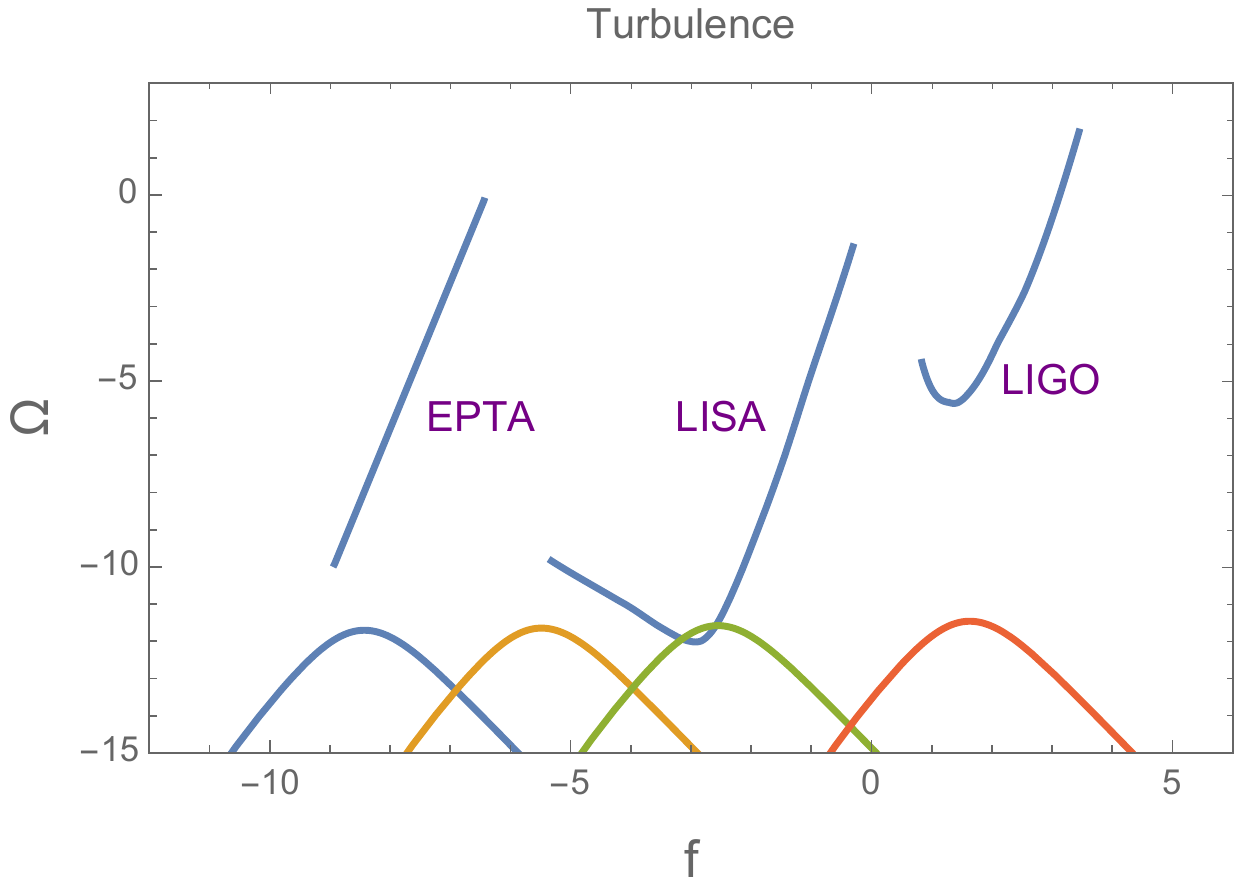}
    \caption{Gravitational wave detectors probing the scale of a phase transition. Sound wave (top panel), collision (bottom left) and turbulence (bottom right) contributions to a gravitational wave source with $\alpha=v_w=1$, $(\kappa _{\rm col},\kappa _{\rm turb}) = (0.3,0.02)$ and $\beta/H*=1.3 \log[T_*/M_{\rm pl}]$ for phase transitions occuring at a scale $T_n = (10^{-5},10^{-2},10,3\times10^5)$ GeV respectively against initial sensitivities of LIGO/VIRGO/Virgo \cite{TheLIGOScientific:2014jea,TheVirgo:2014hva}, LISA \cite{AmaroSeoane:2012km} and the European Pulsar Timing Array (EPTA) \cite{Desvignes:2016yex}. After integrating over frequency the sensitivity improves by several orders of magnitude \cite{Thrane:2013oya}. Also see Ref. \cite{Dev:2016feu}}
    \label{Fig:GWscale}
\end{figure}
\begin{figure}
    \centering
    \includegraphics[width=0.48\textwidth]{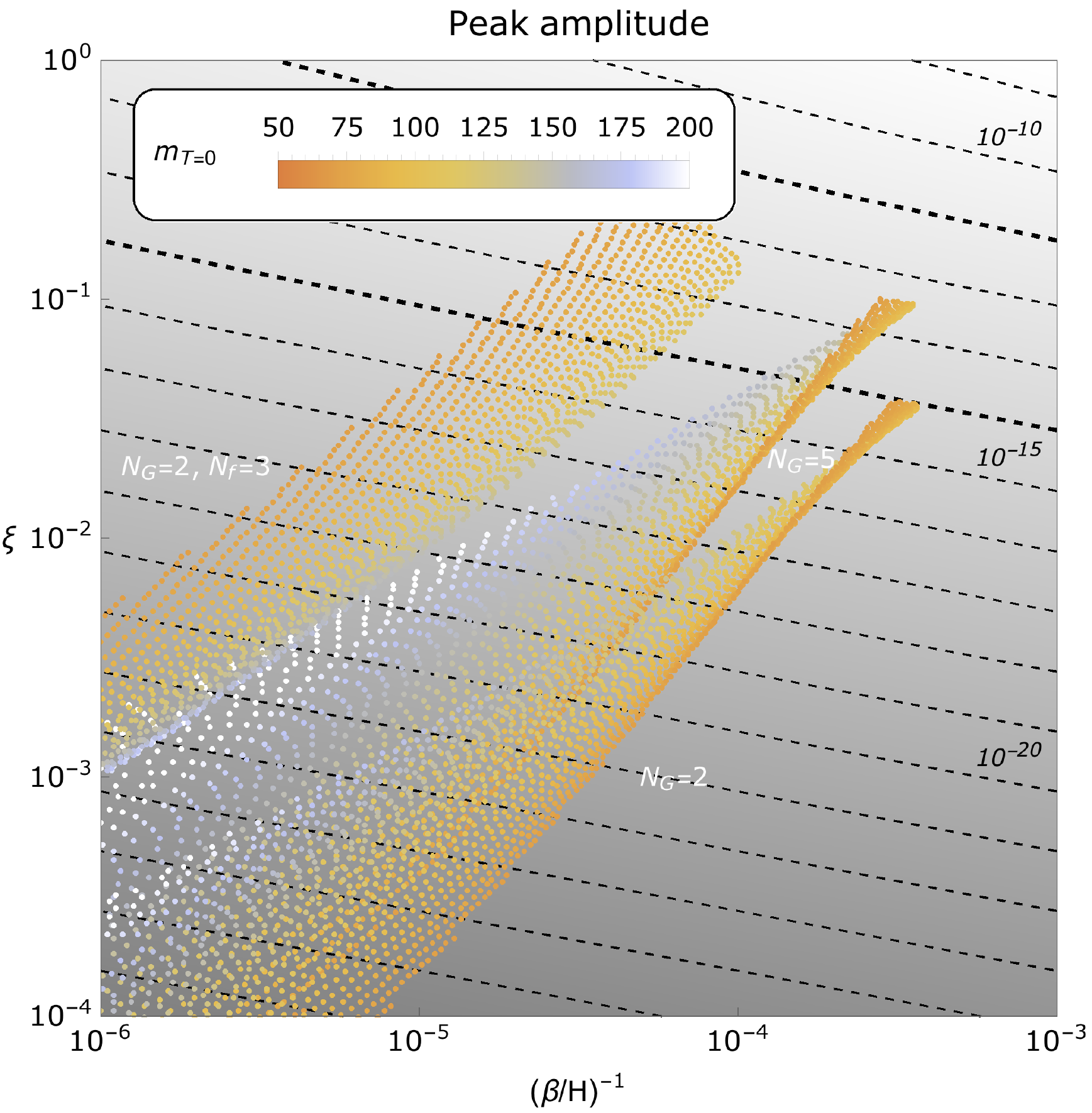}
     \includegraphics[width=0.48\textwidth]{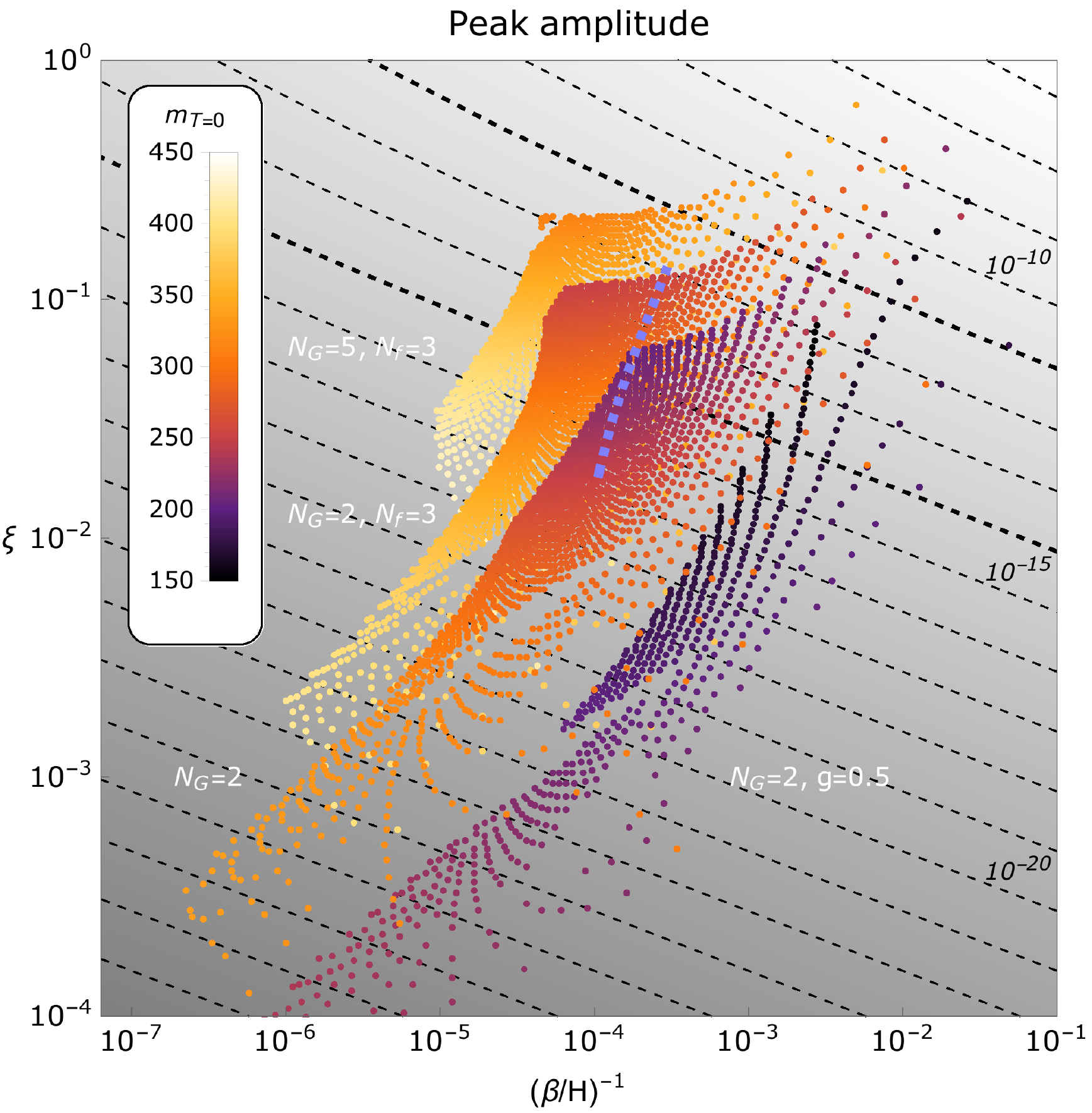}
    \caption{Thermal parameters from a dark Higgs with (right panel) and without (left panel) non-renormalizable operators for various models. In the above $N$ denotes the rank of the group and $N_F$ denotes the number of fermions coupled with unity Yukawa coupling. The plot points are coloured by their effective zero temperature mass. Note that $\xi$ in the above denotes the usual thermal parameter $\alpha$.
    The dashed contours in the plots correspond to the GW amplitude $\Omega_{\rm sw}$, with $v_w = 0.5$. The upper thicker contour corresponds to the LISA 1-year peak sensitivity \cite{Moore:2014lga}. The lower thicker dashed contour corresponds to the LISA for a power-law spectrum (integrated over frequency), taken from \cite{Thrane:2013oya}.
 Figure taken from \cite{Croon:2018erz}}
    \label{Fig:GWmodel}
\end{figure}

\begin{figure}
    \centering
    \includegraphics[width=.4\textwidth]{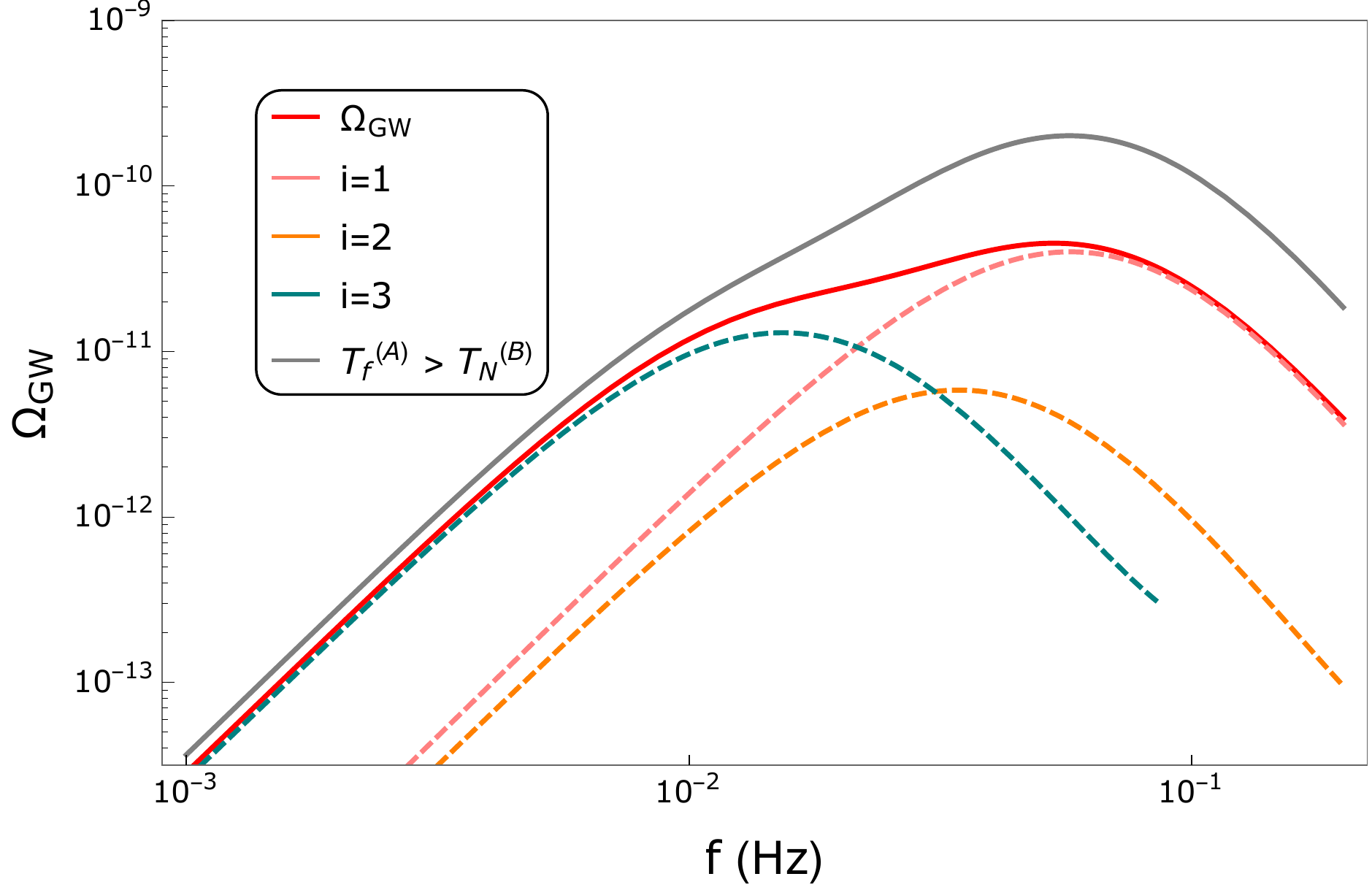}
    \includegraphics[width=.4\textwidth]{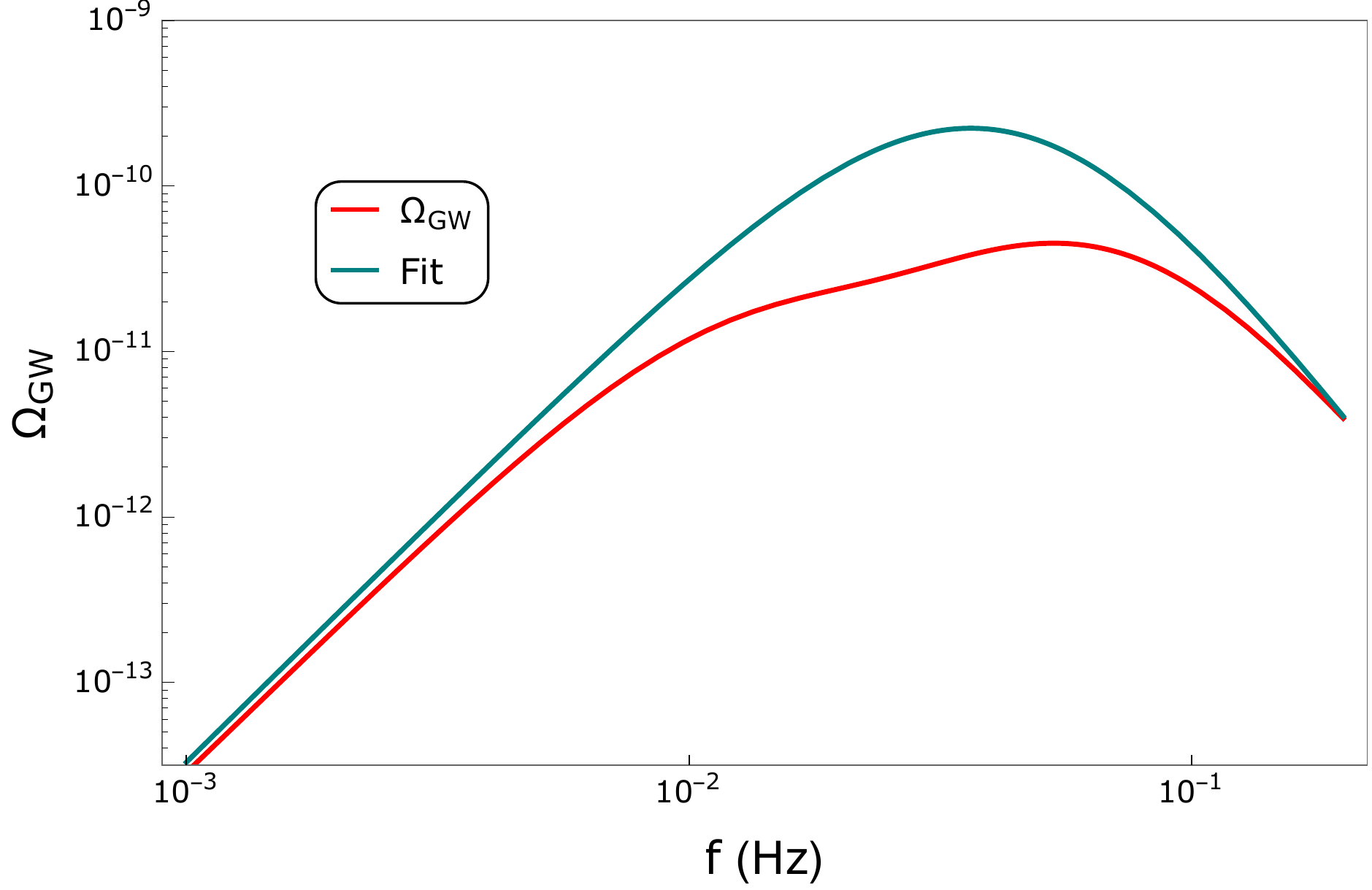} \\
    \includegraphics[width=.4\textwidth]{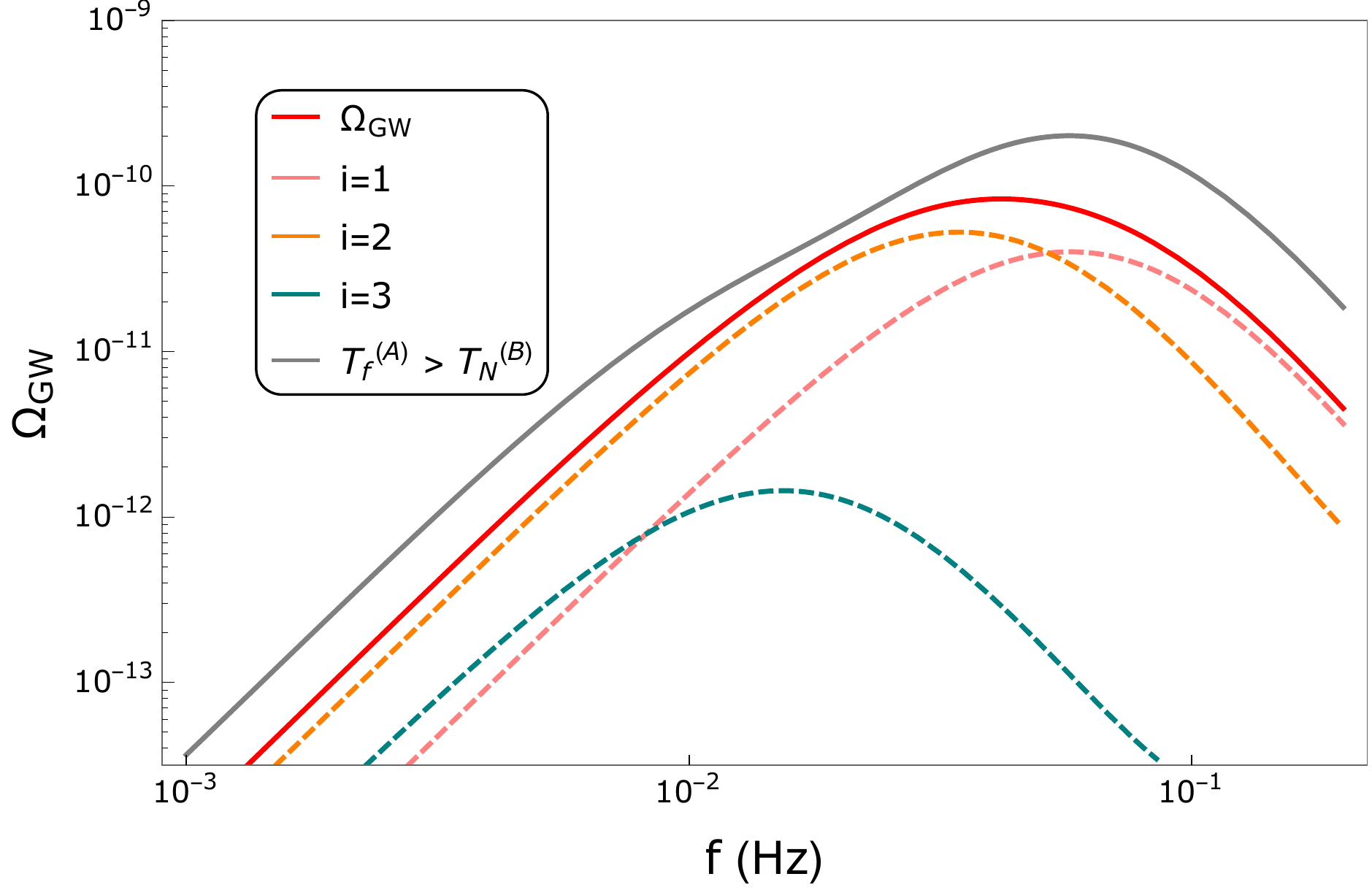}
    \includegraphics[width=.4\textwidth]{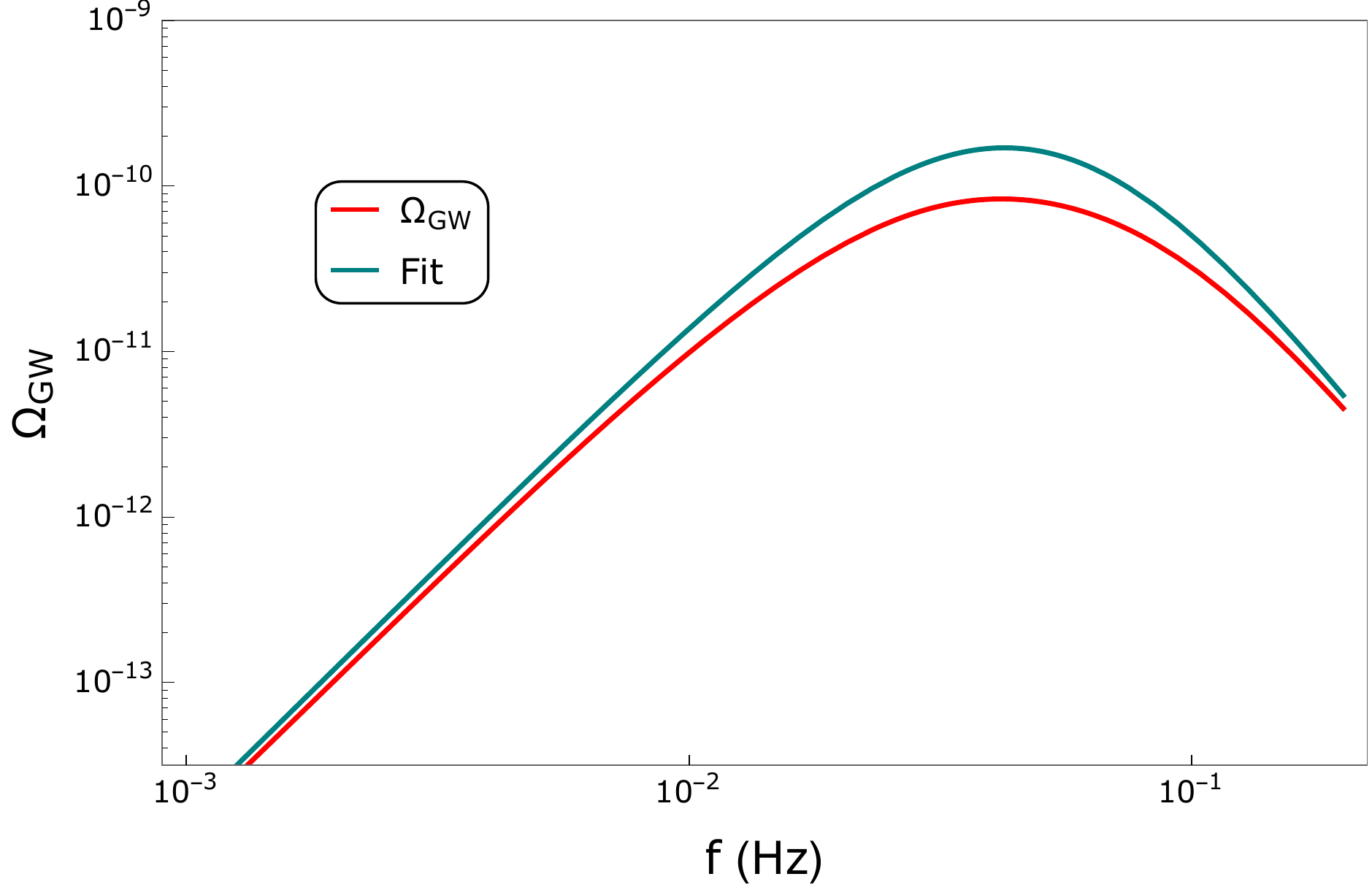}
    \caption{Left panels: the spectrum due to two simultaneous PTs, given by the red line, leads to a different spectrum than consecutive transitions. This can be seen by comparison with the gray line, which is predicted by the same thermal parameters. Right panels: the spectrum from simultaneous PTs can not be fitted to the spectrum than from a single PT.  Figure taken from \cite{Croon:2018new}}
    \label{Fig:doublebubble}
\end{figure}
 Of all the thermal parameters only the nucleation temperature depends strongly on the scale. $\beta/H\sim \log[T_*/M_{\rm pl}]$ also depends weakly on the scale but is more strongly influenced by the ratio of scales $v/\Lambda$ as is $\alpha$ and the wall velocity. The transition temperature also controls the peak frequency. Therefore the scale of the new physics can be directly linked to the peak frequency. As each gravitational wave detector probes a different frequency, each probes a different scale of physics \cite{Grojean:2006bp}. At the very low frequency one has pulsar timing arrays which probes phase transitions at the sub GeV scale. Lisa probes the electroweak phase transition and LIGO/VIRGO as well as KAGRA probes a scale of around $10^7$ GeV. KAGRA will soon be online and is expected to break a degeneracy in testing polarization \cite{Hagihara:2018azu}. The precise scale of new physics that is probed depends on the thermal parameters and varies for the soundwave, turbulent and collision contributions to the total spectrum. In figure \ref{Fig:GWscale} we show the scales probed for $\alpha=1$, $v_w=1$ and $\beta/H^* =1.3\log[T_*/M_{\rm P}$ for all three contributions. A phase transition with a peak frequency visible by LIGO/VIRGO can be motivated by vacuum stability \cite{Balazs:2016tbi}, split supersymmetry \cite{Demidov:2017lzf}, a Pati-Salam transition \cite{Croon:2018kqn} or neutrino masses \cite{Okada:2018xdh,Brdar:2018num}. Lisa probes both the electroweak phase transition \cite{Ashoorioon:2009nf,Huang:2015izx,Cao:2017oez,Huang:2017rzf,Chao:2017vrq,Bian:2017wfv,Alves:2018oct,Chao:2014hya,Huber:2007vva,Figueroa:2018xtu,Chala:2018ari,Matsui:2017ggm,Weir:2017wfa,Leitao:2012tx,Apreda:2001tj,Vaskonen:2016yiu,Beniwal:2017eik,Chiang:2017nmu,Kakizaki:2015wua,Huang:2016odd}, dark phase transitions \cite{Jaeckel:2016jlh,Baldes:2017rcu,Croon:2018erz,Schwaller:2015tja,Baldes:2017ygu,Hashino:2018zsi,Huang:2017kzu,Madge:2018gfl} other low scale symmetry breaking \cite{Chao:2017ilw,Huang:2018aja,Huang:2017laj,Marzola:2017jzl}, multistep transitions \cite{Matsui:2017ggm} and multistep phase transitions \cite{Matsui:2017ggm,Croon:2018new} whereas pulsar timing arrays can probe supercooled electroweak phase transitions and the QCD phase transition \cite{Chen:2017cyc,Ahmadvand:2017tue,Ahmadvand:2017xrw}. To probe the scale in between Lisa and LIGO/VIRGO, several other experiments have been proposed  including Magis \cite{Graham:2017pmn}, BBO \cite{Grojean:2006bp} and Decigo \cite{Kawamura:2006up}. \par 
 Beyond the scale of new physics more information can be garnered from the combined spectrum. Fig. \ref{Fig:3peakGW} shows the combined spectrum against Lisa sensitivity curves. Note that the combined gravitational wave spectrum does not necessarily look like a multipeaked spectrum, instead one might see a shoulder where the power law is broken away from the absolute peak. If the peak frequency and amplitude from any two of the peaks can be both detected and discerned from the background, one has four parameters from which one can in principle reconstruct the four thermal parameters. Comparing this to the simplest extension of the standard model - a real singlet extension - even a reconstruction of the four thermal parameters is a mapping of 5 free Lagrangian parameters to 4 thermal parameters. Moreover, one cannot gaurantee which scalar extension is responsible for the phase transition without complimentary collider searches probing the same scale.  Even still, recent work by \cite{Croon:2018erz} showed a non-trivial level of model discrimination for a generic dark Higgs with an SU(N) gauge symmetry. They mapped the thermal parameter space for different rank groups with and without the introduction of non-renormalizable operators and strongly coupled fermions. Unsurprisingly there was significant overlap between different models. Nonetheless there is nontrivial model discrimination as can be seen in Fig. \ref{Fig:GWmodel}. \par
 In the case of multistep phase transitions, one can have a striking signal of having more than three peaks which may overlap \cite{Matsui:2017ggm}. That is, for example, the sound wave contribution from one phase transition may have a higher peak frequency than the collision term of the phase transition that occurs at a higher scale. Remarkably, it appears to be possible that for the case where a phase transition occurs very slowly, even more than 6 peaks are possible as bubbles of a new phase can nucleate both in the high and intermediate temperature vacuum. The viability of such a scenario may depend on the precise details of reheating and a precise numerical simulation is yet to be attempted, but a cursory calculation indeed gives an intriguing signature which can in principle be discerned from both single and consecutive transitions \cite{Megevand:2007sv,Croon:2018new}.  

More information about the underlying physics that produced a primordial gravitational wave signal can be gleaned from measuring primordial anisotropies that result from a strongly first order phase transition. Work by \cite{Geller:2018mwu} analysed  phase transitions occurring between $1-1000$ TeV and demonstrated that we will obtain new anisotropies that can affect the CMB. One can then check to see if it is a dark sector or visible sector phase transition by checking correlations of $\delta \rho/\rho$ with the CMB. If $\delta \rho /\rho$ is uncorrelated with the CMB one knows that the Universe had a dark sector phase transition.



\subsection{Baryogenesis}
A triumph of modern cosmology is that two different measurements of the baryon to entropy ratio have concordance \cite{Canetti:2012zc}. The first is through BBN constraints where deuterium in particular is sensitive to the initial ratio of the baryon to entropy density \cite{Agashe:2014kda},
\begin{equation}
    Y_B=\frac{n_b-\bar{n}_b}{s}=7.3 \pm 2.5 \times 10^{-11}\,.
\end{equation}
Furthermore Planck measurements of oscillations in the CMB power spectrum give an overlapping estimate of the baryon yield \cite{Ade:2015xua}
\begin{equation}
    Y_B=8.59 \pm 0.11 \times 10 ^{-11} \,.
\end{equation}
This is unlikely to be an initial condition in any cosmology involving inflation. Although there exists, in the authors words \cite{Krnjaic:2016ycc}, an ``ugly and inelegant'' exception, inflation tends to wash out any initial baryon asymmetry. To produce a baryon asymmetry in a CPT conserving theory one needs to satisfy three conditions known as the Sakharov conditions \cite{Sakharov:1967dj}\footnote{There are models that violate CPT and achieve succesful baryogenesis without fulfilling the Sakharov conditions \cite{Cohen:1987vi,Davoudiasl:2004gf,Kusenko:2014lra,Yang:2015ida}}
\begin{itemize}
    \item[1] C and CP violation (one or the other is insufficient)
    \item[2] Violation of baryon number conservation
    \item[3] a departure from thermal equilibrium.
\end{itemize}
Electroweak baryogenesis  \cite{Kuzmin:1985mm,Shaposhnikov:1987tw,McLerran:1990zh,Farrar:1993sp,Rubakov:1996vz,Morrissey:2012db,White:2016nbo,Cohen:1993nk,Cohen:1990it,Cline:2006ts}  generates this during the electroweak phase transition where topological processes known as sphalerons efficiently produce both baryons and anti-baryons in the symmetric phase. If the electroweak phase transition is strongly first order, CP violating interactions with bubbles of electroweak broken phase biases the sphalerons to produce more baryons than anti baryons. As the bubbles of broken phase expands, some of the net asymmetry is swept up into the low temperature phase and makes up the present asymmetry. If the electroweak phase transition is strongly first order, the initial baryon asymmetry produced during the transition will not be washed out by the very sphalerons which formed them. \par 
The standard model fails to produce a sufficiently large baryon asymmetry. The standard model falls short on two Sakharov conditions, for a Higgs mass of $125$ GeV the departure from equilibrium is too weak as the electroweak transition is actually a crossover transition. Furthermore, the CP violation in the CKM matrix is far too feeble to sufficiently bias the electroweak sphalerons. Therefore if electroweak baryogenesis is part of our cosmic history, one needs to extend the standard model to accommodate both Sakharov conditions. The required extensions to the standard model are in principle probable by experiment with both particle colliders and gravitational wave observatories probing the ingredients for a strongly first order electroweak phase transition while searches for permanent electric dipole moments probe sources of CP violation. The fact that electroweak baryogenesis is both testable and minimal makes it one of the most attractive paradigms. \par 
Calculating the baryon asymmetry during a cosmic phase transition is a difficult problem. One usually calculates the overall left handed number density produced through CP violating interactions and then assume those processes are fast compared to weak sphaleron processes. In this case one can uncouple the dynamics of the baryon asymmetry production from the dynamics of the production of a chiral asymmetry. In this case the baryon asymmetry is given by \cite{Lee:2004we}
\begin{equation}
    D_Q \rho _B^{\prime \prime} (z)-v_w \rho _B ^\prime (z)-\Theta (z) {\cal R} \rho _B= \Theta (-z) \frac{n_F}{2} \Gamma _{ws}n_L (z) 
\end{equation}
Solving the above equation one finds that the baryon asymmetry is proportional to the sphaleron rate divided by the entropy which is the same order of magnitude of the observed baryon asymmetry. Therefore electroweak baryogenesis naturally produces the correct order of magnitude for the baryon asymmetry. The more challenging task is calculating $n_L$ which is the result of solving multiple coupled Boltzmann equations. The challenge in solving such equations lies in the fact that the mass basis evolves with both space and time during the phase transition. It is therefore customary to follow one of two approximate treatments: the first a semi-classical treatment using WKB methods \cite{Cline:2000nw,Cline:1997vk}. The second is known as the vev-insertion framework where one makes the assumption that the bulk of baryon production occurs immediately outside the bubble wall where vev is small so we can use the degrees of freedom and mass basis of the symmetric phase \cite{Lee:2004we}. The vev insertion paradigm utilizes the closed time path formalism and captures resonance and memory effects which can substantially boost the overall asymmetry and has the advantage that it can be solved analytically \cite{Lee:2004we,White:2015bva}.  The vev insertion paradigm neglects flavour oscillation effects which can dampen the resonance \cite{Cirigliano:2009yt,Konstandin:2005cd,Konstandin:2005cd}. Including gradient effects appears to recover some of the resonance \cite{Cirigliano:2011di}. When various approaches to calculating CP violating sources is valid remains an open problem in the field \cite{Konstandin:2013caa}. \par 
Since the standard model fails on two accounts to satisfy the Sakharov conditions, it is typical to extend the standard model by two sectors - one sector which catalyzes the electroweak phase transition, and another which is responsible for CP violating interactions with the bubble wall. If both sectors are heavy compared to the weak scale then one can in principle use an effective field theory approach \cite{deVries:2017ncy,Balazs:2016yvi}. More common is to look at the case where the new physics sectors are weak scale themselves. For example, in the MSSM, if one had a stop lighter than the standard model top it could catalyze a strongly first order electroweak phase transition. The CP violation can then occur through stop-Higgs interactions or gaugino-Higgsino-Higgs interactions \cite{Lee:2004we,Chung:2009qs,Cirigliano:2006wh,Chung:2009cb,Cirigliano:2006dg,Menon:2009mz,Cirigliano:2009yd,Li:2008ez,Kozaczuk:2011vr}. The existence of colored scalars in the plasma also provide substantial drag on the bubble wall making the wall velocity naturally small which tends to make baryon production more efficient (though also makes the gravitational waves from the electroweak phase transition less visible). Unfortunately the light stop mechanism for catalyzing the electroweak phase transition is in serious conflict with collider constraints \cite{Liebler:2015ddv}. Indeed the EWBG within the MSSM was starting to look unviable even in the early LHC era \cite{Curtin:2012aa}. Furthermore, EDM limits make both sources of CP violation severely constrained. Extending the MSSM by a gauge singlet (that is the NMSSM), one can catalyze a strongly first order electroweak phase transition with the additional scalar singlet \cite{Pietroni:1992in,Davies:1996qn,Carena:2011jy,Menon:2004wv,Huber:2006wf,Balazs:2013cia,Kozaczuk:2014kva,Bian:2017wfv,Huber:2006ma} and the source of CP violation can be Singlino - Higgsino - Higgs interactions \cite{Cheung:2012pg,Akula:2017yfr}. Alternatively one can extend the MSSM by effective operators that catalyze the CP violation \cite{Blum:2010by}. It is worth commenting that the minimal model of baryogenesis probably requires two additions to the standard Model to be viable - an addition that provides a source of CP violation and a source that catalyses a strong first order electroweak phase transition. Some examples of such minimal models include the standard model with an CPV effective operator and the addition of an effective operator \cite{deVries:2018tgs} or an additional scalar \cite{Cline:2012hg} to catalyze the transition. Alternatively it has been shown that the addition of two additional fermions is sufficient \cite{Egana-Ugrinovic:2017jib}. \par

Within the minimal supersymmetric standard model (MSSM) and 2HDM (Higgs doublet model) using the vev insertion frame work, one finds that the strength of CP violating sources for tree level interactions with the bubble wall are suppressed by a factor of $\Delta \beta \sim 10^{-2}$ where $\tan \beta (z)$ is the space time dependent ratio of the vevs $v_u(z)/v_d(z)$ and $\Delta \beta$ is its maximal variation. A study of the NMSSM showed that the addition of a gauge singlet can boost $\Delta \beta$, and therefore the baryon asymmetry, by an order of magnitude \cite{Kozaczuk:2014kva}. By contrast, if CP violation is a loop effect (for example the term $H f_R \bar{f}_L(a+\frac{b}{\Lambda ^2} |H|^2)$ can contain a relative phase), one no longer has a $\Delta \beta$ suppression but instead supressed by a factor $v^2/\Lambda^2$. Therefore the scale of CPV physics can be reasonably high. Furthermore, tree level CP violating interactions result in a baryon asymmetry that is essentially independent of the bubble wall width in contrast to the case where the CP violation is loop induced where a strong dependence on the bubble wall width results. Finally we note that the electroweak phase transition need not be weak scale. Indeed if the phase transition proceeds through a multistep procedure either through an intermediate transition that breaks another symmetry \cite{Ramsey-Musolf:2017tgh}, or through a two step electroweak phase transition \cite{Inoue:2015pza}, the scale of new physics required can be at the multi-TeV level and are best probed by gravitational wave observers and future colliders.

Outside of supersymmetry, Baryogenesis can also be linked with the production of dark matter \cite{Petraki:2011mv,Cline:2017qpe,Cline:2013bln,Cline:2012hg,Fornal:2017owa,Balazs:2004ae,Menon:2004wv,Carena:2011jy,Kozaczuk:2012vx} and has been explored in extended scalar sectors \cite{Huber:2000mg,Cline:2011mm,Cline:2017qpe,Cline:2013bln,Cline:2012hg} and other low scale phase transitions \cite{Long:2017rdo}. It has also been proposed that baryogenesis occurs spontaneously during the electroweak transition \cite{Cohen:1991iu}. One can also use CP violation in the lepton sector to produce enough baryon asymmetry \cite{Guo:2016ixx}. We end this section by noting that even if the baryon asymmetry is produced through leptogenesis, it may still involve a phase transition \cite{Pascoli:2018cqk}.\\

\noindent
{\bf Acknowledgements:}  We would like to thank Mark Hindmarsh, Djuna Croon, Pratika Dayal, Ruth Durrer, Hiren Patel, Michael Ramsey Musolf, 
Jim Cline, Lachlan Morris and Jonathan Kozaczuk. AM's research is financially supported by Netherlands Organisation for Scientific Research (NWO) grant number 680-91-119.
Triumf receives federal funding via a contribution agreement with the National Research Council of Canada. This work was performed in part at Aspen Center for Physics, which is 
supported by National Science Foundation grant PHY-1607611.
\\
\\

\bibliographystyle{mybibstyle}

\bibliography{references}

\end{document}